\documentclass[12pt]{article}
\pdfoutput=1
\usepackage{dsfont}
\usepackage{graphicx}
\usepackage{amsfonts}
\usepackage{amssymb}
\usepackage{amsmath}
\usepackage{bbold}
\usepackage[mathscr]{eucal}
\usepackage{booktabs}

\def\href#1#2{#2}   
\newif\ifdraft
\draftfalse	  
\reversemarginpar   
\makeatletter
\let\mlabel=\label
\let\adkendequation=\endequation%
\def\endequation{\adkendequation\adklabel\global\@ignoretrue}
\let\adkendeqnarray=\endeqnarray%
\def\endeqnarray{\adkendeqnarray\adklabel\global\@ignoretrue}
\newbox\marglabbox
\def\adklabel{\ifvoid\marglabbox\else\marginpar{\unhbox\marglabbox}\fi}
\def\label#1{\ifdraft\ifmmode%
  \global\setbox\marglabbox=\hbox{\hfill\fbox{\tiny\verb*~#1~}}%
  \else\ifinner\else\marginpar{\hfill\fbox{\tiny\verb*~#1~}}%
  \fi\fi\fi \mlabel{#1}}
\makeatother
\ifdraft%
\fi
\def\bb{\mathbb}
\def\eusm{\mathscr}
\font\twelvefrak=eufm10 scaled 1200
\font\tenfrak=eufm10
  \newfam\frakfam
  
  \textfont\frakfam=\twelvefrak
  \scriptfont\frakfam=\tenfrak
  \scriptscriptfont\frakfam=\scriptfont\frakfam
\def\sqr#1#2{{\vcenter{\hrule height.#2pt
   \hbox{\vrule width.#2pt height#1pt \kern#1pt
      \vrule width.#2pt}
   \hrule height.#2pt}}}

\def\bsqr#1#2{{\vrule width #1pt height#2pt}}
\def\bsquare{{\mathchoice\bsqr66\bsqr66\bsqr33\bsqr33}}
\def\badbreak{\penalty1000}

\def\Trb{\mathop{\rm Tr}}		    
\def\floor{\mathop{\rm floor}}		 
\def\union{\cup}                                  
\def\intersection{\cap}                         
\def\N{{\bb N}}				    
\def\Z{{\bb Z}}				    
\newcommand{\muhat}{{\hat \mu}}         
\newcommand{\gmu}{\gamma_{\mu}}             
\newcommand{\gfive}{\gamma_{5}}        
\newcommand{\cC}{{\cal C}}                    
\newcommand{\cP}{{\cal P}}                    
\newcommand{\psibar}{{\bar\psi}}             
\def\Reff{{\cal R}}                                        
\def\rer{\epsilon}                                         
\def\Rer{\eusm E}                                       
\def\aer{{\delta}}                                          
\def\Aer{\Delta}                                            
\def\Aero{\Aer_0}                                         
\def\Prf{B}                                                    
\def\rc{{\rho}}                                                
\begin{document}

\begin{center}
{\Large{\bf Locality and Efficient Evaluation of Lattice Composite}} \\
\vspace*{.14in}
{\Large{\bf Fields: Overlap-Based Gauge Operators}} \\
\vspace*{.24in}
{\large{Andrei Alexandru$^1$ and Ivan Horv\'ath$^2$}}\\
\vspace*{.24in}
$^1$The George Washington University, Washington, DC, USA\\
$^2$University of Kentucky, Lexington, KY, USA

\vspace*{0.25in}
{\large{Dec 28 2016}}

\end{center}

\vspace*{0.05in}

\begin{abstract}

\noindent
We propose a novel general approach to locality of lattice composite fields, which in case
of QCD involves locality in both quark and {\em gauge} degrees of freedom.  The method 
is applied to gauge operators based on the overlap Dirac matrix elements, showing for 
the first time their local nature on realistic path-integral backgrounds. The framework entails 
a method for efficient evaluation of such non-ultralocal operators, whose computational cost is 
volume-indepenent at fixed accuracy, and only grows logarithmically as this accuracy 
approaches zero. This makes computation of useful operators, such as overlap-based 
topological density, practical. The key notion underlying these features is that of exponential 
insensitivity to distant fields, made rigorous by introducing the procedure of 
{\em statistical regularization}. The scales associated with insensitivity property are useful 
characteristics of non-local continuum operators.

\end{abstract}

\vspace*{0.04in}

\section{Introduction}
\label{sec:intro}

Local quantum field theories are quantum field systems with dynamics  
prescribed by local action density. When such theories serve to describe 
particle interactions then, in addition, $n$-point correlation functions 
of local operators encode most of the interesting observables. The notion 
of a {\em local operator} is thus deeply engrained in these descriptions. 

Locality is rarely viewed as problematic or subtle in formal continuum 
considerations. Indeed, the presence of space-time derivatives invokes 
confidence that field variables separated by non-zero distance are not 
explicitly coupled by the operator, consistently with the intuitive meaning 
of locality. However, the concept becomes richer once an actual definition 
of the theory, such as via lattice regularization which we follow 
here, is carried out.

A common approach to formulating lattice-regularized systems is to replace
space-time field derivatives with nearest-neighbor field differences. More 
generally, operators that only depend on field variables within fixed 
lattice distance away from each other are referred to as {\em ultralocal}.
However, lattice operators with couplings extending to arbitrary distances  
naturally arise in Wilson's renormalization group considerations. Moreover, 
chirality-preserving Dirac operators of Ginsparg-Wilson type~\cite{Gin82A} 
are all of such {\em non-ultralocal} variety~\cite{Hor98A,Hor00A}. Locality 
becomes a more subtle notion in these situations, and requires some care.

In this work, we consider non-ultralocal operators associated with overlap 
Dirac matrix $D \equiv \{ D_{x,y}(U)\}$~\cite{Neu98BA} in the context of QCD. 
Here the color-spin indices are implicit and $U \equiv \{U_{x,\mu}\}$ is the 
SU(3) lattice gauge field. There are at least two relevant circumstances 
to consider. Firstly, since $\sum_{x} \psibar_x (D \psi)_x$ prescribes 
interactions of quarks and gluons, it is required to be a sum of local 
contributions. Operator $(D \psi)_x$ thus has to be local with respect to both 
fermionic and gauge variables. Secondly, there are interesting gauge operators 
based on overlap matrix elements, such as topological charge 
density~\cite{Has98A,Nar95A}, gauge action density~\cite{Hor06C,Hor06D,Ale08A} 
or gauge field strength tensor~\cite{Hor06C,Liu07A} constructed from 
$D_{x,x}(U)$. To be used in well-founded QCD calculations, these objects need 
to be local in gauge potentials.

While fermionic locality of the overlap operator was studied in some
detail~\cite{Her98A}, the aspects of gauge locality have barely been considered.
In fact, they were not studied at all for realistic gauge fields of lattice 
QCD ensembles. The theme of the present work is to examine this issue, especially 
in relation to the above non-ultralocal gauge operators. The novelty of our
approach is that it naturally connects locality of an operator to efficiency 
achievable in its evaluation. Consequently, the results that follow have direct 
bearing on the practical use of these computationally demanding objects.

To start describing our approach, recall that the modern notion of locality
for lattice-defined operators includes their exponentially decreasing sensitivity 
to distant field variables. Standard treatment formalizes this into exponential 
bound on the corresponding field derivatives. For example, fermionic locality 
of $(D \psi)_x$ then simply requires sufficiently fast decay of $D_{x,y}$ 
as $y$ is taken increasingly far away from $x$. Albeit less elegant due 
to gauge field entering in a more complicated manner, this prescription can 
also be followed to study gauge locality of $(D \psi)_x$ or locality of 
$D_{x,x}(U)$.

However, for our purposes it is fruitful to replace the above ``differential'' 
treatment of dependence on distant fields with a direct ``integral'' approach. 
In other words, we ask how well is it possible to know the value of a composite 
operator $O_x$ when the knowledge of fundamental fields is restricted to some 
neighborhood of $x$. Exponentially suppressed sensitivity to distant fields is 
then formalized as the existence of estimates whose precision exponentially 
improves with the linear extent of these neighborhoods. 

\begin{figure}[t]
\begin{center}
    \centerline{
    \hskip 0.00in
    \includegraphics[width=6.2truecm,angle=0]{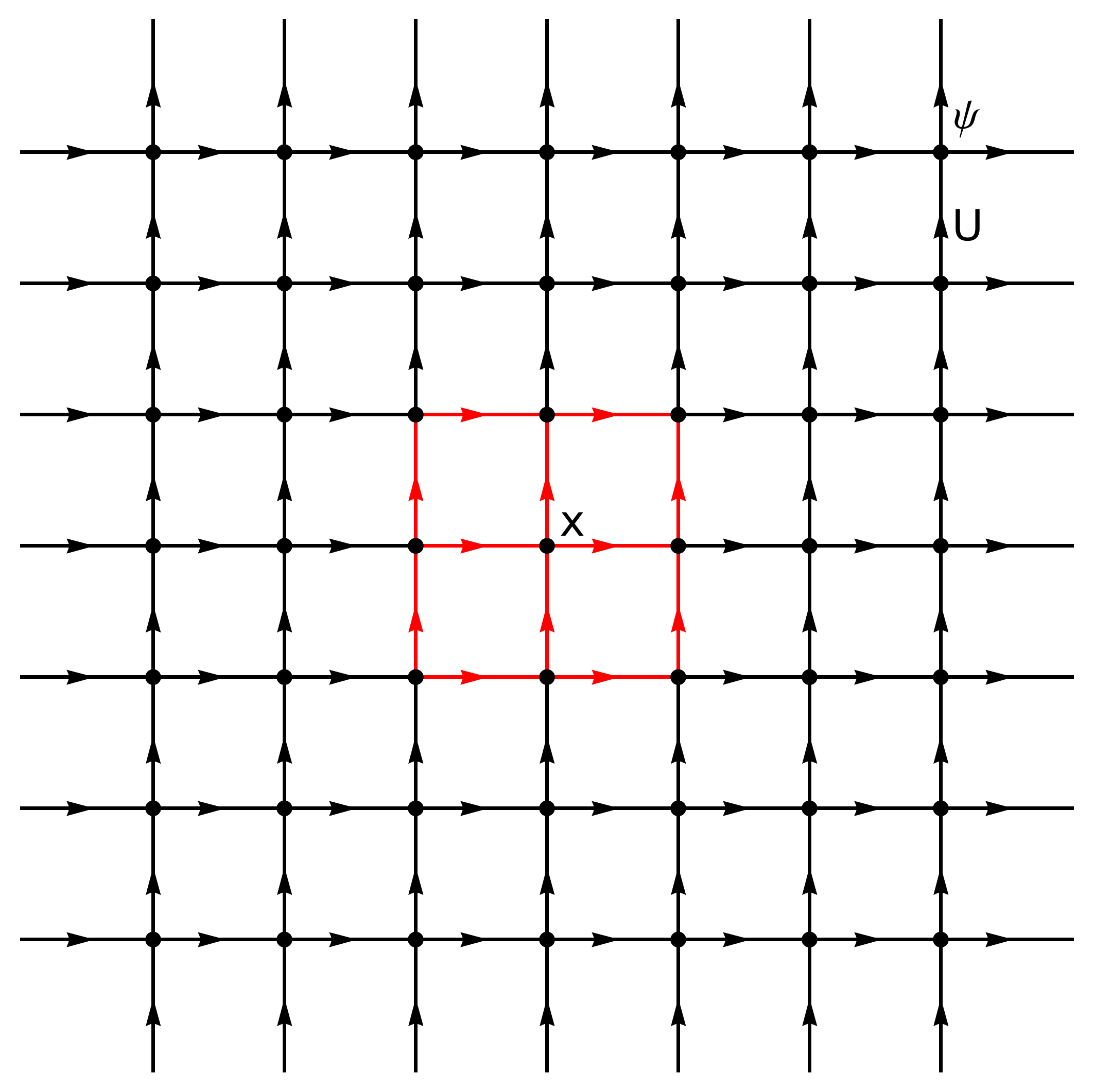}
    \hskip 0.50in
    \includegraphics[width=6.2truecm,angle=0]{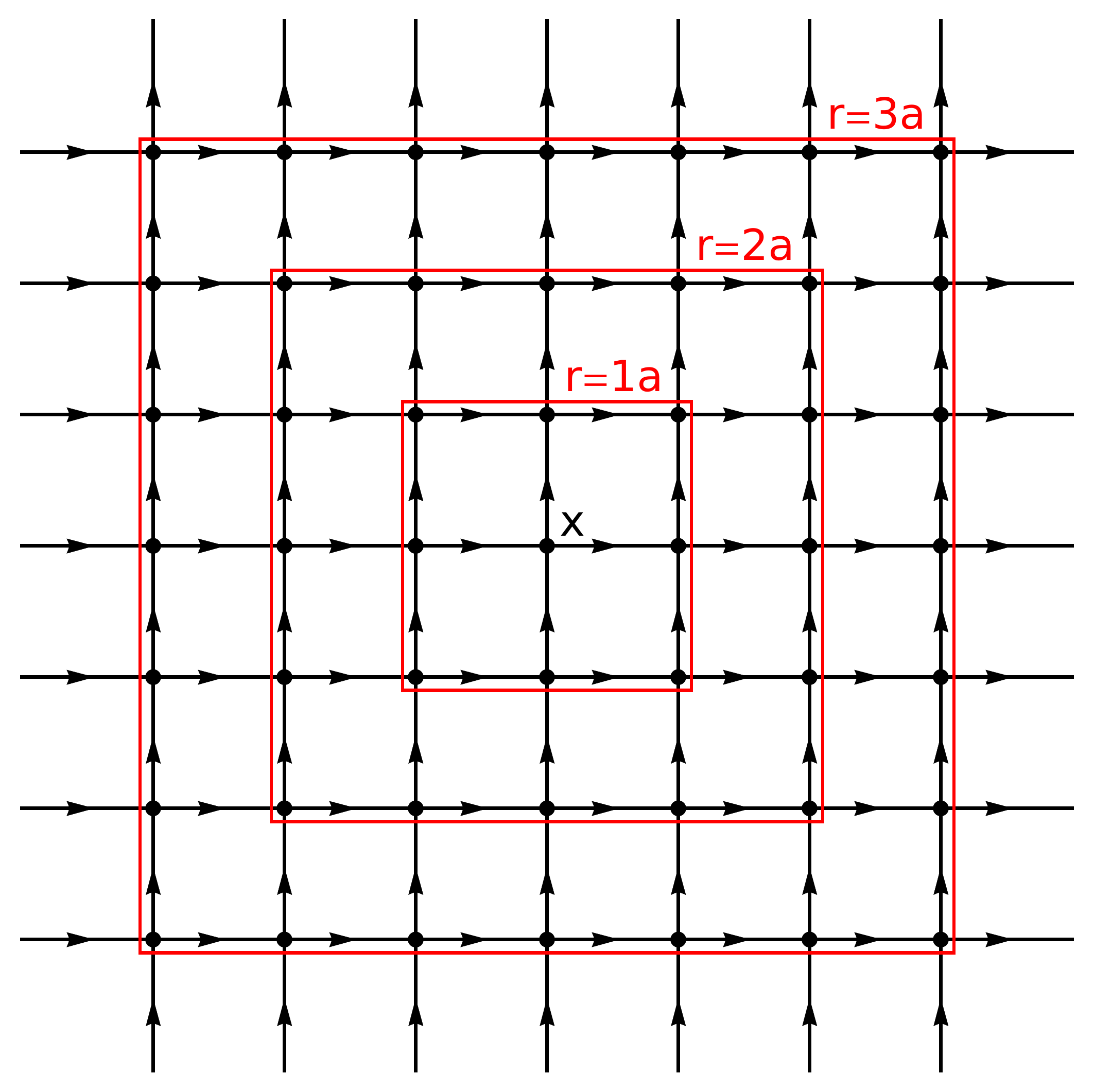}
     }
     \vskip -0.0in
     \caption{Gauge field patch involved in simplest gauge covariant
operators $O_x$ (left) and the geometry of hypercubic neighborhoods with 
radius $r$ (right).}
     \label{fig:setup}
    \vskip -0.40in
\end{center}
\end{figure}

To explain this in more detail, consider operators $O_x$ that only depend 
on the gauge field\footnote{This is the context for which we develop the method
in detail here. Including other fundamental fields is conceptually analogous 
with specifics, especially as it relates to fermions, forthcoming.}, such as covariant 
discretizations of $F\tilde{F}(x)$. 
The simplest ultralocal option is based on minimal gauge loops (plaquettes), 
but general $O_x(U)$ may couple fields everywhere. We will use hypercubic 
neighborhoods of $x$ with ``radius'' $r$ to define patches $U^{x,r} \subset U$ 
of the field, as illustrated in Fig.~\ref{fig:setup}. Note that $r \!=\! j a$ is discrete 
at finite lattice cutoff $a$. Faced with the task of estimating $O_x(U)$ given an 
incomplete knowledge ($U^{x,r}$) of its argument, one is led to construct 
approximants $O_x^r(U)=O_x^r(U^{x,r})$ depending only on variables in 
the patch. If $\aer(r,U) \equiv \| O_x (U) - O_x^r (U) \|$ denotes the associated 
error, one may choose to formalize {\em exponential insensitivity to distant fields} 
of $O_x$  by requiring the existence of $O_x^r$ and finite positive constants 
$\Prf$, $R$, $s$ such that
\begin{equation}
   \aer(r,U) \,\le\, \Prf  \exp\left( -r/R \,\right)  
   \quad\; , \;\quad  r \ge s
   \quad\; , \;\quad  \forall \,U   
   \label{eq:10}
\end{equation}
But the notion so construed is unnecessarily strong since violations of \eqref{eq:10} 
involving configurations $U$ that are statistically irrelevant in the path integral, are 
inconsequential in the context of corresponding quantum theory. We thus 
replace condition \eqref{eq:10} by 
\begin{equation}
   \aer(r,p) \,\le\, \Prf  \exp\left( -r/R \,\right)  
   \quad\; , \;\quad  r \ge s
   \quad\; , \;\quad  \forall \,p<1   
   \label{eq:12}
\end{equation}
where $\aer(r,p)$ is a statistical construct representing ``error with probability $p$", 
effectively demanding that the bound is satisfied up to events of probabilistic measure 
zero. In the resulting procedure of {\em statistical regularization}, the bound is examined
at fixed certainty $p$, and this cutoff is eventually lifted via the appropriate
$p \!\to\! 1$ limit (Sec.\ref{sec:insensitivity}). Note that tying exponential insensitivity 
of an operator to path integral in which it is used implies that minimal sensitivity range 
$R\!\to\!R_0$ can have non-trivial dependence on the lattice spacing.  Locality then requires, 
among other things, that $R_0(a)$ vanishes in the continuum limit.

The notion of exponential insensitivity, outlined above, admits arbitrary 
$O_x^r$ to be considered in Eq.~\eqref{eq:12}. However, if the property 
holds for $O_x$, the facilitating approximant is clearly not unique, and different 
options may involve widely varied degrees of computational complexity. 
In fact, albeit coupling fewer degrees of freedom, computational demands for 
most precise choices of $O_x^r$ can be as large or larger than those for $O_x$ 
itself. However, our goal here is not to determine the best approximation. 
Rather, we are interested in finding $O_x^r$ whose computational complexity 
scales with $r/a$ in qualitatively the same manner as that of $O_x$ 
with $L/a$. 
Here $L \!\gg\! r$ is the size of the system. Apart from demonstrating 
exponential insensitivity, the existence of such approximants would offer 
great computational advantage. Indeed, if the program computing 
$O_x$ has no a priori knowledge about insensitivity, then single 
evaluation incurs cost growing at least with the lattice 4-volume for 
non-ultralocal operators of interest here. This is reduced as
\begin{equation}
   \Bigl(\, \frac{L}{a} \,\Bigr)^{4+\alpha}  \quad \longrightarrow \quad
   \Bigl(\, \frac{R}{a} \, \log \frac{\Prf}{\aer} \Bigr)^{4+\alpha'}
   \quad , \quad \alpha, \alpha' \ge 0
   \label{eq:20}
\end{equation}
for the approximant that guarantees absolute precision $\aer$. Computation 
could thus be performed at a constant cost (independent of the volume) that 
only depends logarithmically on the desired precision. 

Our suggestion for constructing generic and practical approximants $O_x^r$ 
of the above type is to treat the neighborhood containing $U^{x,r}$ as 
a finite system of its own. Indeed, definition of non-ultralocal $O_x$ 
implicitly involves a sequence of operators: one for each space-time 
lattice involved in the process of taking the infinite-volume limit. Considering
$L^4$ lattices and making the $L$-dependence explicit for the moment 
($O_x \rightarrow O_{x,L}$), the replacement
\begin{equation}
   O_{x,L}(U)  \quad \longrightarrow \quad  O_{x,L\to 2r+a}(U^{x,r})
   \; \equiv \; O_x^r(U^{x,r})   \,=\, O_x^r(U)
   \label{eq:30}
\end{equation} 
offers a generic scheme for obtaining candidate approximants. Note that 
$L/a$ is defined to count the sites and hence $2r\!+\!a$ is the ``size" of 
the lattice system contained in hypercubic neighborhood with radius $r$. 
Variations on this prescription discussed in the body of the paper correspond 
to different treatment of boundaries in the subsystem associated with the patch. 
We refer to approximants of type \eqref{eq:30} as {\em boundary approximants} 
since they test sensitivity to the boundary created by the restriction 
$U \rightarrow U^{x,r}$. If they exhibit the behavior \eqref{eq:12}, then $O_x$ 
is exponentially insensitive to distant fields, and a stronger notion of locality 
({\em boundary locality}) can be built around this concept.

The paper is organized as follows. In Sec.~\ref{sec:insensitivity} we introduce 
the concept of exponential insensitivity to distant fields via statistical 
regularization. Given its pivotal role in the present discussion, this is carried
out  in detail so that relevantly distinct behaviors are discerned, and the 
subtleties known to us are all accounted for. A notable feature of the resulting 
framework is that the removal of lattice and statistical cutoffs necessitates 
not only a non-divergent exponential range $R_0$, but also a non-divergent 
auxiliary (non-unique) scale $\rc_0$ representing the threshold distance for 
validity of the bound. This part concludes with connecting the locality to exponential 
insensitivity and defining it correspondingly. In Sec.~\ref{sec:boundary} the stronger 
notions of boundary insensitivity and boundary locality are put forward, 
emphasizing their practical relevance. In particular, the consequences of this property
for efficient evaluation of computable non-ultralocal operators is discussed in detail.
The possibility that suitably constructed boundary approximant can also serve as 
a standalone ultralocal operator, interesting in its own right, is suggested here as well. 
In Sec.~\ref{sec:overlapops} we investigate the properties of basic overlap-based 
gauge operators in the proposed framework. Our numerical results in pure glue theory 
readily show the weak form of insensitivity (for any fixed statistical cutoff) at the lattice 
level, and the corresponding 
weak form of  locality in the continuum. They also lend an initial support to full 
insensitivity (statistical cutoff removed) sufficiently close to the continuum limit, and 
the associated locality.\footnote{The precise meaning of these qualifications on 
insensitivity/locality is given in the body of the paper.} To illustrate the practical 
aspects of exponential insensitivity, we discuss in Sec.~5 the use of boundary
approximants for efficiently computing the ``configurations" of overlap-based 
topological density. This lattice topological field was crucial
for identifying the low-dimensional long-range topological structure in QCD 
vacuum~\cite{Hor03A}. The insensitivity properties tested here make large scale 
computations of this type practical since, at fixed accuracy, the cost per configuration
is simply proportional to the volume, similarly to the case of generic ultralocal 
gauge operators.

\vfill\eject

\vspace*{0.08in}

\section{Exponential Insensitivity to Distant Fields}
\label{sec:insensitivity}

The concept of exponential insensitivity to distant fields is central 
to this work, and we thus begin by working out the necessary details. 
The issues in need of attention arise mostly due to the quantum setting 
we are dealing with. To see this, consider some extended operator 
(functional) $O_x[A]$ of gauge fields $A \!\equiv\! \{ A_\mu(y) \}$ 
in the continuum space-time. 
The analog of Eq.~\eqref{eq:10}, involving spherical neighborhoods 
of $x$ in Euclidean space, is intended to characterize exponential 
insensitivity of $O_x$. However, the required bound may not hold, 
or even be meaningful, for arbitrary $A$, and yet be satisfied 
by a subset of fields relevant to the situation at hand. Indeed,
the proper definition requires specifying the class of fields in 
question. Problems involving classical dynamics of $O_x$ naturally 
come with needed analytic restrictions since they deal with fields 
$A$ obeying classical equations of motion.\footnote{Mere definition 
of $O_x[A]$ directly in the continuum often forces $A$ to satisfy 
certain analytic properties.}

However, in field theory regularized and quantized via lattice path 
integral, there is no a priori restriction on the fundamental fields 
(``configurations'') $U$: there is only a hierarchy induced by their 
statistical weights. While this forces one to adopt the notion of 
exponential insensitivity involving all lattice fields in principle,
it also makes room for the concept to be viable even when there 
are sufficiently improbable configurations violating any exponential 
bound. Indeed, we are led to a statistical approach respecting the path 
integral hierarchy of fields and, at the same time, facilitating the systematic 
separation of potential outliers in the process we refer to as
{\em statistical regularization}. Its idea is to replace the sharp construct 
of least upper bound with the statistical 
``least upper bound with probability $p$''. Before defining the concept
in Sec.\ref{ssec:statreg}, we need to discuss some preliminaries.

\subsection{The Setup}

Unless stated otherwise, the position coordinates of an underlying infinite 
system are spanned by entire $d$-dimensional hypercubic lattice with 
cutoff $a$, i.e.
\begin{equation}
    {\cal H}_\infty \equiv 
    \{\, x \mid \, x_\mu \!=\! a n_\mu \,,\, n_\mu \in \Z \,,\, \mu=1,2,\ldots,d \,\}
    \label{eq:32}
\end{equation}
The general case where infinite volume involves a subset of points in
${\cal H}_\infty$ is described in Appendix~\ref{app:A}, and allows e.g. for
applying our methods to arbitrary spatial geometry, or to theories at finite 
temperature. The hypercubic neighborhood of point $x$ with radius $r$ is 
\begin{equation}
    {\cal H}^{x,r} \equiv 
    \{\; y \in {\cal H}_\infty  \;\mid\; | x_\mu - y_\mu | \le r  \;,\; \forall \mu \;\}
    \qquad\qquad
     r \in \{\, ja  \,\mid \, j=1,2,3,\ldots \,\} \equiv \N_a    
    \label{eq:34}
\end{equation}

We consider operator $O_x\!=\!O_x(U)$, composed of gauge field 
$U \!\equiv\! \{ U_{y,\mu} \}$, in theory defined by $S(U)$, namely the
effective gauge action after integrating out quark fields, if any. This
quantum dynamics is infrared-regularized on symmetric lattices of $(L/a)^4$ 
sites centered around $x$, i.e. the center
of the lattice is at $x$ for $L/a$ odd, and at $x + (a/2,\ldots,a/2)$ for $L/a$ 
even. Each ${\cal H}^{x,r}$ contained within given finite system is assigned
a hypercubic field patch
\begin{equation}
    U^{x,r} \equiv \{\; U_{y,\mu} \in U \; \mid\; \; 
    y,y\!+\!\muhat \, \in \,  {\cal H}^{x,r}  \;\}\;,   \qquad\;  2r+a < L
    \label{eq:app:36}
\end{equation}
Note that $U^{x,r}$ doesn't include links ``dangling" with respect to 
${\cal H}^{x,r}$. For example, the set of red links in Fig.\ref{fig:setup} (left) 
is $U^{x,r=a}$. Any operator $O_x^r$ with range in the normed space of
$O_x$ and satisfying $O_x^r(U) \!=\! O_x^r(U^{x,r})$, will be referred to as 
{\em approximant} of $O_x$ with radius $r$.

\subsection{Kinematics of Exponential Bounds}
\label{ssec:kinematics}

Important technical aspect of the analysis that follows involves sufficiently 
detailed description of exponential bounds which we now discuss.  Generic 
real-valued ``error function"
\begin{equation}
     0 \le \aer(r)  <  \infty     \qquad,\qquad     r \in \N_a                        
     \label{eq:k10}
\end{equation}
is said to be {\em exponentially boundable} if there are finite positive 
$R$, $\Aer$, $s$ such that 
\begin{equation}
   \aer(r)  \,\le\,  \Aer \, \exp\left( -\frac{r-s}{R} \,\right)   
   \quad\; , \;\quad  r \ge s \in \N_a
   \label{eq:k20}
\end{equation}
Note that, in this form, the prefactor $\Aer$ specifies the bound at threshold distance 
$s$. Our goal is to identify a region in parameter space $R \!>\! 0$, $\Aer \!>\! 0$, 
$s \!\in\! \N_a$, if any,  where \eqref{eq:k20} holds. 

Explicit solution to this problem is obtained by rewriting \eqref{eq:k20} as
\begin{equation}
      \Aer  \;\ge\;  \exp\left( - s/R \right) \;\,
      \sup\, \{\, \aer(r) \, \exp\left( r/R \right)  \,\mid \, r \ge s \,\}  \;\equiv\; \Aero(R,s) 
     \label{eq:k30}
\end{equation}
Thus, for given $(R,s)$, the indicated range of valid $\Aer$ materializes, as long 
as $\Aero(R,s) < \infty$. At the same time, finiteness of $\Aero(R,s)$ only
depends on $R$. Indeed, $s \to s'$ induces at most a finite change in $\Aero$ 
since the sets involved in suprema only differ by finite number 
of finite elements. Consequently, the following defines an $s$-independent object
\begin{equation}
     R_0 \;\equiv\;  \inf\, \{\, R>0  \,\mid \, \Aero(R, s)  < \infty \,\}   \;\ge\; 0
     \label{eq:k40}
\end{equation} 
namely the {\em effective range} of $\aer(r)$. Exponential boundability of 
$\aer(r)$ is equivalent to $R_0 < \infty$, and the parameter domain of validity
for \eqref{eq:k20} is $R>R_0$, $s \in \N_a$, $\Aer \ge \Aero(R,s)$. We will
work with description of this domain in which the threshold error $\Aer$, rather 
than threshold distance $s$, enters as an unconstrained free parameter. 
Since $\Aero(R,s)$ is decreasing in $s$, and $\lim_{s\to\infty} \Aero(R,s)=0$, 
its ``inverse" defines the desired representation, namely
\begin{equation}
   R > R_0   \quad\qquad   \Aer > 0   \quad\qquad   s \ge s_0(R,\Aer)   
   \;\equiv\; \min\, \{\, r  \,\mid\,  \Aer_0(R,r)  \le \Aer \,\}
   \label{eq:k50}
\end{equation}
Function $s_0(R,\Aer)$ is a {\em core size} outside of which a bound with 
desired $(R,\Aer)$ sets in. 

This detailed kinematics of exponential bounds acquires relevance in quantum setting, 
where ultraviolet and statistical regularizations produce error functions depending on 
associated cutoffs. It turns out that monitoring cutoff dependence of $R_0$ alone is not 
sufficient to ensure requisite bounds upon regularization removal, and information in 
$s_0$ is also needed. To that end, it is useful to treat $s_0$ as a continuous entity 
(like $R_0$) so that trends can be detected even for small changes in the cutoffs. We 
thus extend $\Aero(R,s)$ at fixed $R$, into a decreasing continuous map $\Aero(R,\rc)$ 
from $\rc \!\in\! (0,\infty)$ onto $(0,\infty)$. Omitting the $R$-dependence, the exponential 
behavior of $\Aero(s)$ motivates a practical choice
\begin{equation}
    \log \Aero(\rc)  \equiv  \begin{cases} 
    \text{\tt Lin} \, \bigl( \, \log \Aero(s) \,,\, \rc \,\bigr) & \text{for }  \rc \geq a \\
     -1 + a/\rc  + \log \Aero(a) 
    & \text{for } \rc < a
  \end{cases} 
   \label{eq:k60}   
\end{equation}
where  {\tt Lin} $\!( f(s), \rc )$ denotes linear interpolation of $f(s)$ via variable $\rc$. 
The arbitrary completion for $\rc \!<\! a$ only serves to realize the desired range. 
With $\Aero(R,\rc)$ so fixed, the associated core-size function $\rc_0(R,\Aer)$ is 
uniquely defined via $\Aero(R,\rc_0) \!=\! \Aer$, guaranteeing that
\begin{equation}
   \aer(r)  \,\le\,  \Aer \, \exp\left( -\frac{r-\rc}{R} \,\right)   
   \quad\; \text{for} \quad\;  r \ge \rc
   \quad \quad  R > R_0
   \quad \quad  \Aer > 0      
   \quad \quad  \rc \ge \rc_0(R,\Aer)   
   \label{eq:k70}
\end{equation}

Note that setting $R\!=\!R_0$ to optimize the bound at large distances is not always 
possible since $\rc_0(R,\Aer)$ may diverge for $R \!\to\! R_0^+$. This occurs when 
$\aer(r)$ decays as an exponential modulated by an unbounded function, e.g.
$\aer(r) \propto r \exp (-r/R_0)$. Such cases require $R$ in \eqref{eq:k70} that 
may be arbitrarily close but larger than $R_0$, as indicated. However, setting
$\rc \!=\! \rc_0(R,\Aer)$, whenever finite, always produces a valid bound
\begin{equation}
   \aer(r)  \,\le\,  \Aer \, \exp\left( -\frac{r-\rc_0(R,\Aer)}{R} \,\right)   
   \qquad , \qquad  r \ge \rc_0(R,\Aer)
   \label{eq:k80}
\end{equation}
Vice versa, when $\rc_0(R,\Aer)$ is ill-defined (infinite), there is no exponential bound
involving range $R$ and threshold error $\Aer$. The core-size function is thus a master 
construct containing complete information on exponential bounds of $\aer(r)$.
The bounds in 2-parameter family \eqref{eq:k80} are {\em optimal} in that they cover 
the maximal range of distances for given $(R,\Aer)$. 
 
\subsection{Statistical Regularization}
\label{ssec:statreg}

Given a composite field $O_x(U)$ and its approximant $O_x^r(U)=O_x^r(U^{x,r})$, 
the properties of the latter are described by distribution $P_r(\aer)$ of its errors 
over the path integral ensemble. The corresponding cumulative probability 
function $F_r(\aer)$ is explicitly given by
\begin{equation}
   F_r(\aer) \equiv 
   \Big\langle \, 
   H \, \Big(\,\aer - \| O_x (U) - O_x^r (U) \| \,\Big)
   \,\Big\rangle   \qquad\qquad 
   P_r(\aer) \equiv \frac{d}{d\aer} F_r(\aer) 
   \label{eq:40}
\end{equation}
with $H$ denoting a Heaviside step function and $\langle \ldots\rangle$
the path integral average specified by $S(U)$. This information can 
be recast into error bounds $\aer(r,p)$ satisfied by fractions $p$ of the overall 
population with smallest deviations by imposing
\begin{equation}
   F_r \Big( \aer(r,p) \Big) \equiv p
   \label{eq:50}
\end{equation}
i.e. by inverting $F_r(\aer)$. By construction, the meaning of $\aer(r,p)$ 
is that of an ``error bound with probability $p$'':  if $O_x^r$ is used to 
estimate $O_x$, the expected error is less than $\aer(r,p)$ with probability 
$p$. This family of error functions is central to the analysis proposed here. 

\begin{figure}[t]
\begin{center}
    \centerline{
    \hskip 0.00in
    \includegraphics[width=7.0truecm,angle=0]{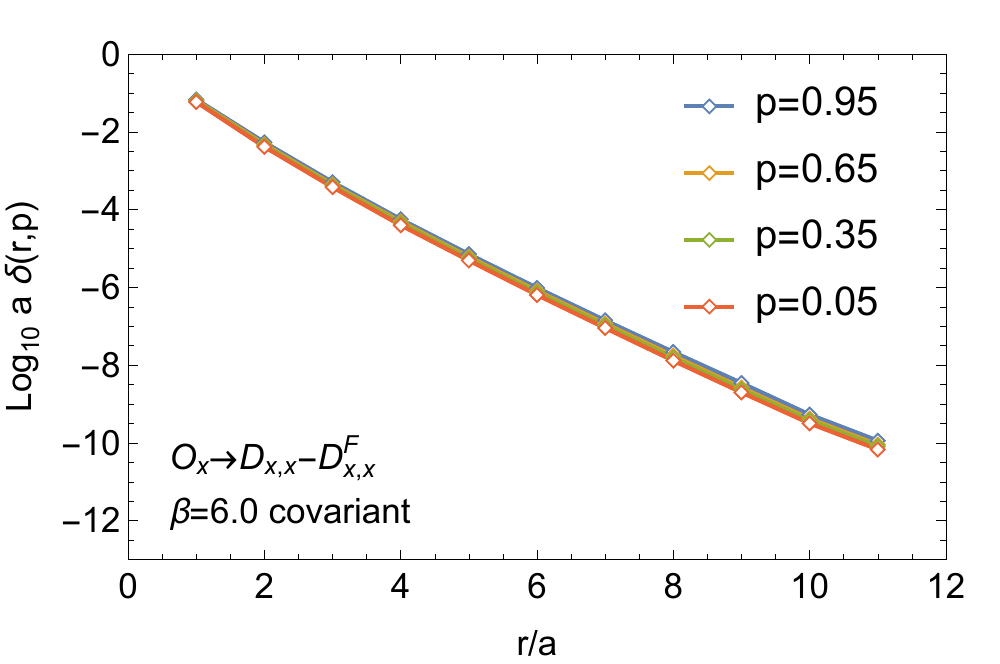}
    \hskip -0.00in
    \includegraphics[width=7.0truecm,angle=0]{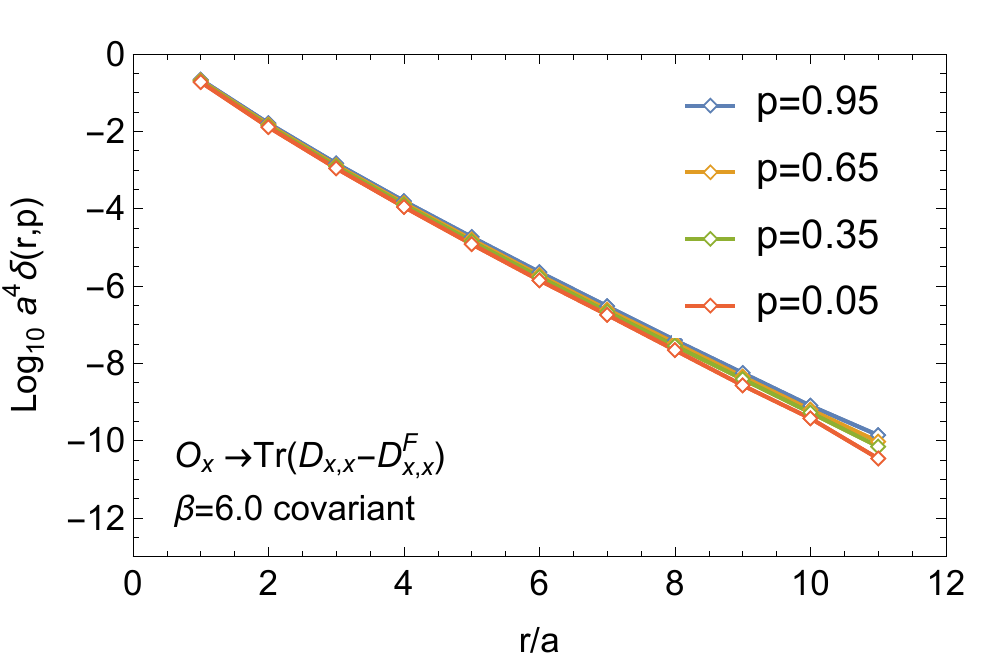}
     }
     \vskip -0.0in
    \centerline{
    \hskip 0.00in
    \includegraphics[width=7.0truecm,angle=0]{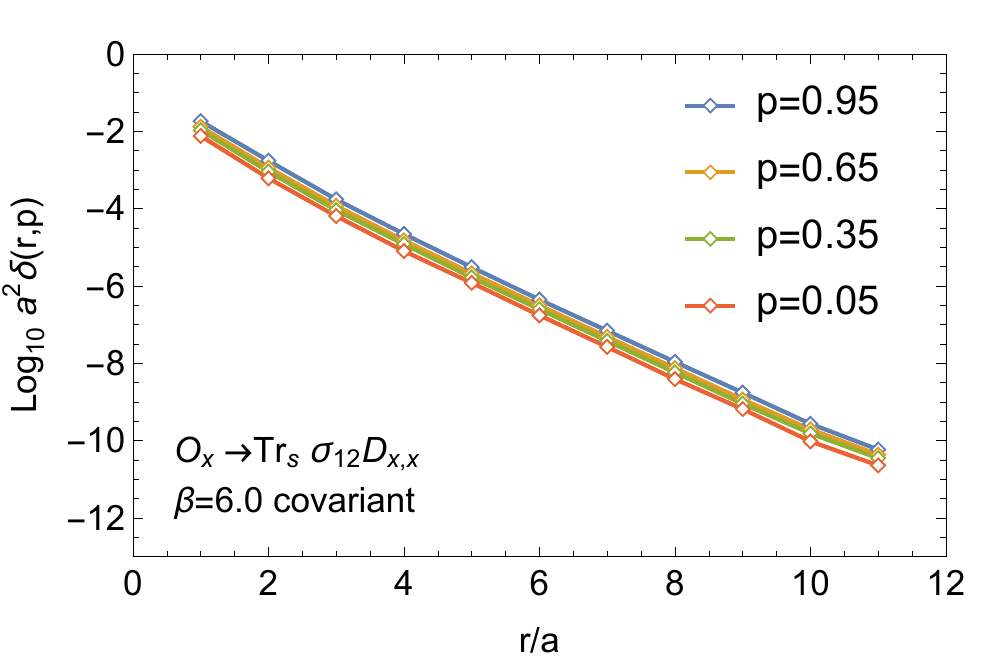}
    \hskip -0.00in
    \includegraphics[width=7.0truecm,angle=0]{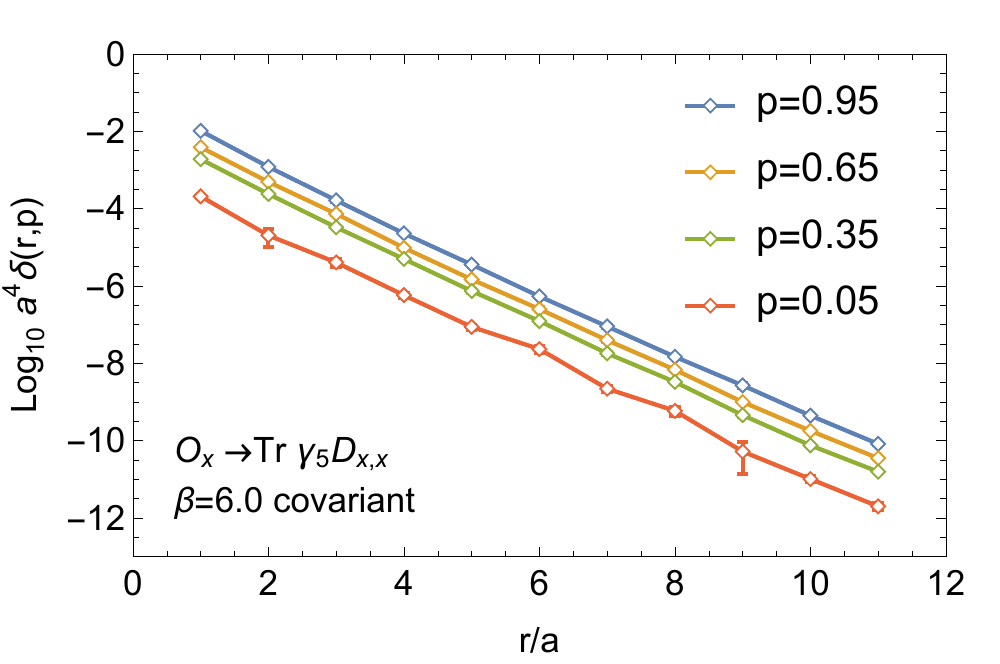}
     }
     \vskip -0.1in
     \caption{The behavior of $\aer(r,p)$ at fixed $p$ for overlap-based gauge operators 
     (Sec.$\,$\ref{sec:overlapops}). Note that $\Trb \gfive D_{x,x} \propto q(x)$ (topological 
     density) has the strongest dependence on $p$.}
     \label{fig:stat_reg}
    \vskip -0.40in
\end{center}
\end{figure}

For $O_x$ to exhibit exponential insensitivity to distant fields, statistically regularized by 
fixed $0\!<\!p\!<\!1$, we require that $\aer(r,p)$ decays at least exponentially at asymptotically 
large $r$, i.e. that it is exponentially boundable. The advertised ``separation of potential outliers'' 
is thus accomplished by fixing the degree of certainty $p$ in examining the influence of distant 
fields on error. Examples of measured $r$-dependencies for several overlap-based operators are 
shown in Fig.~\ref{fig:stat_reg}, with details described in Sec.~3. The observed behavior is clearly 
compatible with exponential falloff at rates depending very weakly (if at all) on $p$.

Exponential insensitivity at any non-zero $p$ is in itself (without $p\!\to\!1$ 
considerations) a notable feature of non-ultralocal lattice operator. Indeed, at minimum, it 
signals the existence of insensitive subpopulation in the path integral, which can be useful 
computationally and otherwise. We formalize this lattice concept as follows.

\vspace{0.15in}

\noindent {\bf Definition 1} {\sl (exponential insensitivity at fixed $p$)}

\smallskip
\noindent {\sl Let $O_x(U)$ be the operator with values in normed space and $S(U)$ 
the action of Euclidean gauge theory, both defined on hypercubic lattices of arbitrary 
size $L$. We say that $O_x$ is exponentially insensitive to distant fields with respect
to $S$ at probability 
$p$, if there is an $r$-dependent approximant $O_x^r(U)=O_x^r(U^{x,r})$ such that} 
\begin{description}
\item[{\sl (i)}] {\sl The infinite-volume limit 
$\aer(r,p) \equiv \lim_{L\to \infty} \aer(r,p,L)$ of its error function exists.}
\item[{\sl(ii)}] {\sl $\aer(r,p)$ is exponentially boundable.} 
\end{description}
{\sl Here $U^{x,r}$ is the patch of $U$ contained in hypercubic neighborhood of $x$ 
with radius $r$}. $\quad \Box$

\vspace*{0.15in}

\noindent This definition categorizes lattice operators at given position in terms 
of their dependence on remote fields.  Abbreviating exponential insensitivity to distant 
fields as ``insensitivity'', if there is no $p\!>\!0$ at which $O_x$ is insensitive, then 
$O_x$ is {\em sensitive}, while in the opposite case it is said to contain an 
{\em insensitive component}. When the latter holds for all $p\!<\!1$ then $O_x$ is referred 
to as {\em weakly insensitive} provided that the ``outliers" do not contribute finitely 
to $O_x$ in $p \!\to\! 1$ limit.\footnote{The second requirement bars a logical 
possibility that samples defying any exponential bound would finitely influence $O_x$ 
albeit forming a set of measure zero. The property expressing the absence of such 
singular behavior is formulated in Appendix~\ref{app:D} and will be referred to as 
{\em regularity}. It is automatically satisfied by bounded lattice operators with bounded 
approximants, such as those studied here.} 
Weakly insensitive operator is under exponential control with any
preset probability short of certainty which, among other things, can provide a powerful 
advantage for its evaluation. Finally, if removing statistical cutoff ($p\!\to\!1$) in weakly 
insensitive $O_x$ leaves some exponential bounds in place, we speak of
{\em insensitive} operator. However, the process of cutoff removal needs to be 
specified and discussed in some detail.

\subsection{The Removal of Statistical Cutoff}
\label{ssec:removal}

Imposing the statistical cutoff $p$ turns a quantum situation, involving path integral
over fields, into classical-like setting specified by single error function $\aer(r,p)$. Regularized 
exponential insensitivity to distant fields is synonymous with its exponential boundability 
which, in turn, is equivalent to the associated effective range $R_0(p)$ being finite. 
It is thus tempting to conclude that  
$\lim_{p \to 1} R_0(p) \!<\! \infty$ 
is the proper requirement for weakly insensitive operator $O_x$ to be fully insensitive.

However, this cutoff-removal prescription is not sufficient because it doesn't guarantee 
the existence of $p$-independent exponential bound. For example, consider 
the family of error functions $\aer(r,p)$ taking constant value $\Aer$ for $r \!\le\! r_0(p)$,
and decaying as pure exponential of range $R_0$ for $r \!>\! r_0(p)$. With $\Aer$ 
and $R_0$ being $p$-independent, if radius $r_0(p)$ of the constant core grows 
unbounded as the cutoff is lifted ($\lim_{p \to 1} r_0(p) \!=\! \infty$) then there is no exponential
 bound valid for all $p\!<\!1$, albeit the condition of finite limiting range is readily satisfied. 

The possibility of such behavior should not be too surprising in light of our analysis in 
Sec.~\ref{ssec:kinematics}, and its result \eqref{eq:k70}. Indeed, the subset of parameter 
space $(R, \Aer, \rc)$ describing valid exponential bounds is determined not only by $R_0(p)$ 
but also by the core size $\rc_0(R,\Aer,p)$. Finiteness of both is needed in $p \!\to\! 1$ limit,
namely
\begin{equation}
     \lim_{p \to 1} \, R_0(p) < \infty       \qquad\quad  \text{and}  \qquad\quad    
     \exists \; R>0 \;,\; \Aer > 0  \quad : \quad 
     \lim_{p \to 1} \, \rc_0(R,\Aer,p) <  \infty 
     \label{eq:r10}	
\end{equation}
The example of previous paragraph violates the second condition which is, strictly speaking, 
alone sufficient for insensitivity since it implies finite limiting range.\footnote{Indeed, finite 
limiting core size implies $R_0(p) \!\le\! R$, $\forall \, p \!<\! 1$, 
because $\rc_0(R,\Aer,p)$ is non-decreasing in $p$. Moreover, since $R_0(p)$ is also 
non-decreasing, the $p \!\to\! 1$ limit exists and satisfies 
$\lim_{p\to 1} R_0(p) \!\le\! R \!<\! \infty$.} 
However, keeping both requirements explicit is more reflective of steps involved in 
determination of insensitivity in practice. The equivalent formal definition, given below, 
closely mimics this process and is tailored for the eventual step of ultraviolet cutoff 
removal. Instead of $(R,\Aer)$, this formulation specifies the bounds of $\aer(r,p)$ 
via relative parameters $(\kappa,\Rer)$, namely 
\begin{equation}
    R = \kappa\, R_0(p) \;,\; \kappa > 0     \qquad\quad
    \Aer = \Rer \, \langle  \, \| O_x \| \, \rangle      \;,\; \Rer > 0   \qquad\quad
    \rc_0(R,\Aer,p) \rightarrow \rc_0(\kappa,\Rer,p)
    \label{eq:r20}    
\end{equation}
and monitors the finiteness of $\rc_0(\kappa,\Rer,p)$ in $p \!\to\! 1$ limit. At fixed
$\kappa \!>\! 1$, i.e. $\kappa R_0(p) \!>\! R_0(p)$, this limiting process is manifestly 
well-defined for weakly insensitive operator. 
Introduction of $\Rer$ corresponds to measuring the error in units of typical magnitude 
of the operator, which is inconsequential at fixed ultraviolet cutoff but essential for taking 
the continuum limit. Indeed, since the operator values depend on lattice spacing, 
a meaningful assessment of insensitivity is to be performed at fixed $\Rer$.  
Reparametrization \eqref{eq:r20} puts optimal bounds \eqref{eq:k80} of $\aer(r,p)$ 
into the form
\begin{equation}
   \delta(r,p) \,\le\, \Rer \, \langle  \| O_x \| \rangle  \,  
   \exp\bigg(\! -\frac{r-\rc_0(\kappa,\Rer,p)}{\kappa\, R_0(p)} \,\bigg)  \qquad,\qquad 
   r \ge \rc_0(\kappa,\Rer,p) 
   \label{eq:r30}
\end{equation}
and the aforementioned definition of exponential insensitivity is as follows.

\vspace{0.15in}

\noindent {\em\bf Definition 2} {\sl (exponential insensitivity)}

\smallskip
\noindent {\sl Let $O_x(U)$ be a weakly insensitive operator with respect to 
$S(U)$, implying the existence of approximants $O_x^r(U)=O_x^r(U^{x,r})$ 
characterized by finite length scales $R_0(p)$, $\rc_0(\kappa,\Rer,p)$, 
for all $0 \!<\! p \!<\! 1$, $\kappa \!>\!1$, $\Rer \!>\!0$. If there is $O_x^r$ and 
$\kappa$, $\Rer$ for which the finite limits below exist} 
\begin{equation}
   \lim_{p\to 1} \, R_0(p) \,\equiv\, R_0 < \infty \qquad\; \text{and} \qquad\; 
   \lim_{p\to 1} \, \rc_0(\kappa,\Rer,p) \,\equiv\, \rc_0(\kappa,\Rer) < \infty  
   \label{eq:120}
\end{equation}
{\sl we say that $O_x$ is exponentially insensitive with respect to $S$}. 
$\quad \Box$

\vspace{0.15in}

\noindent
Note that the $p$-independent optimal bound for given
$(\kappa,\Rer)$ is obtained by inserting $R_0$ and 
$\rc_0(\kappa,\Rer)$ into formula \eqref{eq:r30}.
There are two points regarding Definition 2 we wish to emphasize.

\vspace{0.12in}

\noindent
(i) It is shown in Appendix~\ref{app:B} that, if $\rc_0(\kappa,\Rer) \!<\! \infty$, 
then $\rc_0(\kappa',\Rer') \!<\! \infty$ for all $(\kappa',\Rer')$ with $\Rer' \!>\! 0$ 
and $\kappa' \!\ge\! \kappa$. Thus, the threshold relative error $\Rer$ remains 
an unconstrained free parameter to keep fixed in $p \!\to\! 1$ 
limit: its choice is purely a matter of practical convenience. However,
Appendix~\ref{app:B} also shows that
finiteness of $\rho_0$ is not guaranteed for $1 \!<\! \kappa' \!<\! \kappa$.
As a practical consequence, examining a single value $\rc_0(\kappa,\Rer)$ 
is not always sufficient to determine insensitivity. Indeed, if $\rc_0(\kappa,\Rer)$ 
is infinite, there may be $\kappa' \!> \kappa$ for which  
$\rc_0(\kappa',\Rer)$ is finite.  

\vspace{0.12in}

\noindent
(ii) As discussed in Sec.~\ref{ssec:kinematics}, there is a class of error functions 
$\aer(r,p)$ obeying an exponential bound with $R$ set to the effective range 
$R_0(p)$. In this case the $p \!\to\! 1$ limiting procedure at $\kappa \!=\! 1$ can 
be set up and examined. This, however, cannot be assumed in general.

\subsection{The Removal of Ultraviolet Cutoff}
\label{ssec:continuum}

The prescription of monitoring $\rc_0$ at fixed $(\kappa,\Rer)$ as statistical
cutoff is lifted ($p \!\to\! 1$ limit), is directly applicable to the process of ultraviolet 
cutoff removal ($a\!\to\!0$ limit). Indeed, an uncontainable core can emerge in 
the process of continuum limit as well. The importance of fixing $\Rer$ is further 
underlined by the fact that $\langle \, \| O_x \| \,\rangle$ is $a$-dependent, 
making it imperative that the approximation error (and thus core size $\rc_0$) 
relates to this changing typical magnitude in fixed proportion. 

We now formulate this precisely in order to classify {\em continuum} operators 
defined by arbitrary lattice prescriptions in terms of their sensitivity to distant 
fields. Making the dependence on lattice spacing explicit, the error function
$\aer(r,p,a)$ assigned to the pair $O_x$ and $O_x^r$ depends on both cutoffs,
as do the associated characteristics $R_0(p,a)$, $\rc_0(\kappa, \Rer, p,a)$. 
Following the structure of our formalism at fixed ultraviolet cutoff, 
the first step is to define the continuum version of exponential insensitivity
at fixed $p$.

\vfill\eject

\noindent {\bf Definition 3} {\sl (exponential insensitivity at fixed $p$ -- continuum)}

\smallskip
\noindent {\sl Let $O_x(U)$ be the lattice operator exponentially insensitive with respect 
to $S(U)$ at given $p$ and lattice spacings $0\!<\!a\!<\!a_0$ i.e. sufficiently close 
to the continuum limit. Thus, there exist approximants $O_x^r(U)=O_x^r(U^{x,r})$
with finite characteristics $R_0(p,a)$ and $\rc_0(\kappa, \Rer, p,a)$ for all 
$0\!<\!a\!<\!a_0$, $\kappa \!>\! 1$, $\Rer \!>\! 0$. If it is possible to find $O_x^r$ and 
$\kappa$, $\Rer$ for which}
\begin{equation}
   \lim_{a\to 0} \, R_0(p,a) \,\equiv\, R_0^c(p) < \infty \qquad\; \text{and} \qquad\; 
   \lim_{a\to 0} \, \rc_0(\kappa,\Rer,p,a) \,\equiv\, \rc_0^c(\kappa,\Rer,p) < \infty  
   \label{eq:160}
\end{equation}
{\sl we say that the continuum operator $O_x^c$ defined by $O_x$ is exponentially 
insensitive at probability $p$ in the continuum theory $S^c$ defined by $S$.} $\quad \Box$

\vspace*{0.15in}

\noindent There are two points regarding this definition we wish to highlight.

\vspace{0.12in}

\noindent
(i) The existence of indicated $a \!\to\! 0$ limits is a somewhat stronger requirement than 
what is sufficient to capture the concept of exponential insensitivity. Indeed, the latter only 
requires that the characteristics in question are bounded for $a$ sufficiently close to zero.
For example, dependencies bounded on $(0,a_0)$ where increasingly rapid oscillations 
near $a \!=\! 0$ destroy $a \!\to\! 0$ limits, still allow for exponential bound of $\aer(r,p,a)$ 
simultaneously valid for all $0 \!<\! a \!<\! a_0$.
While such behavior is not expected to occur in intended applications, Appendix~\ref{app:C} 
describes the adaptation of the formalism to the most general context. Note that this subtlety 
doesn't arise when removing statistical cutoff at fixed $a$ because $p$-dependence is always 
monotonic, making boundedness and existence of finite limit interchangeable.
 
\vspace{0.12in}

\noindent
(ii) Similarly to the situation with statistical cutoff (see discussion point (i) following 
Definition 2), one can infer finiteness of generic $\rc_0^c(\kappa',\Rer',p)$ from finiteness of 
single $\rc_0^c(\kappa,\Rer,p)$. In particular, $\rc_0^c(\kappa',\Rer',p)$ is guaranteed to be 
finite for all $\Rer' \!>\! 0$ and $\kappa' \!\ge\! \kappa$.\footnote{It is worth emphasizing that, 
in generic situations of interest, the parametric dependence of $\rc_0^c$ is in fact entirely 
universal, i.e. finiteness occurs for any $\Rer' \!>\! 0$ and $\kappa' \!>\! 1$.}  

\vspace{0.12in}

Using Definition 3, continuum operators can be classified via the same scheme we used at 
fixed ultraviolet cutoff. In particular, if there is no $0\!<\!p\!<\!1$ such that $O_x^c$ is exponentially 
insensitive, then it is considered {\em sensitive} to distant fields. 
In the opposite case, $O_x^c$ is said to contain an {\em insensitive component}. 
If $O_x^c$ is insensitive for all $0\!<\!p\!<\!1$ and regularly approximated (Appendix~\ref{app:D}),  
it is referred to as {\em weakly insensitive}. The definition of insensitive operator then 
straightforwardly proceeds as follows. 

\vspace{0.15in}

\noindent {\em\bf Definition 4} {\sl (exponential insensitivity -- continuum)}

\smallskip
\noindent {\sl Let $O_x^c$ be weakly insensitive continuum operator in theory $S^c$. 
Thus, there are approximants $O_x^r(U^{x,r})$ of its defining lattice operator $O_x(U)$, 
with finite length scales $R_0^c(p)$ and $\rc_0^c(\kappa,\Rer,p)$ for all 
$0\!<\!p\!<\!1$, $\Rer \!>\! 0$ and sufficiently large $\kappa \!>\! 1$. If there is 
$O_x^r$, $\kappa$, $\Rer$ such that} 
\begin{equation}
   \lim_{p\to 1} \, R_0^c(p) \,\equiv\, R_0^c < \infty \;\qquad\qquad\;
   \lim_{p\to 1} \, \rc_0^c(\kappa,\Rer,p) \,\equiv\, \rc_0^c(\kappa,\Rer) < \infty  
   \label{eq:180}
\end{equation}
{\sl we say that $O_x^c$ is exponentially insensitive with respect to $S^c$}. 
$\quad \Box$

\vspace{0.15in}

The above leaves us with the option offering the highest degree of control over 
the contribution of distant fields to a non-ultralocally defined continuum operator. 
This can arise when lattice definition is strictly exponentially insensitive 
(Definition 2), thus guaranteeing $p$-independent bound (up to events of 
probabilistic measure zero) at successive ultraviolet cutoffs defining the continuum 
limit. If the associated length scales tend to finite values in this process, then
both $p$ and $a$-independent bounds (at least sufficiently close to the continuum
limit) can be found.  We then speak of {\em strong insensitivity} as formulated below.

\vspace{0.14in}

\noindent {\bf Definition 5} {\sl (strong exponential insensitivity)}

\smallskip
\noindent {\sl Let $O_x(U)$ be the lattice operator exponentially insensitive with respect 
to $S(U)$ for lattice spacings $0\!<\!a\!<\!a_0$. Thus, there are approximants 
$O_x^r(U^{x,r})$ characterized by finite $R_0(a)$ and $\rc_0(\kappa,\Rer,a)$ 
for $0\!<\!a\!<\!a_0$, $\Rer \!>\! 0$ and sufficiently large $\kappa \!>\! 1$. If there is $O_x^r$
and $\kappa$, $\Rer$ such that}
\begin{equation}
   \lim_{a\to 0} \, R_0(a) \,\equiv\, \bar{R}_0^c < \infty  \;\qquad\qquad\; 
   \lim_{a\to 0} \, \rc_0(\kappa,\Rer,a) \,\equiv\, \bar{\rc}_0^c(\kappa,\Rer) < \infty  
   \label{eq:200}
\end{equation}
{\sl we say that the continuum operator $O_x^c$ defined by $O_x(U)$ is strongly 
exponentially insensitive in theory $S^c$ defined by $S(U)$.} $\quad \Box$

\vspace{0.14in}

\noindent
We emphasize that, although defined statistically, the bounds constructed via the process of 
statistical regularization are as consequential as conventional upper bounds. The convenience 
of strongly insensitive operators (Definition 5) is that these bounds exist, and can be taken full 
advantage of, even within the context of ultraviolet-regularized dynamics.  Note that the difference 
relative to insensitive operators (Definition 4) is in the order of limits, namely
\begin{equation}
  \lim_{p\to 1}\, \lim_{a\to 0} \lim_{L \to \infty}  
  \quad \text{(insensitive)} \qquad \text{vs} \qquad
  \lim_{a\to 0}\, \lim_{p\to 1} \lim_{L \to \infty}  \quad \text{(strongly insensitive)}
  \label{eq:210}
\end{equation}

Finally, it is important that the proposed formalism of exponential insensitivity is not only relevant 
for the issues of locality, but also for characterizing non-local operators. Indeed, such operators are 
useful in field theory if they have well-defined scale(s) associated with them (e.g. a spatial Wilson loop 
of fixed physical size). Exponentially insensitive operators with finite effective range are natural objects 
of interest in this regard.

\subsection{Locality}
\label{ssec:locality}

The broadest naive notion of {\em local} composite field (operator) $O_x^c$ in the continuum 
refers to an object that doesn't depend on fundamental fields residing non-zero distance 
away from $x$. However, when the theory and $O_x^c$ are rigorously defined via lattice 
regularization, more detailed considerations come into play. In particular, since the naive 
approach is essentially classical with continuum limit involving smooth fields only, two questions
arise regarding the full quantum treatment. (a)  How to formulate the requirement that
``$O_x^c$ doesn't depend on fields non-zero distance away from $x$" for quantum definition
involving general non-ultralocal lattice operators? (b) Given the variety of possible behaviors, 
do all options consistent with (a), when properly formulated, lead to acceptable definition of 
a local quantum $O_x^c$?

The formalism of statistical regularization and exponential insensitivity, just introduced, 
provides an umbrella for both of these issues. Indeed, the resolution of (a) is the requirement
that the relative error function $\rer(r,p,a)$ defined by
\begin{equation} 
      \aer(r,p,a)  =  \langle \, \| O_x \| \, \rangle_a  \; \rer(r,p,a)
      \label{eq:215}
\end{equation}
vanishes in $a \!\to\! 0$ limit for all $r \!>\! 0$ and $0 \!<\! p \!<\! 1$. Here 
$\langle \ldots \rangle_a$ is the expectation at lattice spacing $a$, and $O_x$ is the lattice
operator defining $O_x^c$.

With regard to (b), the chief worry is that non-ultralocal operator may couple distant fields in 
a way that can mimic massless behavior in correlation functions. In other words, power law 
decays could be introduced by virtue of operator's explicit couplings, rather than dynamics 
of fundamental fields, which can be very dangerous to the universality of quantum definition. 
Thus, the basic ``quantum" requirement beyond the properly formulated naive one is that the 
influence of distant fields decays (at least) exponentially with distance in the regularized 
operator. In other words, locality is expected to be safely realized within the realm 
of exponentially insensitive operators, where vanishing contribution from fields 
at non-zero distances translates into vanishing of the corresponding characteristic 
length scales in the continuum limit. Our formalism then leads to the following hierarchy.

\vspace{0.15in}

\noindent {\bf Definition 6} {\sl (statistical degrees of locality)}

\smallskip
\noindent {\sl Let $O_x^c$ be a continuum operator in continuum theory $S^c$, both defined 
via the process of lattice regularization. We say that}

\vspace{0.05in}
\noindent {\sl (a) $O_x^c$ has {\bf local component} with respect to $S^c$ if it has an exponentially 
insensitive component and there is $p \!<\! 1$, $\kappa \!>\! 1$, $\Rer \!>\! 0$  for which
the associated characteristics vanish}
\begin{equation}
   R_0^c(p) = 0       \qquad\qquad\qquad    \rc_0^c(\kappa,\Rer,p) = 0 
   \label{eq:220}
\end{equation}
\noindent {\sl (b) $O_x^c$ is {\bf weakly local} with respect to $S^c$ if it is exponentially insensitive and}
\begin{equation}
   R^c_0 = 0           \qquad\qquad\qquad     \rc_0^c(\kappa,\Rer) = 0 
   \label{eq:240}
\end{equation}
\noindent {\sl (c) $O_x^c$ is {\bf local} with respect to $S^c$ if it is strongly exponentially 
insensitive and}
\begin{equation}
   \bar{R}_0^c = 0   \qquad\qquad\qquad     \bar{\rc}_0^c(\kappa,\Rer) = 0    
   \label{eq:260}
\end{equation}
$\Box$

\vspace{0.15in}

It should be emphasized that locality is a fairly subtle and complicated notion. Indeed, while it would 
be ideal to list all necessary locality-related conditions, and thus identify the maximal set of lattice 
operators defining single continuum dynamics, one is typically only able to formulate generic constraints
expected to be sufficient. In this regard, Definition 6 is quite minimalistic in its requirements.
Additional conditions can certainly be put in place, but they tend to be motivated by convenience 
in working with such operators rather than the notion of locality itself. For example, a well-motivated 
additional restriction is to demand that the relative contribution to $O_x$ from fields beyond distance 
$r$ approaches zero faster than any power of lattice spacing, i.e. faster than expected dynamical 
scaling violations in ultralocal operators. One can easily inspect that this is guaranteed when 
$R_0(a)$ is boundable by a positive power of $a$ sufficiently close to $a \!=\!0$.

Finally, barring certain non-generic singular cases, the property of weak locality in Definition 6 
is already expected to be sufficient for universality. Nevertheless, given the qualitatively different 
levels of exponential control over dynamical degrees of freedom for listed cases, we find it useful 
to keep the corresponding statistical distinctions in place.

\section{Boundary Insensitivity and Boundary Locality}
\label{sec:boundary}

Determination of exponential insensitivity for given non-ultralocal lattice 
operator $O_x(U)$ and action $S(U)$ is a computable problem if both $O_x(U)$ 
and $S(U)$ are computable. In fact, the task doesn't require a search through
possible approximants: there exists an algorithm, albeit computationally demanding, 
that directly evaluates $\delta^{op}(r,p)$ of the optimal approximant to arbitrary 
precision.\footnote{To make this statement precise, several additional ingredients need 
to be defined. These details and the associated construction are outside 
the main line of this paper, and will be addressed elsewhere~\cite{Hor16C}.}
However, as advertised in the Introduction, our chief goal here is to describe
a version of exponential insensitivity which, albeit somewhat stronger, produces
a scheme that is computationally efficient and generic in an essential way.
The merit of such concept will clearly depend on whether it captures 
the meaning of insensitivity to distant fields in sufficiently robust physical terms.

The premise underlying our approach is that exponential insensitivity of $O_x(U)$,
as defined in Sec.~\ref{sec:insensitivity}, should translate into exponential insensitivity 
to presence of a distant boundary: a concrete physical requirement. While this is expected 
to hold generically, it is not strictly guaranteed. At the same time, to single it out as 
a stronger defining feature is fruitful in that, as explained below, ``boundary effects'' are 
associated with simple approximants whose error functions just need to be computed 
and examined. 

To explain the meaning of  ``boundary'' in this context, recall that our infrared-regularized 
setup involves the sequence of triples $({\cal L}^x_L, S_L, O_{x,L})$ specifying the 
$L^4$ lattice with $x$ at its center, the theory, and the operator respectively. This means 
that, working at fixed $L$, the definition of the operator provides not only 
$O_x(U) \!\equiv\! O_{x,L}(U_L)$ for gauge field $U \!\equiv\! U_L$ on ${\cal L}^x_L$, 
but also a finite sequence of values 
\begin{equation}
    O_x(U)  \;\longrightarrow\; O_x^\ell(U)   \,\equiv\, O_{x,\ell}(U_\ell)  \;, \qquad   \ell < L
    \label{eq:290}     
\end{equation}
Here $U_\ell\subset U$ is the restriction of $U$ to a valid gauge field on 
${\cal L}^x_\ell \subset {\cal L}^x_L$, namely on a congruent system of smaller size $\ell$. 
The operation
${\cal L}^x_L \rightarrow {\cal L}^x_\ell$,  $U \rightarrow U_\ell$, $O_{x} \rightarrow O_{x,\ell}$ 
eliminates distant degrees of freedom by introducing an artificial boundary 
(that of ${\cal L}^x_\ell$) just to probe the operator $O_x(U)$.\footnote{Even when finite setup 
is boundary-free, such as with periodic boundary conditions, $U \rightarrow U_\ell$ still
creates a boundary effect by bringing together gauge variables in $U_\ell$ that were initially 
far apart in $U$.} This motivates referring to $O_x^\ell(U)$ as {\em boundary approximant} of 
$O_x(U)$, although ``finite-size" or ``smaller-size" approximant would be equally fitting.

We propose that boundary approximants are central objects for addressing the efficiency of 
computations with non-ultralocal lattice operators. As such, they are interesting in their own right, 
and Appendix~\ref{app:A} describes them in general lattice setting. They also provide means for 
constructing relevant hypercubic approximants $O_x^r(U)$. The needed connection is straightforward 
in the symmetric lattice setup employed here. 
Indeed, first note that 
\begin{equation}
      {\cal H}^{x,r} = {\cal L}^x_{\ell(r)}  \;\qquad 
      U^{x,r} \subseteq U_{\ell(r)}          \;\qquad  \ell(r) \equiv 2r + a
      \label{eq:300}
\end{equation}
with the second relation arising since, depending on the boundary conditions on
${\cal L}^x_{\ell(r)}$, field $U_{\ell(r)}$ may contain dangling links (relative to 
${\cal L}^x_{\ell(r)} \!=\! {\cal H}^{x,r}$), while $U^{x,r}$ does not.\footnote{
In general, when representing finite system $({\cal L}_\ell^x, S_\ell)$ by its embedding 
in ${\cal H}_\infty$, as done here, there can be links that do not connect pair of points from 
${\cal L}_\ell^x$ that are neighbors in ${\cal H}_\infty$, but rather connect non-neighbors 
via a boundary condition: these are {\em dangling links} relative to ${\cal L}_\ell^x$. }
This means that hypercubic approximant can be constructed from boundary 
approximant on ${\cal L}^x_{\ell(r)}$ by freezing dangling links of $U_{\ell(r)}$ to values 
independent of $U$. In particular,
\begin{equation}
     U_{\ell(r)} \,=\, U^{x,r}   \,\union\,  \{\, U_b \,\}
     \quad\longrightarrow\quad  
     U_{\ell(r)}^h \,\equiv\,  U^{x,r}  \,\union\, \{\, \bar{U}_b \,\}
     \label{eq:320}
\end{equation}
where $\{\, U_b \,\}$ abbreviates the subset of dangling links and $\{\, \bar{U}_b \,\}$
the choice of their fixed values. The {\em hypercubic boundary approximant} associated 
with this $\{\, \bar{U}_b  \,\}$ is then 
\begin{equation}
      O_x(U)  \;\longrightarrow\;  O_x^r(U) \,\equiv\, O_{x,\ell(r)} \bigl( U_{\ell(r)}^h \bigr)  
      \,=\, O_x^r(U^{x,r}) 
     \label{eq:340}
\end{equation}
If there is a systematic choice of $\{\, \bar{U}_b  \,\}$, i.e. prescription specifying it as 
$r$ changes and $L$ scales to infinity, such that the error function $\delta(r,p)$ of the 
approximant \eqref{eq:340} shows the types of exponential behavior discussed in 
Sec.~\ref{sec:insensitivity}, then corresponding forms of {\em boundary insensitivity} 
to distant fields arise. The notion of {\em boundary locality} is defined accordingly. 

For practical purposes, it is useful to define standard choices of boundary approximant 
so that different tests of boundary insensitivity for various operators can be directly 
compared. To that end, we single out two simple systematic options wherein all dangling 
links are set to identical values.   

\smallskip
\noindent {\bf (i) Universal Form}:  $\bar{U}_b= \mathbf{1}$ ($3\!\times\!3$ identity matrix)
\smallskip

\noindent Here $O_x^r(U)$ is guaranteed to be well-defined and computable if $O_x(U)$
is. Indeed, since $\mathbf{1} \!\in\! SU(3)$, configuration $U_{\ell(r)}^{h}$ represents a valid 
SU(3) field, and the above follows from the very definition and computability of $O_x(U)$. 

\smallskip
\noindent {\bf (ii) Covariant Form}: $\bar{U}_b= \mathbf{0}$ ($3\!\times\!3$ zero matrix)
\smallskip

\noindent Here $O_x^r(U)$ may be ill-defined for certain artificially constructed operators 
$O_x(U)$, since ${\mathbf 0} \!\notin\! \text{SU(3)}$. For example, 
$O_{x,\ell(r)}$ could become singular as $U_{\ell(r)}$ is deformed to $U_{\ell(r)}^h$.
When $O_x^r(U)$ is well-defined (a generic situation), it is also computable. 
Indeed, one option is to use the computer program evaluating $O_{x,\ell(r)}(U_{\ell(r)})$, 
and add a parallel branch that applies every operation performed on input $U_{\ell(r)}$ also 
on input $U_{\ell(r)}^h$. The output from the parallel branch is read off whenever the original 
branch, driving the execution of the code, halts.\footnote{This, in fact, is one of the universal 
ways to define the approximant in covariant form.} 

\medskip

\noindent If $O_x(U)$ admits hypercubic boundary approximant in covariant form, it is by 
definition its default hypercubic approximant, while the universal form is assigned otherwise. 
To motivate this choice, note that for operators of physical relevance, it is desirable that 
$O_x^r(U)$ can be treated as another valid regularized operator on ${\cal L}^x_L$, i.e. on equal 
footing with $O_x(U)$. Its transformation properties should thus match those of $O_x(U)$ to 
the largest extent possible.  However, if $U_{\ell(r)}$ contains dangling links 
($\{\, U_b  \,\} \neq \emptyset\,$), such as with periodic boundary conditions, the universal 
$O_x^r(U)$ may not inherit gauge covariance of $O_x(U)$ (if any). Indeed, when 
$O_{x,\ell(r)}(U_{\ell(r)})$ is covariant in theory $(\, {\cal L}^x_{\ell(r)}, S_{\ell(r)} \,)$, it only 
involves a combination of closed loops on ${\cal L}^x_{\ell(r)}$ with its periodic definition. 
But closed loops containing dangling links generically translate into open line segments on 
${\cal L}^x_L$, thus spoiling gauge covariance of universal $O^r_x(U)$ in theory 
$(\, {\cal L}^x_L, S_L \,)$. The use of covariant form, however, eliminates this issue: when 
a dangling link is set to zero, any closed loop that contains it becomes zero as well, and such 
non-covariant terms simply do not occur.

\begin{figure}[t]
\begin{center}
    \centerline{
    \hskip 0.00in
    \includegraphics[width=6.2truecm,angle=0]{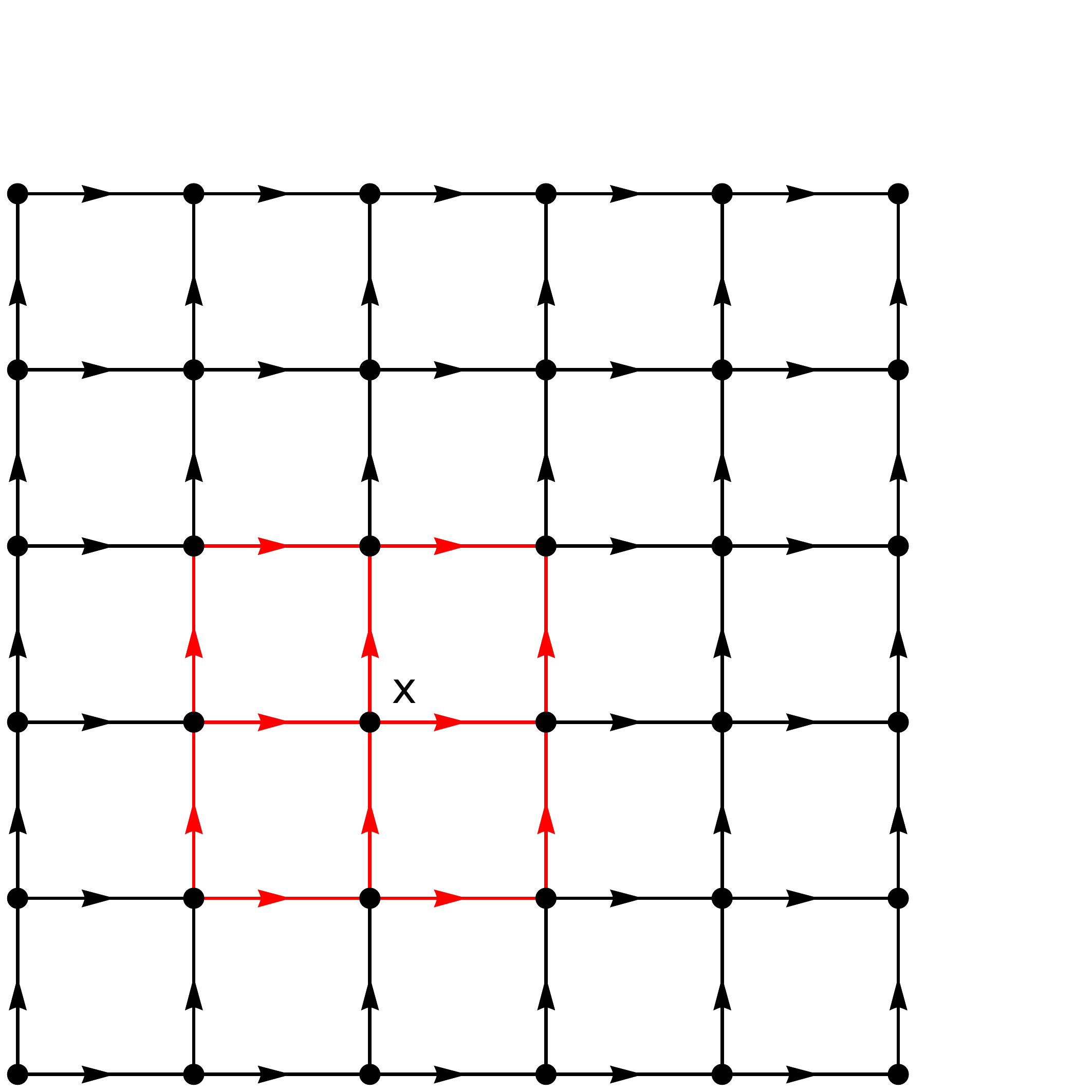}
    \hskip 0.65in
    \includegraphics[width=6.2truecm,angle=0]{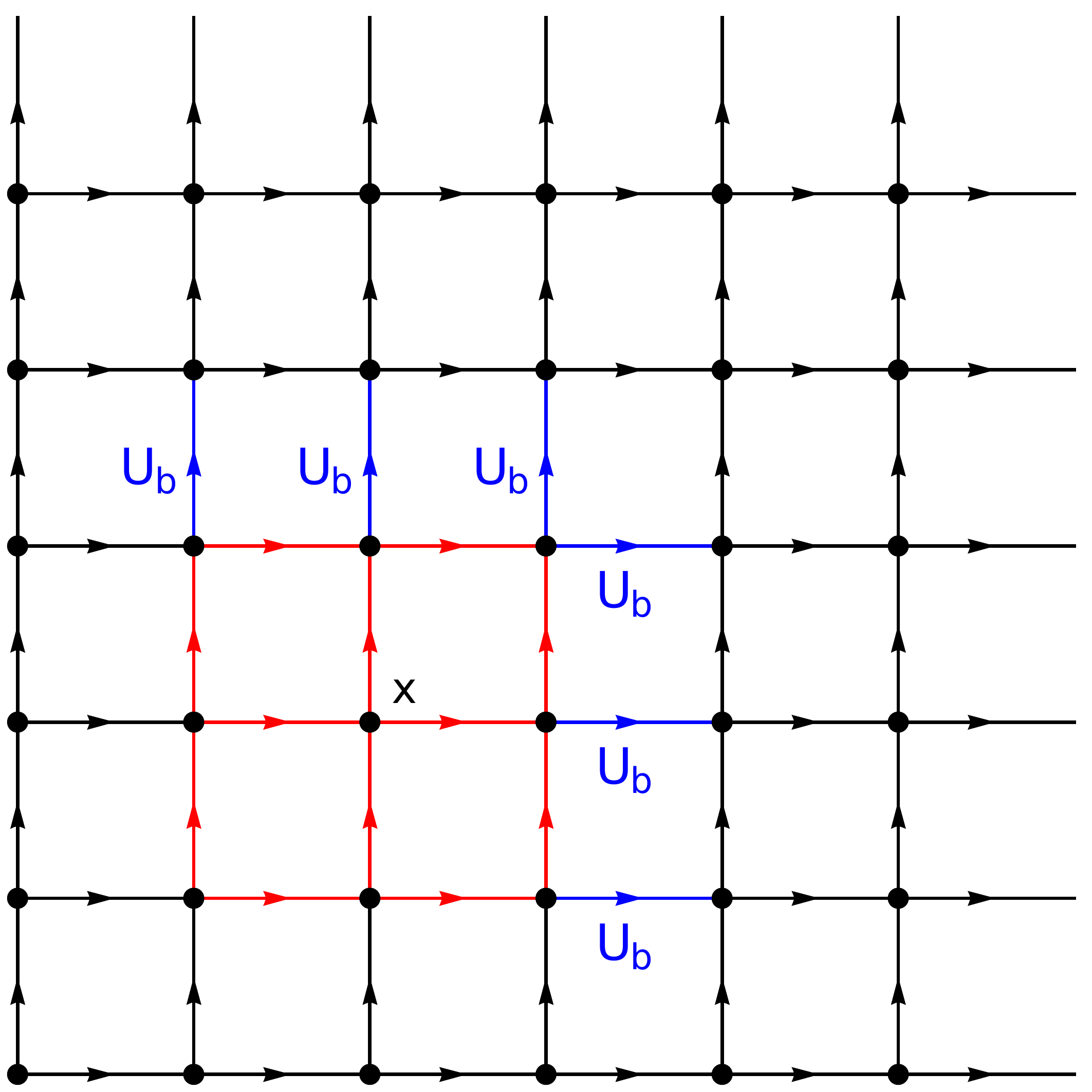}
     }
     \vskip -0.0in
     \caption{The lattice arrangement of gauge links for fixed ``large" lattice ${\cal L}^x_L$ 
     ($L\!=\!6$) and its ``small" sublattice ${\cal L}^x_\ell$ ($\ell \!=\! 3, r\!=\!1$) in case of open 
     boundary conditions (left) and periodic boundary conditions in all directions (right).}    
     \label{fig:openvsper}
    \vskip -0.40in
\end{center}
\end{figure} 

To further exemplify the boundary approach, Fig.$\,$\ref{fig:openvsper} illustrates two common 
situations. First, let the defining sequence of triples $({\cal L}^x_L, S_L, O_{x,L})$ be such that 
the prescription for action $S_L \!=\! S_L(U_L)$ doesn't involve any dangling links (left plot). 
This is the case of open boundary conditions and, since $\{\, U_b  \,\}  \!=\! \emptyset\,$, there 
is a single hypercubic boundary approximant $O_x^r(U)$ which is directly in covariant form. 
Secondly, assume that boundary conditions in $S_L \!=\! S_L(U_L)$ are fully periodic in 
all directions (right plot). The subset of dangling links is now maximal as are the options for 
possible hypercubic boundary approximants.  We emphasize that the sequence of triples
$({\cal L}^x_L, S_L, O_{x,L})$ represents entire information apriori known about $O_x$. 
Given that, constructing covariant form $O_x^r(U)$ in periodic case can be viewed as extending 
the definition of $O_x(U)$ to the case with open boundary conditions. 

\subsection{Computational Efficiency}
\label{ssec:efficiency}

Important feature of the boundary approach is that it connects insensitivity to distant fields 
of an operator (and thus ultimately locality), to efficiency of its evaluation. More precisely,
our main message in this regard is that insensitive operators generically have an 
efficient computer implementation provided by the boundary construction itself.  

To formulate this, recall again that the operator $O_x(U)$ is defined via a sequence of triples 
$({\cal L}^x_L, S_L, O_{x,L})$, and is assumed to be computable in this section. This
guarantees the existence of a program $\cP$ that, given $\{ L/a, U\}$, outputs $O_{x,L}$ 
to arbitrary accuracy. For simplicity, it is understood that the corresponding error is arranged 
to be much smaller than any other accuracy measure in the problem. In that sense
\begin{equation}
     \cP \lbrace L/a, U \rbrace \,=\, O_{x,L}(U)
     \label{eq:360}     
\end{equation}  
Note that we deal with a lattice situation and $a$ is not essential: with eventual continuum 
considerations in mind, we just chose to denote the parameter encoding the size of input 
field $U$ as $L/a$. Let $\cC(\cP, L/a, U)$ be the cost of running $\cP \{L/a, U \}$ measured 
in required number of arithmetic operations.\footnote{Measuring cost in arithmetic 
operations rather than elementary bit operations avoids dealing with cost issues
stemming from bit size of reals. Thus, the cost of $a \!\times\! b$ is one in the former, 
but diverges at least as $\log 1/\delta$, due to increasing bit size of representing 
real $a$, $b$ so that precision $\delta$ in $a \!\times\! b$ is achieved.} 
The average cost function for $O_x$ in realization $\cP$ is then
\begin{equation}
     \cC_\cP(L/a) \,\equiv\,  
     \bigl \langle\,  \cC(\cP, L/a, U)  \,\bigr\rangle_{S_L}
     \label{eq:380}          
\end{equation}
At the same time, the cost function of boundary approximant $O_x^\ell$ of 
Eq.$\,$\eqref{eq:290} is
\begin{equation}
     \cC_\cP'(\ell/a)  \,\equiv\,  
     \bigl \langle\,  \cC(\cP, \ell/a, U_\ell)  \,\bigr\rangle_{S_\ell'}
     \label{eq:400}          
\end{equation}
Here $S_\ell'(U_\ell)$ specifies the probability of fields $U_\ell$ on ${\cal L} ^x_\ell$,  
arising when distribution of $U$ on ${\cal L} ^x_L$  (given by $S_L(U)$) is marginalized 
via the $U \rightarrow U_\ell$ restriction.\footnote{Thus, $S_\ell'$ and $\cC_\cP'$ 
also weakly depend on $L$ which can be thought of as already taken to infinity 
in \eqref{eq:400}.} Thus, when replacing the operator with its boundary approximant, 
the cost of evaluating it via $\cP$ changes as
\begin{equation}
     O_x  \;\longrightarrow\; O_x^\ell   \qquad  :  \qquad\quad
     \cC_\cP(L/a)  \;\longrightarrow\;  \cC_\cP'(\ell/a)
     \label{eq:420}          
\end{equation}

The situation is obviously analogous in case of hypercubic boundary approximants, except 
that $\ell \rightarrow \ell(r)$, and the action $S_{\ell(r)}'=S_{\ell(r)}'(U^{x,r})$ generally 
involves further marginalization due to the fixing of ``boundary" $\{\, \bar{U}_b  \,\}$. 
If $O_x$ is boundary insensitive, there exists hypercubic approximant $O_x^r$ and 
$\kappa \!>\! 1$, $\Rer \!>\! 0$ for which the bound \eqref{eq:r30} holds, and has $p \to 1$ 
limit with finite $R_0$ and $\rc_0(\kappa,\Rer)$. The bound makes it possible to use 
the above default program $\cP$ for computing $O_x(U)$ to arbitrary preset relative 
precision $\rer \!=\! \delta/\langle \| O_x \|  \rangle$ without invoking the full volume. Indeed, 
it provides the minimal $r\!=\!\Reff_\rer$ for which $O_x^r(U)$ is guaranteed 
to produce error smaller than $\rer$, and we just feed $\cP$ the input parameters 
$\ell(\Reff_\rer)$ and $U^h_{\ell(\Reff_\rer)}$. This running comes at the average 
computational cost
\begin{equation}
     \cC_\cP \left( \frac{L}{a} \right)  \quad\longrightarrow\quad  
     \cC_\cP^{\rer}  \,\equiv\, \cC_\cP' \left( \frac{\ell(\Reff_\rer)}{a} \right)
     \quad\;  ,  \quad\;
     \Reff_\rer = \rc_0 + \max \left\{0, \, \kappa R_0\, \log \frac{\Rer}{\rer} \right\}
     \label{eq:440}               
\end{equation}
Note that $\Reff_\rer \!=\! \rc_0(\kappa,\Rer)$ for all $\rer \!>\! \Rer$. In other words, 
$O_x^{\rc_0}$ is the ``leading approximation", with logarithmic dependence setting in 
for $\rer \le \Rer$. There are several points to highlight.

\smallskip
\noindent {\bf (1)} Formula \eqref{eq:440} is universal in that it applies to arbitrary
setup as described in Appendix~\ref{app:A}. Indeed, various situations only differ 
by the precise form of function $\ell(r)$.

\smallskip
\noindent {\bf (2)} Using the boundary approximant in the above manner 
turns computation whose cost normally strongly depends on volume  
(l.h.s. of \eqref{eq:440}) into one that involves a fixed-size input for given
desired precision $\rer$, and is in that sense 
volume-independent.\footnote{Note that it is implicitly understood here, as is
in \eqref{eq:440}, that $L$ is sufficiently large ($L \gg \Reff_\rer$) so that 
finite-$L$ correction to $\cC_\cP'$ is negligible for $\rer$ in question.}
Importantly, this constant cost only depends logarithmically on $\rer$.  

\smallskip
\noindent {\bf (3)} Note that the formula \eqref{eq:440} doesn't assume 
anything about the nature of program $\cP$. Indeed, starting from any valid
$\cP$, however inefficient, results in a volume-independent computation
at fixed precision.  

\smallskip
\noindent {\bf (4)} It is clear that the cost functions $\cC_\cP(s)$ and  $\cC_\cP'(s)$ are 
not identical\footnote{They are identical if $\cC(\cP, L/a, U)$ is $U$-independent, which is 
usually the case for ultralocal operators.}. 
However, their asymptotic behaviors ($s\rightarrow \infty$) are expected to be of the same 
type, with cost driven mainly by the number of field variables (input size) that have to be 
handled by the program as their abundance grows unbounded. Thus, if $\cC_\cP(s)$ grows 
exponentially or as a power, $\cC_\cP'(s)$ generically grows in the same qualitative manner. 
Nevertheless, the introduction of boundary may affect the evaluation to the extent 
that the two cost functions are not simply asymptotically proportional. Taking 
the relevant case of power growth as an example, we can 
have\footnote{Here ``$\sim$" stands for ``asymptotically proportional".}
\begin{equation}
      \cC_\cP(s) \;\sim\;  s^{d+\alpha}        \qquad\qquad
      \cC_\cP'(s) \;\sim\;  s^{d+\alpha'}       \qquad\qquad  
      s \rightarrow \infty   
     \label{eq:460}                     
\end{equation}
with possibly unequal $\alpha$, $\alpha'$. For all standard ways of taking the infinite volume 
in zero-temperature calculations, function $\ell(r)$ is linear in $r$. For example, $\ell(r) \!=\! 2r+a$ 
for maximally symmetric case (see Eq.~\eqref{eq:300}). With that, we have the asymptotic 
reduction
\begin{equation}
   \left(\, \frac{L}{a} \,\right)^{d+\alpha}  \; \xrightarrow[\scriptstyle \rer \to 0]  \qquad\quad
   \left(\, \frac{\ell(\Reff_\rer)}{a} \,\right)^{d+\alpha'}  \quad \sim \quad   
   \left(\, \frac{R_0}{a} \, \log \frac{\Rer}{\rer} \,\right)^{d+\alpha'}
   \quad , \quad \alpha, \alpha' \ge -d
   \tag{{\ref{eq:20}}'}
   \label{eq:20'}
\end{equation}
which is a generalized version of \eqref{eq:20} in the relative error form. Note that 
$\alpha\!=\!\alpha'\!=\!-d$ for ultralocal operator since the cost is constant, and one 
generically expects $\alpha \ge 0$ for non-ultralocal operator coupled to all field 
variables, such as $D_{x,x}(U)$ of overlap Dirac matrix.

\smallskip
\noindent {\bf (6)} If the choice of input for program $\cP$ was based on parameters 
at given statistical cutoff $p$, i.e. $R_0(p)$, $\rc_0(\kappa,\Rer,p)$, the result of any 
particular run would be good to relative precision $\epsilon$ with probability at least $p$. 
This in itself is a powerful advantage of even weakly insensitive non-ultralocal operators,
where $p$ can be set arbitrarily close to unity.

\section{Overlap-Based Gauge Operators}
\label{sec:overlapops}

In what follows, we apply the framework developed in Secs.$\,$\ref{sec:insensitivity},\ref{sec:boundary}
to study boundary insensitivity and locality properties of the operator
\begin{equation}
   O_x(U) \,\equiv\, D_{x,x}(U) - D_{x,x}^F 
    \label{eq:540}     
\end{equation}
where $D(U)$ is the overlap color-spin matrix and $D^F$ its free subtraction ($U_{y,\mu} \to \mathbf{1}$).
Its relevance is that useful operators of practical interest such as topological density $q(x)$, 
scalar gauge density $s(x)$ and field-strength tensor $F_{\mu\nu}(x)$ are based on it, namely
\begin{equation}
    q(x) \,\propto\, \Trb \gfive O_x   \qquad\qquad
    s(x) \,\propto\,  \Trb O_x    \qquad\qquad           
    F_{\mu \nu}(x)  \,\propto\,  {\Trb}_s \,\sigma_{\mu \nu} O_x  
    \label{eq:560}     
\end{equation}
where $\Trb_s$ denotes the trace over spin indices only. Note that the free subtraction in $O_x$ is 
consequential for scalar density but not for the other two operators. 

It should be remarked here that the obvious role of the above $O_x(U)$ as a master construct in 
definitions \eqref{eq:560} is further underlined by the fact that its potential insensitivity at fixed
$p$ and $a$ descends onto these derived operators of interest. This can be seen from the fact that 
the Frobenius matrix norm $\| M \|^2 \equiv \Trb (M^+M)$, which we will use to quantify 
the differences of matrix-valued operators and define the error functions $\delta(r,p)$, is 
sub-multiplicative, i.e. $\| M_1 M_2 \| \le \|M_1\| \|M_2\|$. Thus, the requisite exponential bounds for 
$O_x(U)$ can be used to produce the associated (albeit non-optimal) bounds for derived operators. 
Nevertheless, the concrete behavior of optimal bounds for individual operators is of practical interest 
since their precise form guides the efficient evaluation of these operators. In particular, the threshold 
distances $\rc_0$ could vary appreciably among them.

\begin{figure}[t]
\begin{center}
    \centerline{
    \hskip 0.00in
    \includegraphics[width=7.0truecm,angle=0]{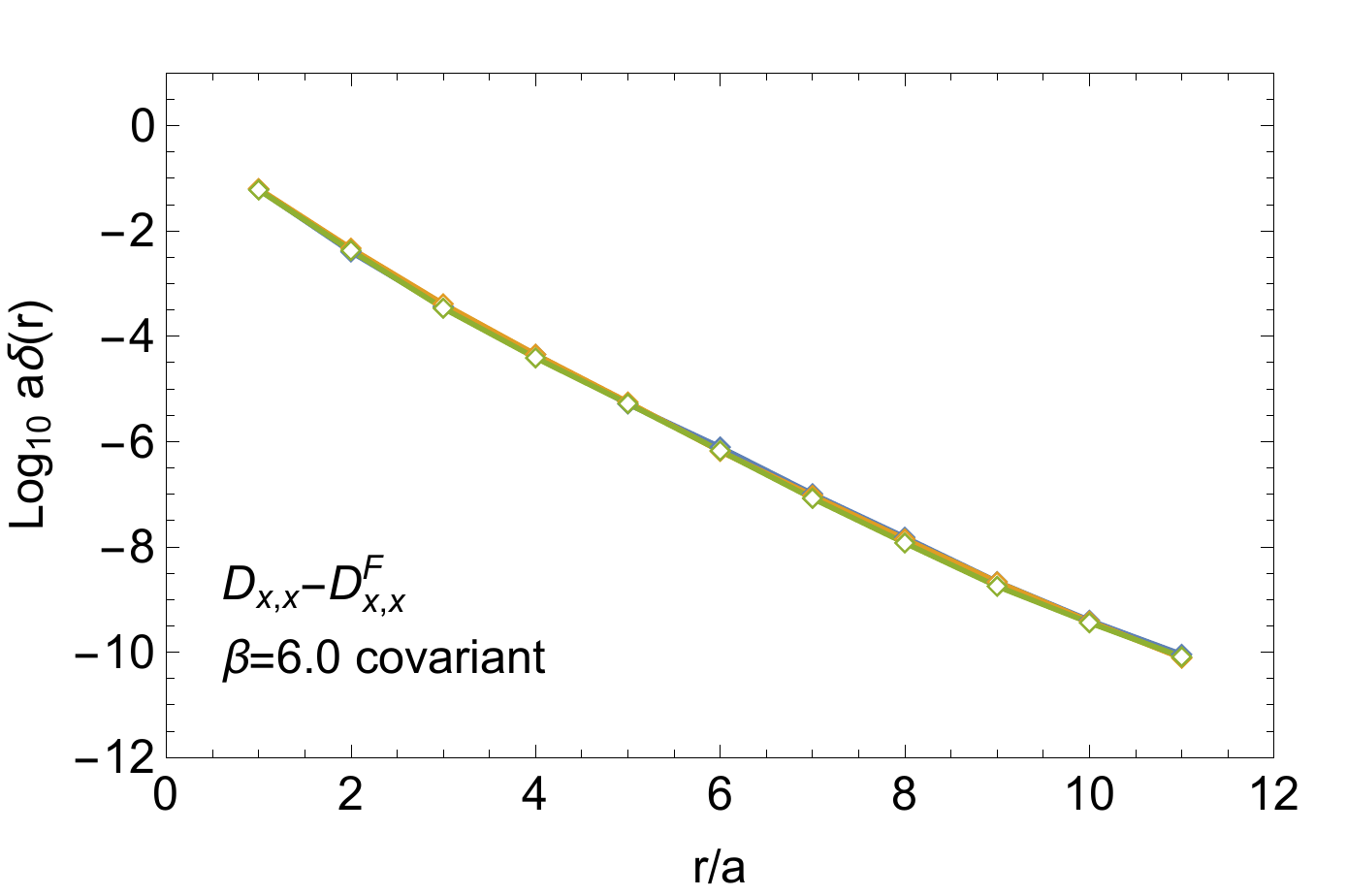}
    \hskip -0.00in
    \includegraphics[width=7.0truecm,angle=0]{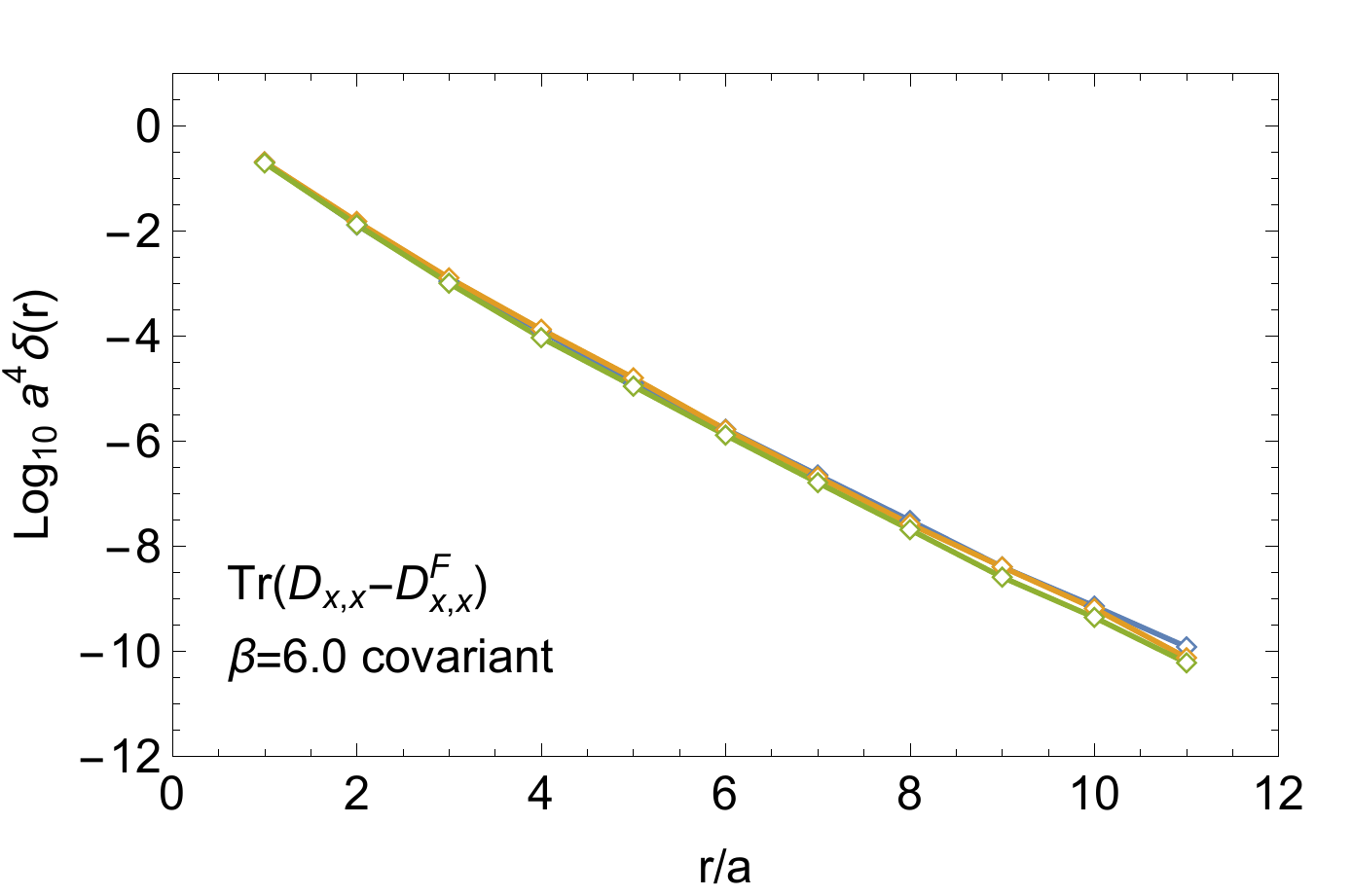}
     }
     \vskip -0.05in
    \centerline{
    \hskip 0.00in
    \includegraphics[width=7.0truecm,angle=0]{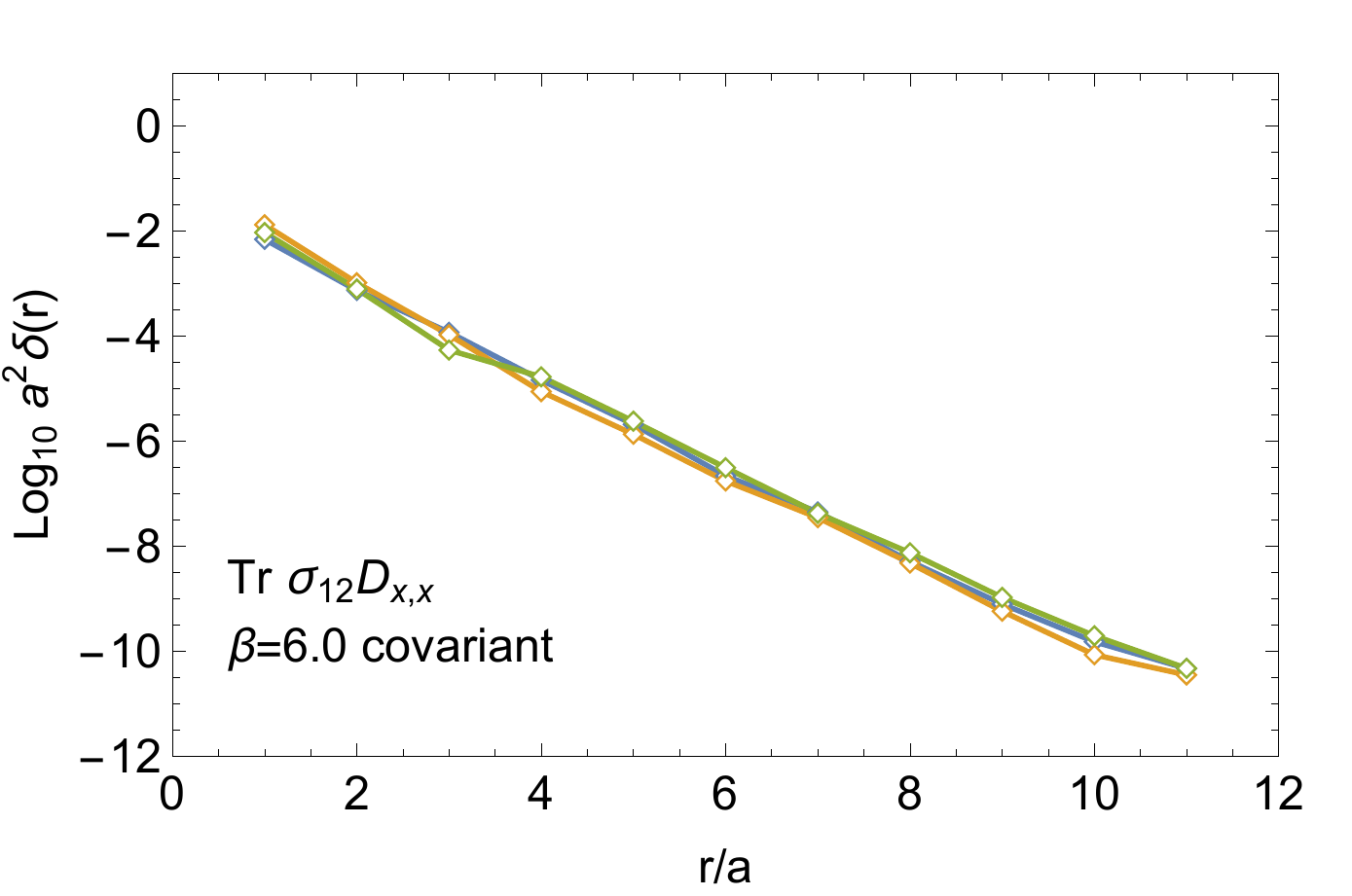}
    \hskip -0.00in
    \includegraphics[width=7.0truecm,angle=0]{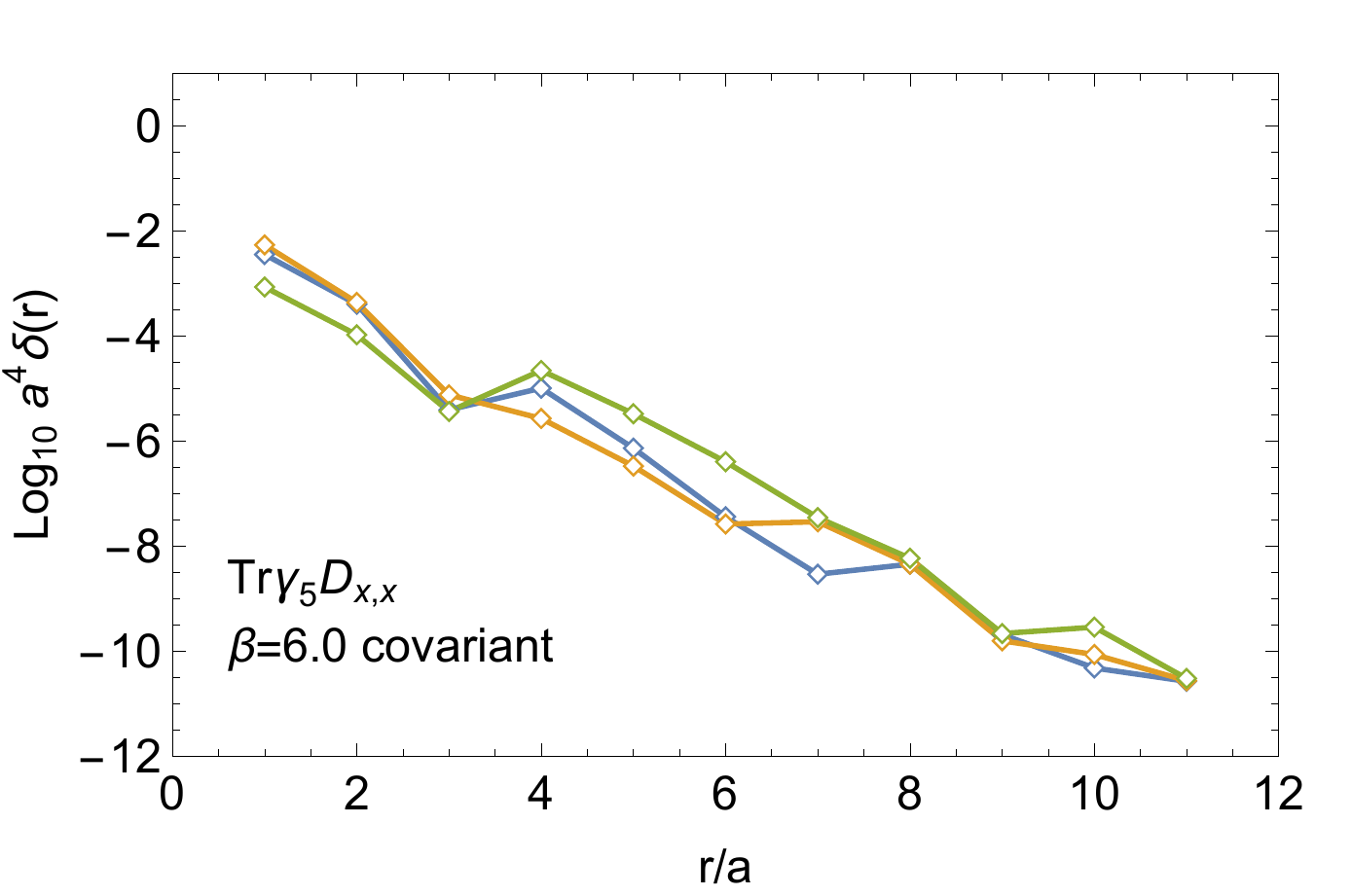}
     }
     \vskip -0.1in
     \caption{The behavior of $\delta(r)$ for covariant boundary approximant at three randomly chosen 
     individual points at $\beta \!=\! 6.0$.}
     \label{fig:rnd_pts}
    \vskip -0.40in
\end{center}
\end{figure}

\subsection{The Setup}

For our numerical work, we use the symmetric $L^4$ lattice geometry adopted as a template for 
the general discussion of Secs.$\,$\ref{sec:insensitivity},\ref{sec:boundary}.  The insensitivity
and locality of the above overlap-based gauge operators will be studied with respect to to pure-glue 
SU(3) theory with Wilson gauge action, and periodic boundary conditions for gauge fields in all 
directions. 

The overlap Dirac operator in \eqref{eq:540} is based on the Wilson-Dirac kernel and the parameter 
$\rho \in (0,2)$, specifying its negative mass, set to $\rho \!=\! 1.368$ ($\kappa \!=\! 0.19$). 
Definition of the operator is given in Appendix~\ref{app:E}. Standard 
boundary conditions for quarks, i.e. periodic in ``space" and antiperiodic in ``time", are used although 
there is no fundamental preference in that regard. In fact, when the sole purpose of utilizing the overlap 
is to define gauge operators, one may opt for periodic boundaries in all directions to maximize 
hypercubic symmetries. The resulting $D_{x,x}(U)$ is gauge covariant and periodic in all directions. 

With regard to the computational realization of the overlap, we follow the MinMax polynomial approach 
of Ref.~\cite{Giu02A} in a specific implementation discussed in Ref.~\cite{Ale11D}. Using deflation
as needed, we explicitly ensured that the accuracy in evaluation of the overlap is always significantly 
better than any quoted error of a boundary approximant.\footnote{Note that the overlap evaluations 
on original lattice of size $L/a$ and on hypercubic subsystems of size $r/a$ are independent computations 
using their own polynomial approximations and deflations as appropriate.}
In other words,  in what follows, Eq.~\eqref{eq:360} can be assumed to be valid without further 
qualifications.\footnote{Rigorous treatment would require specifying the behavior of the program 
for backgrounds with exact zeromodes of $H_W = \gamma_5 D_W$. Given that this occurs on a fixed 
subset of measure zero, it is not consequential.} 

To study the insensitivity and locality properties of the above operators, we generated three ensembles 
of $L/a \!=\! 24$ configurations at $\beta \!=\! 6.0, 6.2, 6.4$. This corresponds to nominal lattice spacings 
of $a \!=\! 0.093, 0.068, 0.051\,$fm respectively, based on $r_0 \!=\! 0.5$ fm.  While the finest 
($\beta \!=\! 6.4$)  lattice system is clearly quite squeezed in physical terms, this is of little consequence 
for the current study. Indeed, as the discussion later in this paper reveals, the finite volume effects on 
observables of our interest are negligible. 

As expected on general grounds, the default hypercubic boundary approximant $O_x^r$ (covariant one) 
is well-defined, and can be computed straightforwardly by running the program designed to evaluate 
$O_{x}$ on input that corresponds to the boundary subsystem. For 20 configurations from each of 
the above ensembles, we computed  $O_x$ as well as its approximants $O_x^r$ for all possible 
distances $r$, at 16 points evenly distributed on the lattice. Note that this calculation produces estimates
for all derived operators \eqref{eq:560}. 

To obtain a preliminary assessment of boundary approximants, we plot in Fig.$\,$\ref{fig:rnd_pts} 
the dependence of error $\aer$ on hypercubic radius $r$ at three randomly chosen points $x$ in 
a given $\beta \!=\! 6.0$ configuration.  As can be seen quite clearly, a straightforward exponential-like 
decay takes place even on an individual point basis. The behavior is least orderly in case 
of pseudoscalar density (bottom right), but the overall trend is quite unmistakable in that case as well. 
A systematic investigation using statistical regularization is thus warranted.

\subsection{Finite Volume}

As emphasized in Sec.$\,$\ref{sec:insensitivity} (see Definition 1), exponential insensitivity is 
an infinite-volume concept. In particular, the property depends on the behavior of statistically 
regularized error function in asymptotically large volumes, i.e. 
$\delta(r,p) \!=\! \lim_{L\to \infty} \delta(r,p,L)$. 
However, in practice, we are bound to infer the exponential behavior from a sequence of finite, 
and usually limited volumes.  To ensure that such estimates are reliable, it needs to be
checked that $\delta(r,p,L)$ is insensitive to $L$ over the range of distances $r$ used in such
calculations.

\begin{figure}[t]
\begin{center}
    \centerline{
    \hskip 0.00in
    \includegraphics[width=7.5truecm,angle=0]{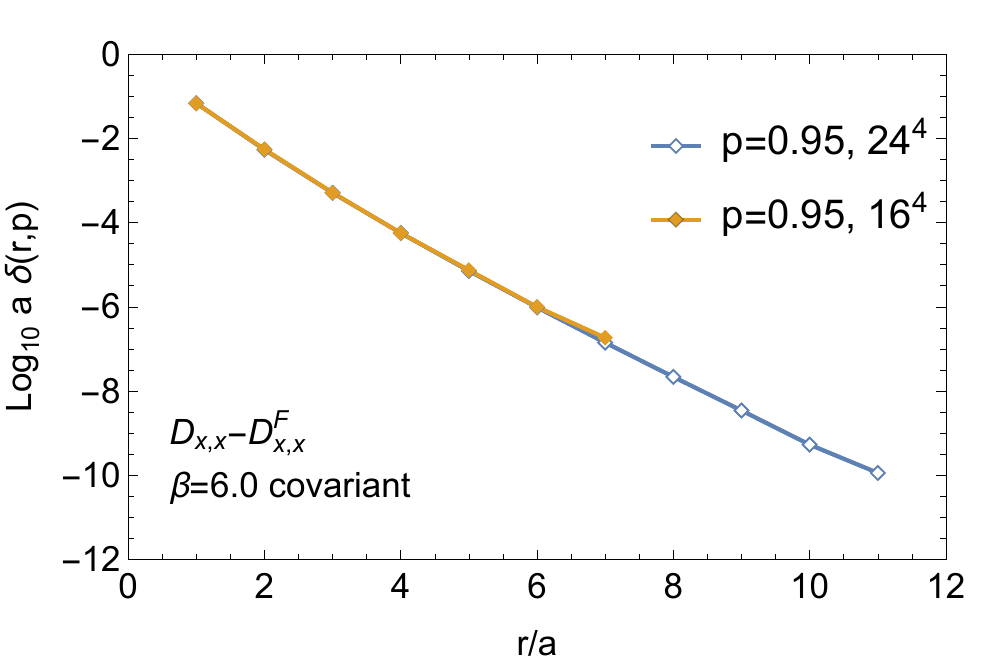}
    \hskip 0.30in
    \includegraphics[width=7.5truecm,angle=0]{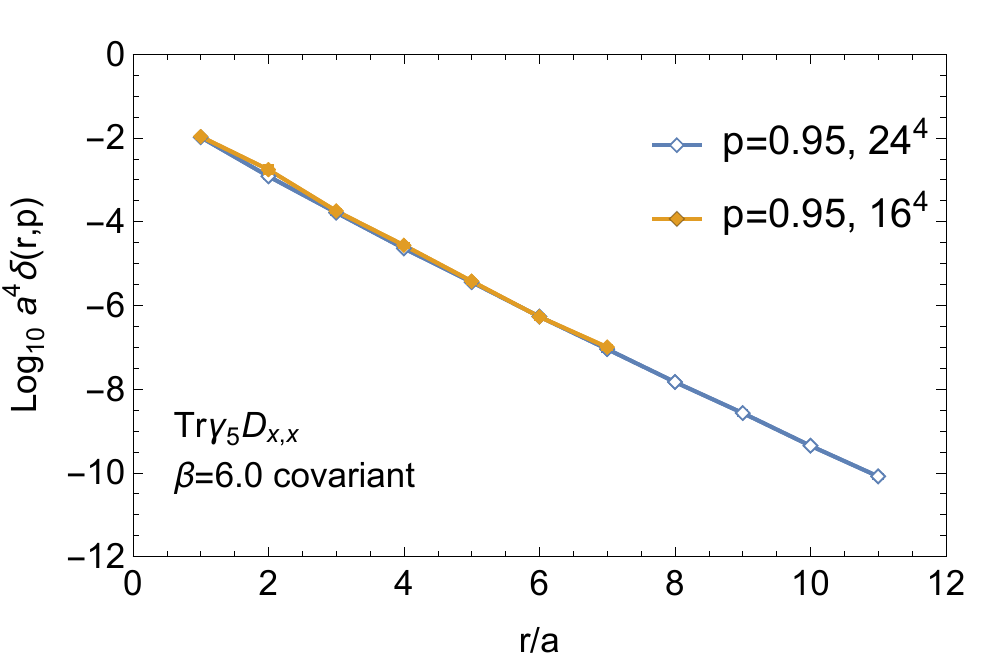}
     }
     \vskip -0.1in
     \caption{The volume comparison at $p\!=\!0.95$ for $O_x$ (left) and $\Trb \gfive O_x$ (right). }
     \label{fig:volcheck}
    \vskip -0.50in
\end{center}
\end{figure} 

To examine this issue, we supplemented the $\beta \!=\! 6.0$ ensemble at $L/a \!=\! 24$ with one 
at $L/a \!=\! 16$. Notice that, since $\ell(r) \!=\! 2r+a$, the maximal hypercubic radius for given $L/a$ 
is $r_m/a \!=\! \floor(L/2a)$, i.e. $r_m \!=\!11a$ and $r_m \!=\! 7a$ correspondingly. 
In Fig.$\,$\ref{fig:volcheck} we compare the results on the two volumes for $O_x$ and 
the pseudoscalar density at $p \!=\! 0.95$. The behavior for other operators is completely analogous. 
It is quite obvious that we have an excellent agreement over the range of common radii.  
Close to $r \!=\! r_m$ one expects some edge effects in principle and, to avoid the possibility of such 
contamination, we only extract the parameters of exponential behavior from distances up to 
$r \!=\! r_m \!-\! a$ in what follows. 

\subsection{Insensitivity and Weak Locality}

The results of the volume study show that error functions of hypercubic boundary approximants 
at $p \!=\! 0.95$ are exponentially boundable. Recalling the more complete results of
Fig.$\,$\ref{fig:stat_reg}, including other operators and the range of cutoffs $p$, it is quite 
obvious that overlap-based gauge operators are insensitive at any fixed $p\!<\!1$. Thus,
within the classification of Sec.$\,$\ref{sec:insensitivity}, they are weakly insensitive
lattice operators at the ultraviolet cutoff in question ($\beta \!=\!6.0$). Results completely
analogous to those of Fig.$\,$\ref{fig:stat_reg} were found also for the other two ensembles 
representing finer lattices. The question then arises whether continuum operators $O_x^c$ defined 
by their overlap-based constructions are themselves weakly insensitive or insensitive. This requires 
demonstrating the insensitivity of $O_x^c$ at any fixed $p$ (see Definition 3). 

\begin{figure}[t]
\begin{center}
    \centerline{
    \hskip 0.00in
    \includegraphics[width=7.0truecm,angle=0]{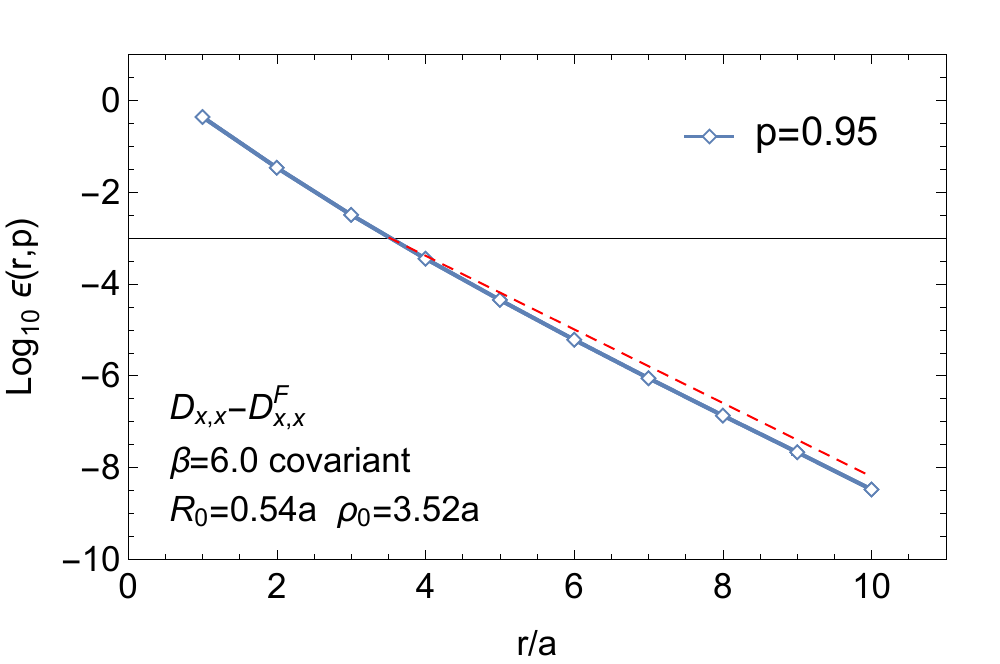}
    \hskip -0.00in
    \includegraphics[width=7.0truecm,angle=0]{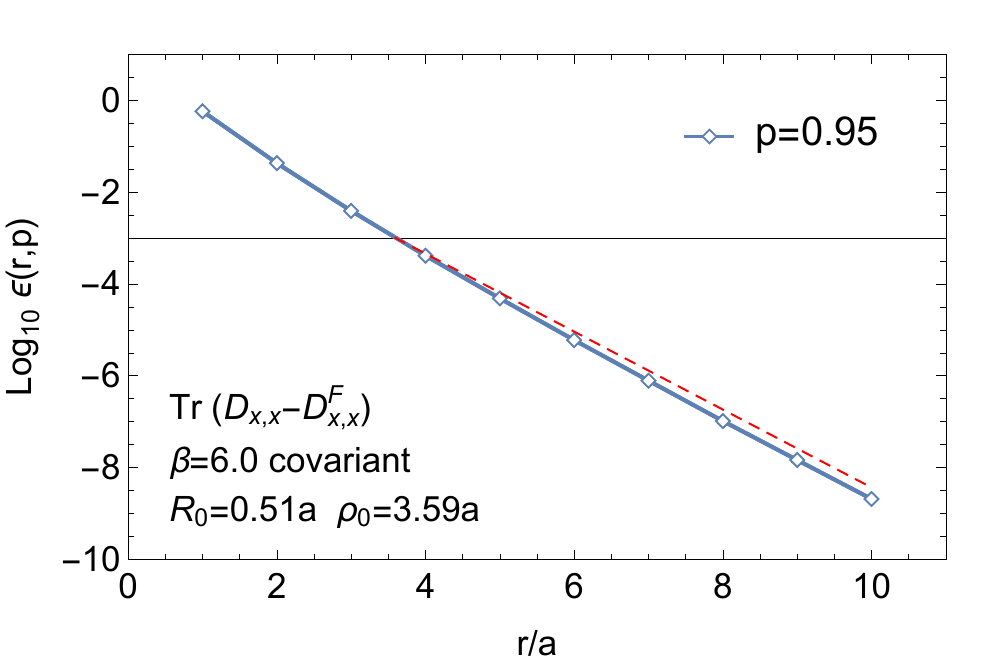}
     }
     \vskip -0.05in
    \centerline{
    \hskip 0.00in
    \includegraphics[width=7.0truecm,angle=0]{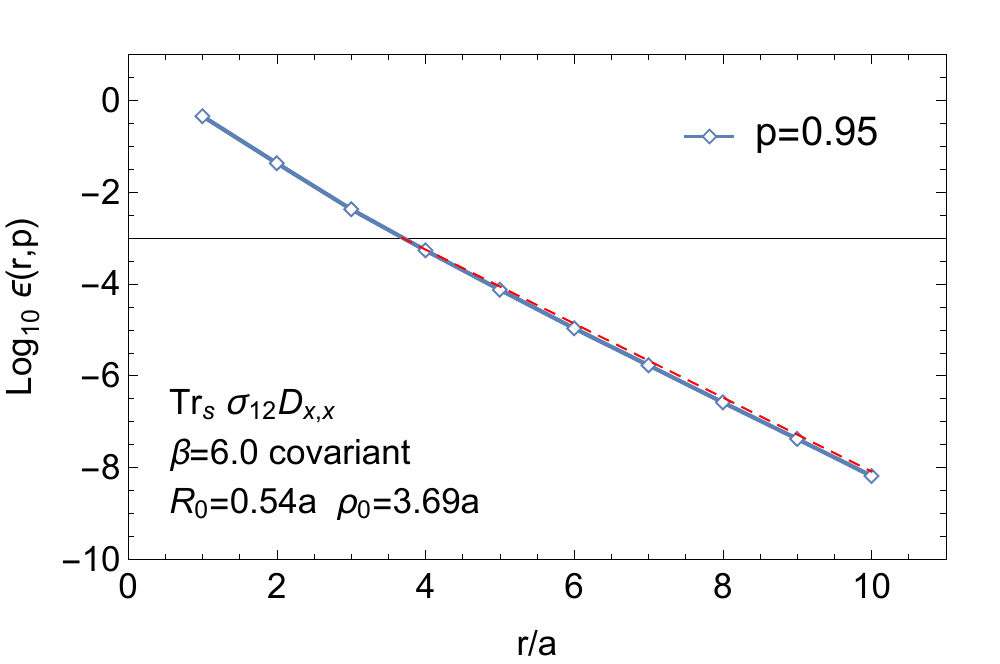}
    \hskip -0.00in
    \includegraphics[width=7.0truecm,angle=0]{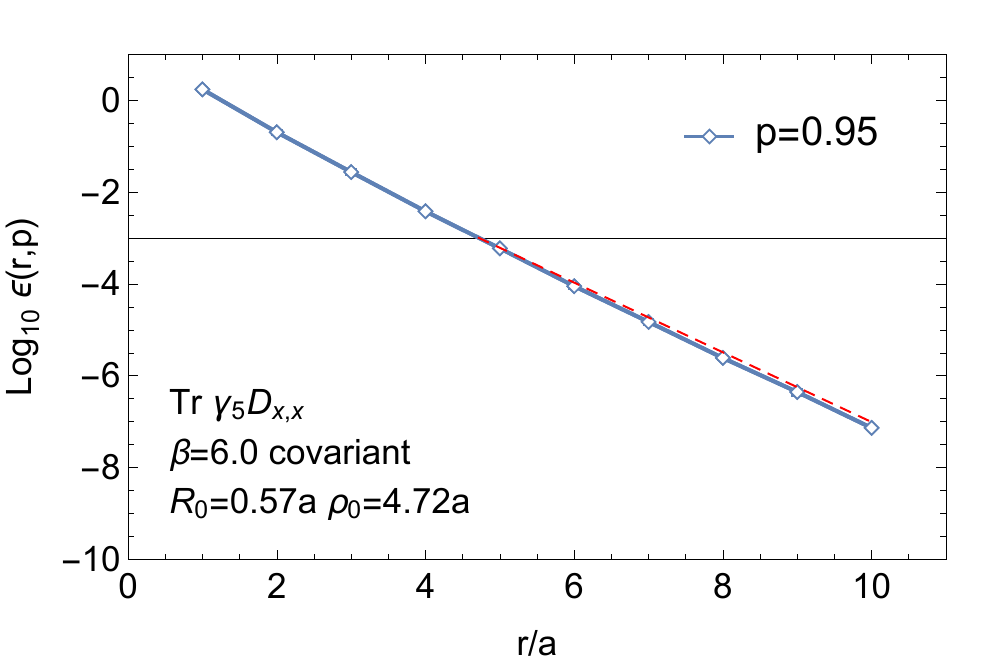}
     }
     \vskip -0.1in
     \caption{The behavior of $\rer(r)$ for covariant boundary approximant at $p \!=\! 0.95$ and 
                    $\beta \!=\! 6.0$. The optimal exponential bounds are shown at $\kappa \!=\!1$ and
                    $\Rer \!=\! 10^{-3}$, indicating the thresholds.}
     \label{fig:epsilon_beta60}
    \vskip -0.40in
\end{center}
\end{figure} 

To undertake this, the relative error data has to be characterized in terms of the corresponding 
bound parameters, as described in Sec.$\,$\ref{ssec:continuum}. In particular, we need to monitor 
the scales $R_0(p,a)$ and $\rc_0(\kappa,\Rer,p,a)$ associated with $\rer(r,p,a)$ 
(see Eq.$\,$\eqref{eq:215}) at arbitrary but fixed $\kappa \!>\! 1$ and $\Rer \!>\! 0$. As seen
from the representative data shown already, $R_0(p,a)$ can be estimated by simply fitting
the measured error functions to exponentials at large distances. In what follows, we extract
these effective ranges using the three largest radii smaller than $r_m$. Our analysis of the available
data does not support the presence of unbounded modulation in asymptotically exponential decays 
of $\rer(r)$. Consequently we can, and always do, determine $\rc_0$ at $\kappa \!=\!1$. 
While the threshold error $\Rer$ can be set to arbitrary positive value, we use $\Rer \!=\! 10^{-3}$ 
here, i.e. the absolute error at $\rc_0$ is one part in a thousand of the average operator magnitude.  
The associated bound then describes the errors for $r \!>\! \rc_0$ quite faithfully and can be used 
to predict the needed hypercubic radii in practical calculations. The situation for all operators of
interest at statistical cutoff $p \!=\! 0.95$ is shown in Fig.$\,$\ref{fig:epsilon_beta60}. 

\begin{figure}[t]
\begin{center}
    \centerline{
    \hskip 0.00in
    \includegraphics[width=7.0truecm,angle=0]{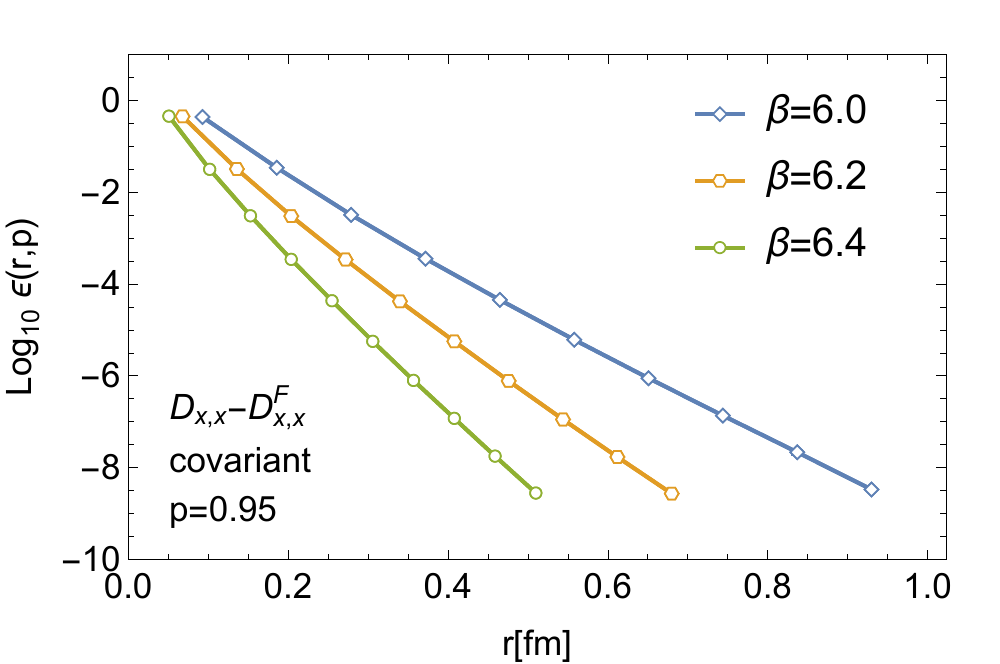}
    \hskip -0.00in
    \includegraphics[width=7.0truecm,angle=0]{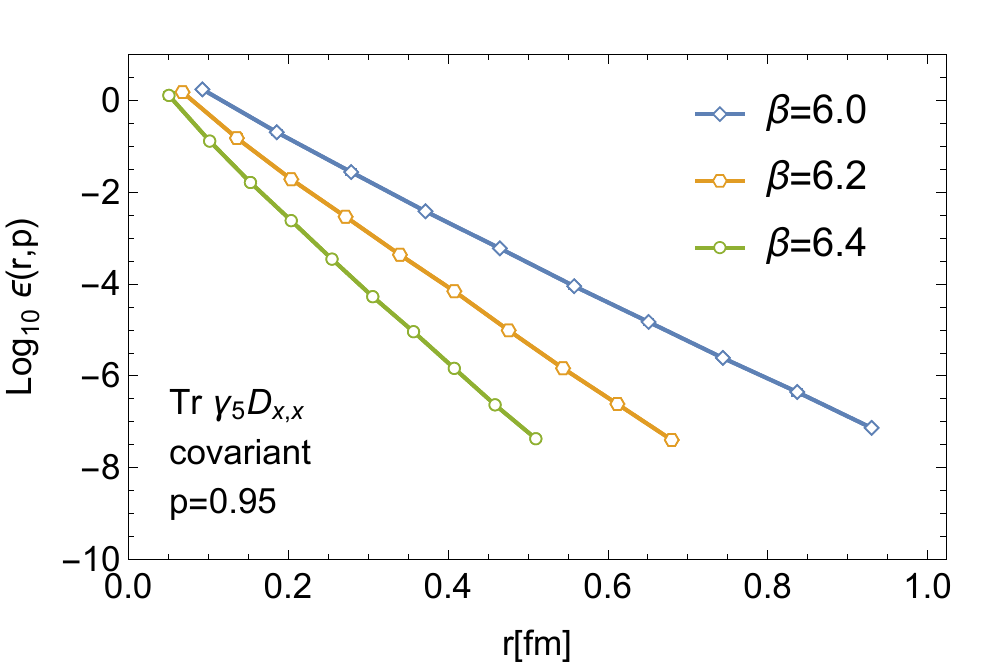}
     }
     \vskip -0.07in
    \centerline{
    \hskip -0.05in
    \includegraphics[width=7.0truecm,angle=0]{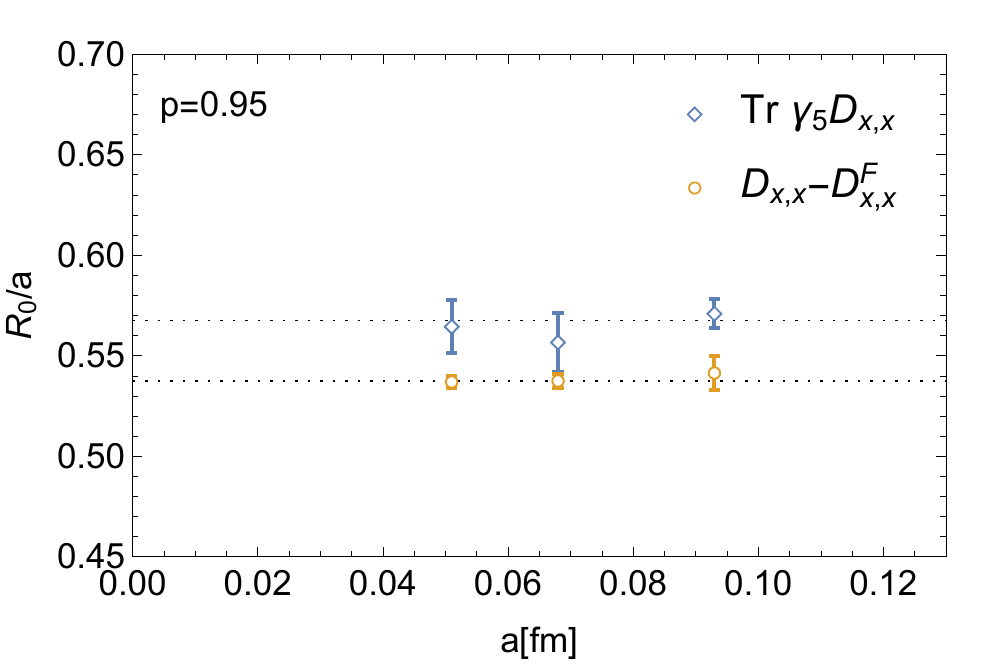}
    \hskip 0.11in
    \includegraphics[width=6.6truecm,angle=0]{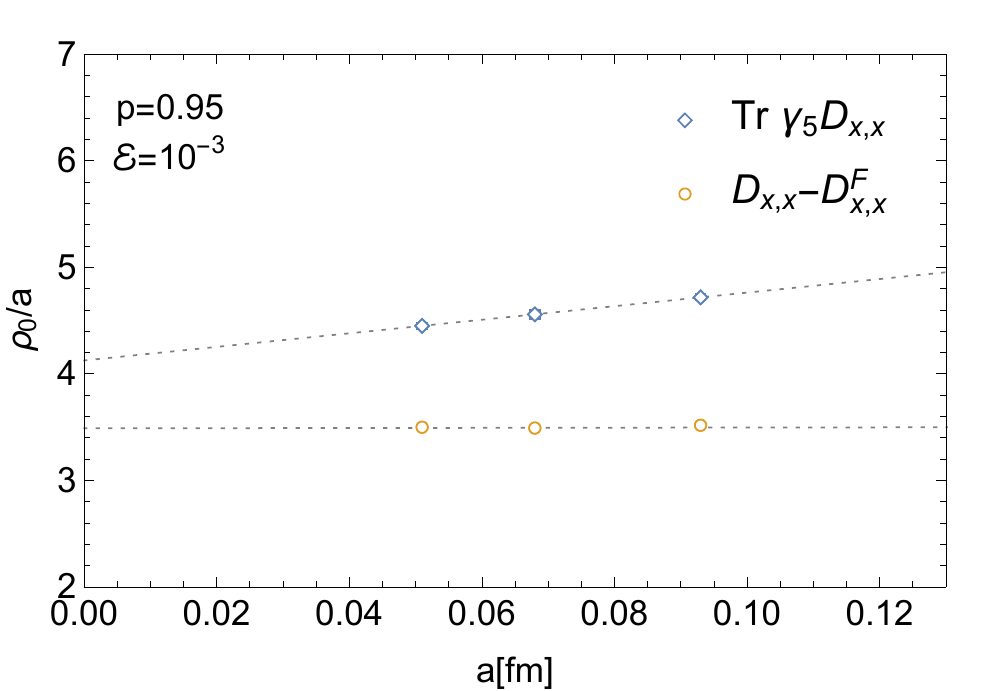}
     }
     \vskip -0.1in
     \caption{The behavior of $\rer(r,p\!=\! 0.95)$ at different ultraviolet cutoffs (top) and the lattice spacing 
     dependence of exponential bound parameters (bottom). Discussion is in the text.}
     \label{fig:eps_cont_p95}
    \vskip -0.40in
\end{center}
\end{figure}

We are now equipped to examine the removal of ultraviolet cutoff at fixed $p$.  
Fig.$\,$\ref{fig:eps_cont_p95} conveys the relevant data for $O_x$ and topological density.
In particular, the top row shows $\rer(r,p\!=\!0.95)$ as a function of physical distance at all three 
ultraviolet cutoffs. The effect of gauge fields beyond fixed distance $r$ clearly decays very 
rapidly and there is thus no doubt that the parameters of the corresponding exponential
bounds $R_0(p\!=\!0.95,a)$ and $\rc_0(p\!=\!0.95,a)$ are decreasing functions of $a$, and can 
only have finite continuum limits $R_0^c(p\!=\!0.95)$, $\rc_0^c(p\!=\!0.95)$. This behavior 
is readily present at all cutoffs $p$ accessible by our statistics, and we conclude that 
the corresponding continuum operators are {\em weakly insensitive}.

Making more restrictive conclusions about insensitivity, or assessing locality, requires cutoff-monitoring 
of the bound parameters. The bottom row of Fig.$\,$\ref{fig:eps_cont_p95} shows their dependence on
ultraviolet cutoff. With constant fits for $R_0/a$ and linear fits for $\rc_0/a$ added to guide the eye, 
our data conveys quite clearly that finite extrapolations in lattice units exist in both cases. This behavior 
is generic in $p$ and we conclude that
\begin{equation}
     R_0(p,a) \propto a    \qquad\quad  \rc_0(p,a) \propto a   \quad\; , \quad\;  \forall \,p<1
     \label{eq:580}            
\end{equation}
for all operators in question. It follows that $R_0^c(p)$ and $\rc_0^c(p)$ are zero, which in turn 
implies that their $p \!\to\! 1$ limits $R_0^c$ and $\rc_0^c$ vanish as well.
In other words, the continuum gauge operators based on the overlap are exponentially 
{\em insensitive} to distant fields (Definition 4) and {\em weakly local} (Definition 6).

\subsection{Strong Insensitivity and Locality}

In the formalism we adopted, weak insensitivity of defining lattice operators 
(sufficiently close to the continuum limit) constitutes a necessary precursor to insensitivity 
and weak locality in the continuum. Indeed, these concepts feature the order 
$\lim_{p \to 1} \lim_{a \to 0}$ in the removal of cutoffs. As discussed in the previous section, 
our analysis indicates that both of these properties in fact materialize in the overlap-based 
gauge operators.

In a similar manner, insensitivity of lattice operators (Definition 2) is a necessary precursor 
to continuum notions of strong insensitivity and locality, since the reverse order 
$\lim_{a \to 0} \lim_{p \to 1}$ of taking limits applies. Note that establishing strict lattice 
insensitivity can be a demanding task. Indeed, here the question whether potential outliers 
of a given bound form a set of measure zero enters most directly, and it is thus crucial that 
the statistics be sufficient to capture the nature of parameter behavior in the vicinity of 
$p \!=\! 1$.

\begin{figure}[t]
\begin{center}
    \centerline{
    \hskip 0.00in
    \includegraphics[width=7.0truecm,angle=0]{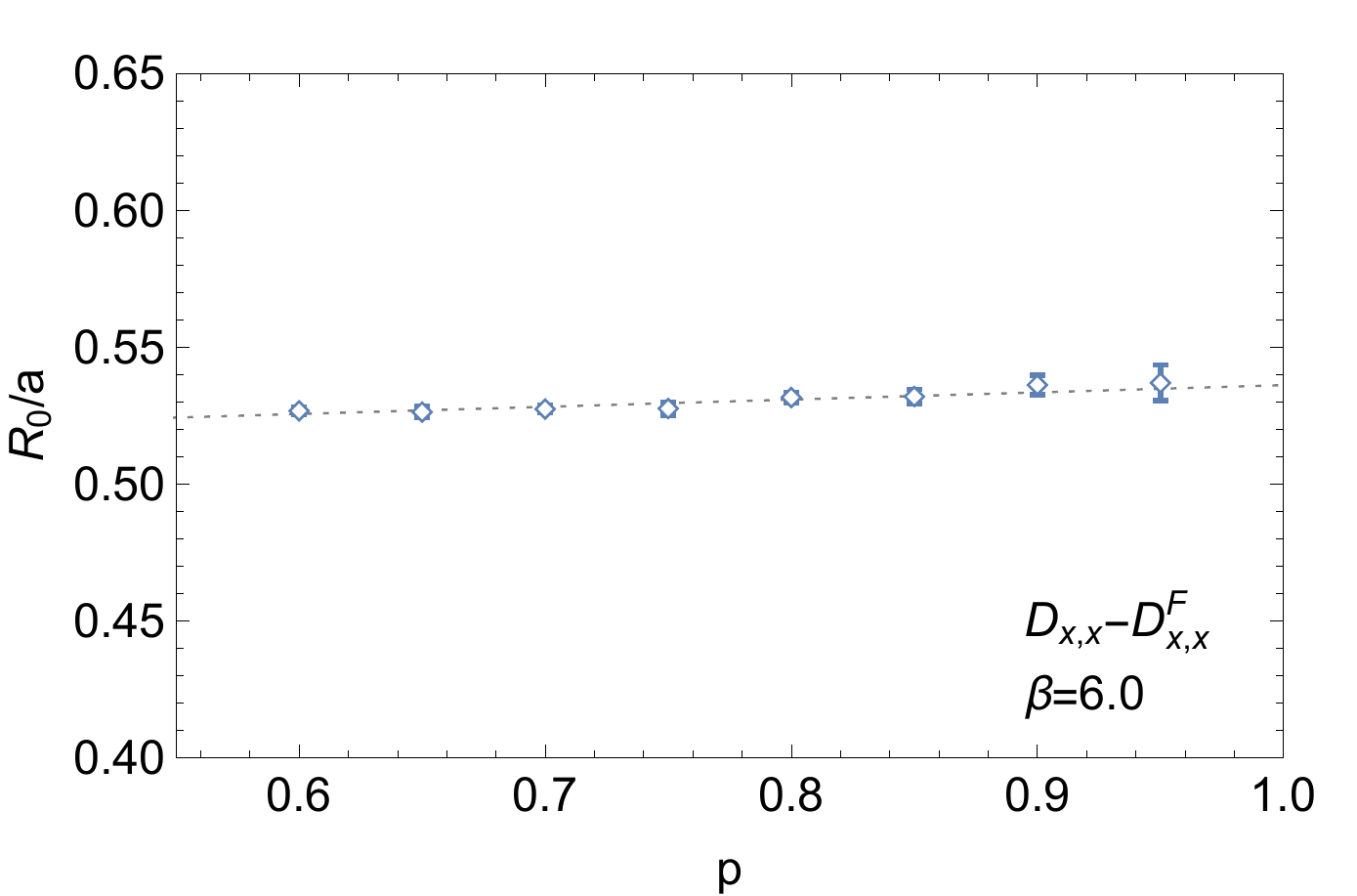}
    \hskip 0.11in
    \includegraphics[width=7.0truecm,angle=0]{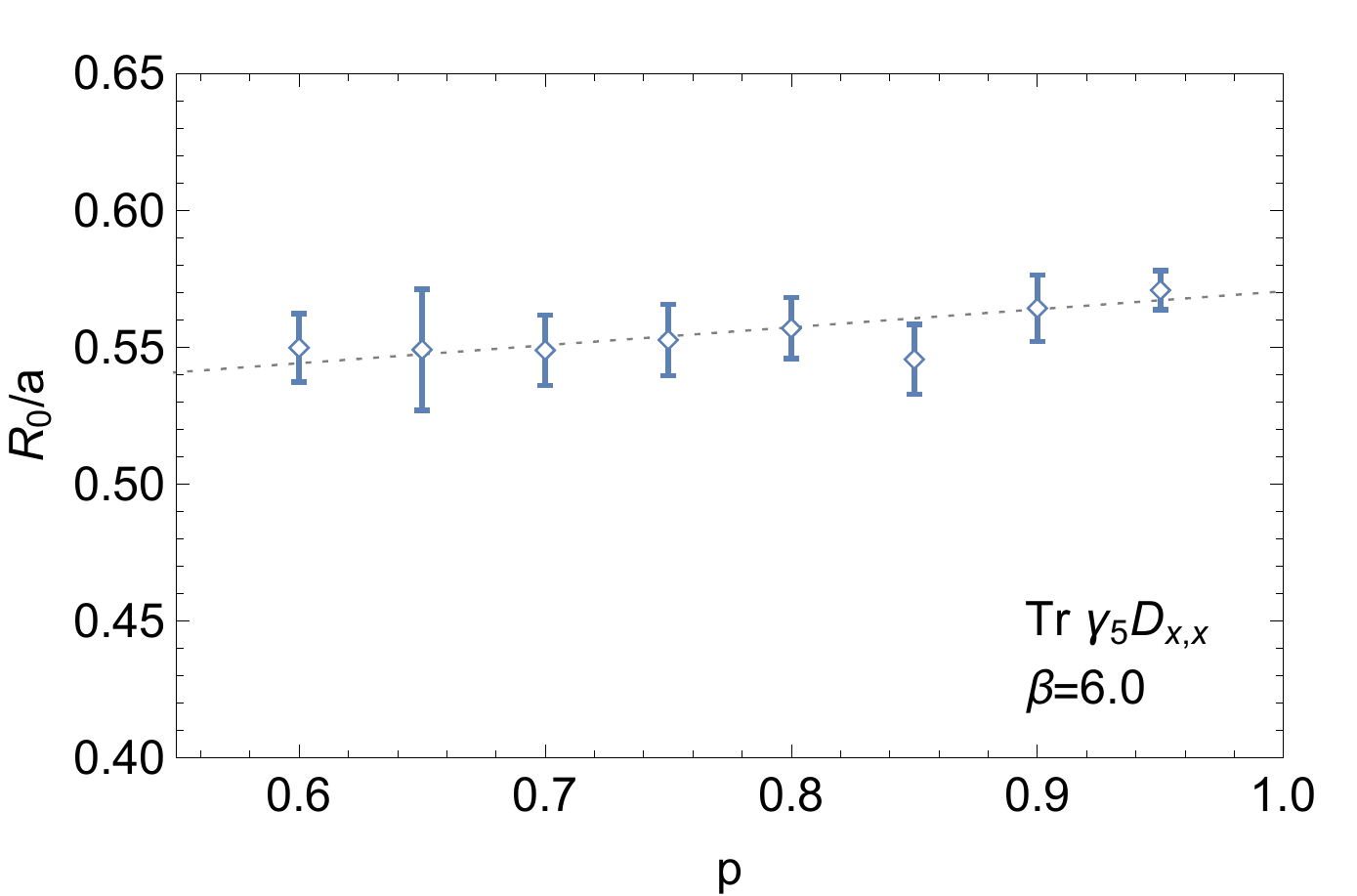}
     }
     \vskip -0.07in
    \centerline{
    \hskip -0.05in
    \includegraphics[width=7.0truecm,angle=0]{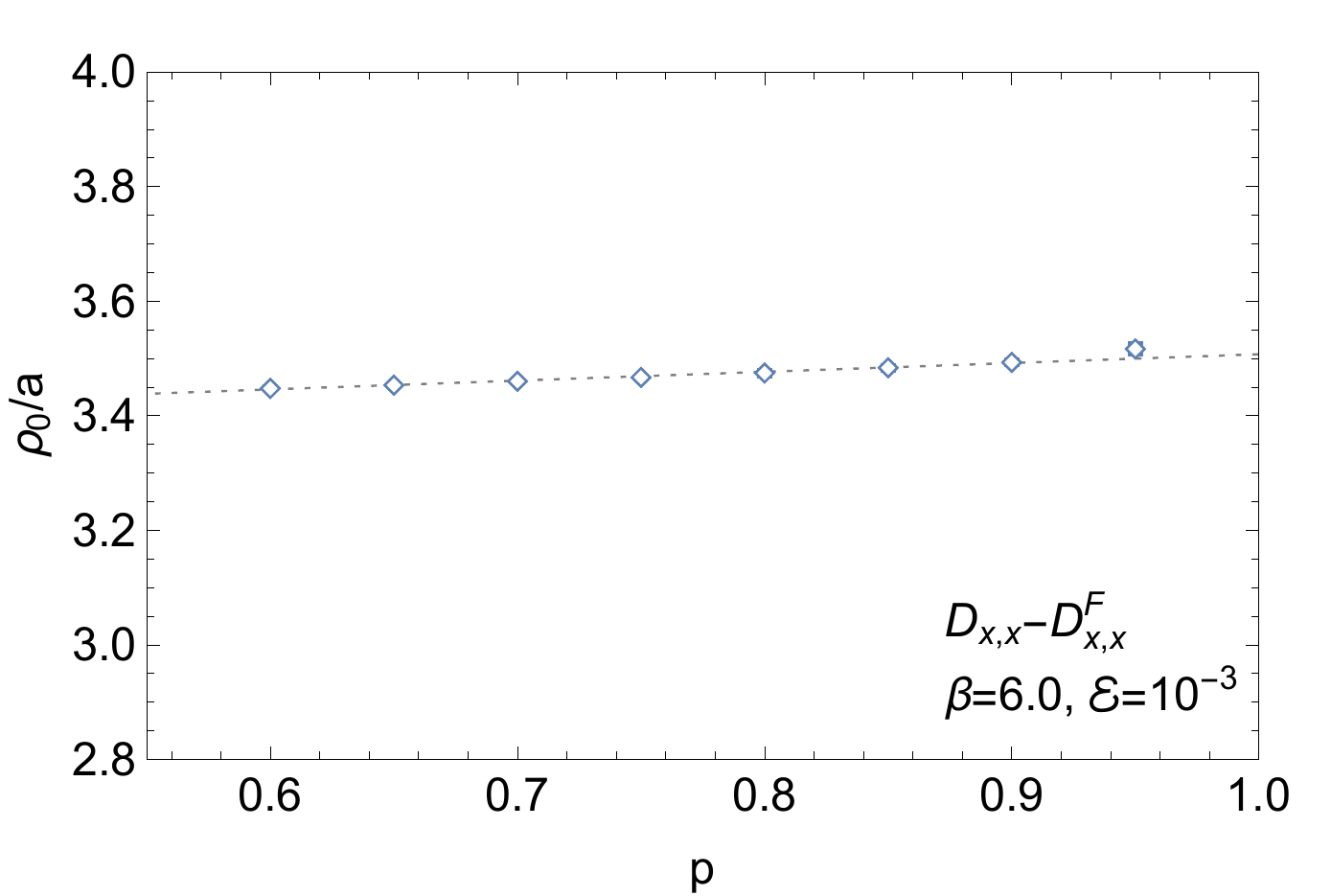}
    \hskip 0.11in
    \includegraphics[width=7.0truecm,angle=0]{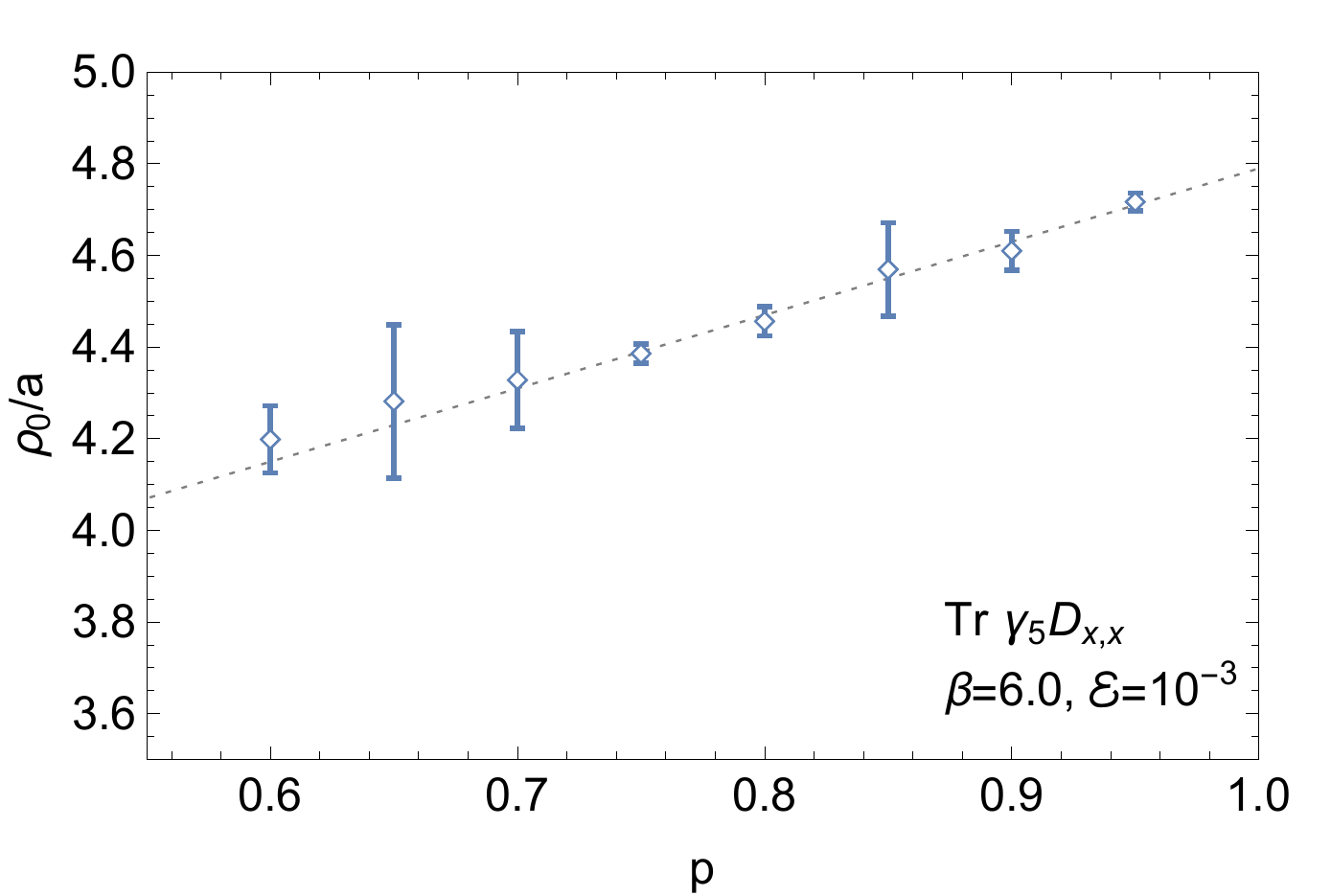}
     }
     \vskip -0.1in
     \caption{The $p$-dependence of exponential bound parameters for $O_x$ (left column) and 
     $q(x)$ (right column) at $\beta \!=\! 6.0$. See the discussion in the text.}
     \label{fig:pars_vsp_beta60}     
    \vskip -0.40in
\end{center}
\end{figure} 

While our statistics is rather limited in this regard, the look at the available data is revealing. 
In Fig.$\,$\ref{fig:pars_vsp_beta60} we show the $p$-dependence of $R_0$ and $\rc_0$ 
in lattice units at $\beta \!=\! 6.0$ for $O_x$ and topological density. First, one should realize
that it follows from definitions of both $R_0$ and $\rc_0$ that they are non-decreasing 
functions of $p$. In case of $R_0$ we only observe an extremely mild trend in this regard, 
well described by the linear behavior fitted to guide the eye. The situation is similar for $\rc_0$, 
albeit the rise in case of topological density is visibly steeper than for $O_x$.

\begin{figure}[t]
\begin{center}
    \centerline{
    \hskip 0.00in
    \includegraphics[width=7.0truecm,angle=0]{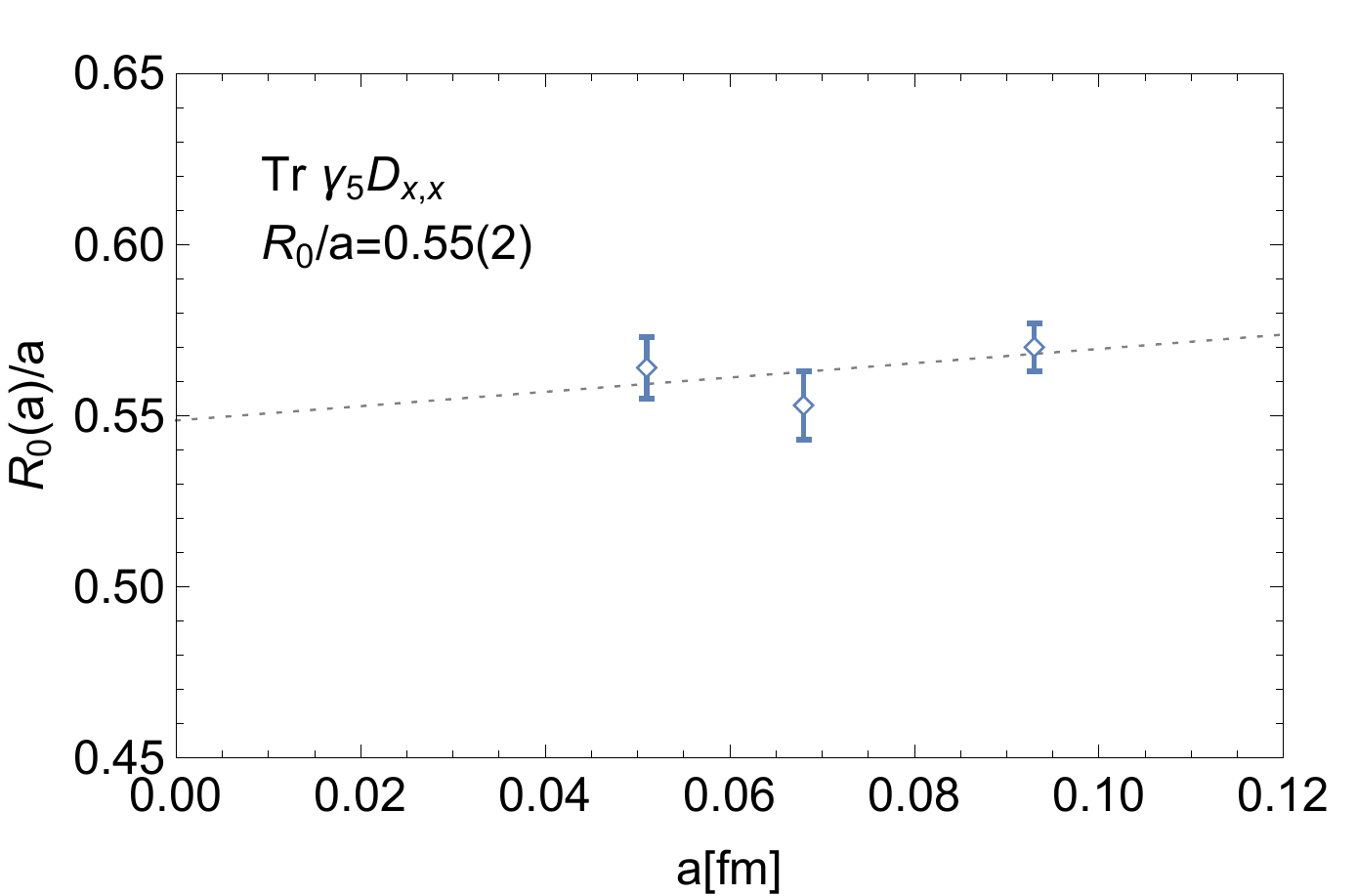}
    \hskip 0.11in
    \includegraphics[width=7.0truecm,angle=0]{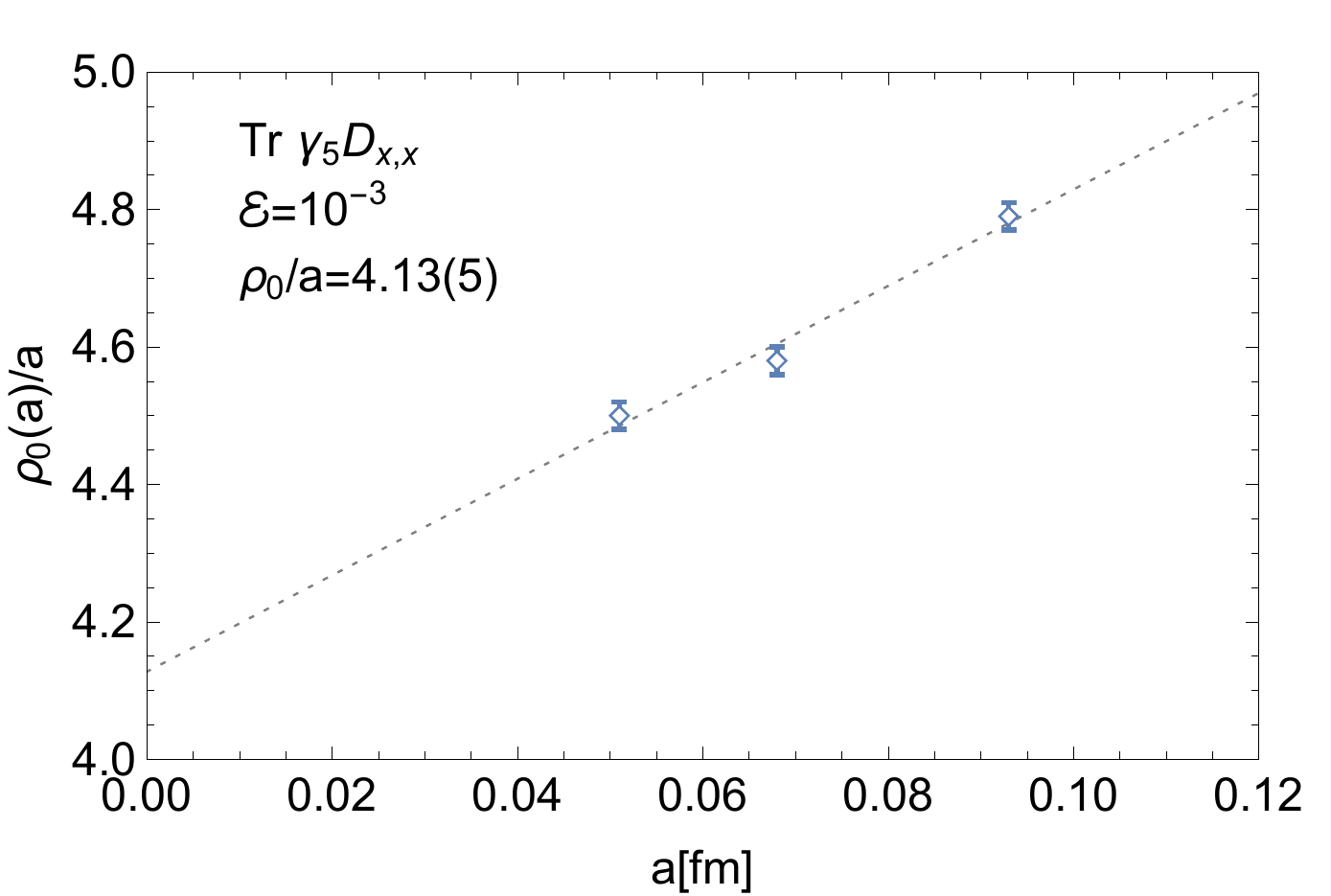}
     }
     \vskip -0.02in
    \centerline{
    \hskip -0.05in
    \includegraphics[width=7.0truecm,angle=0]{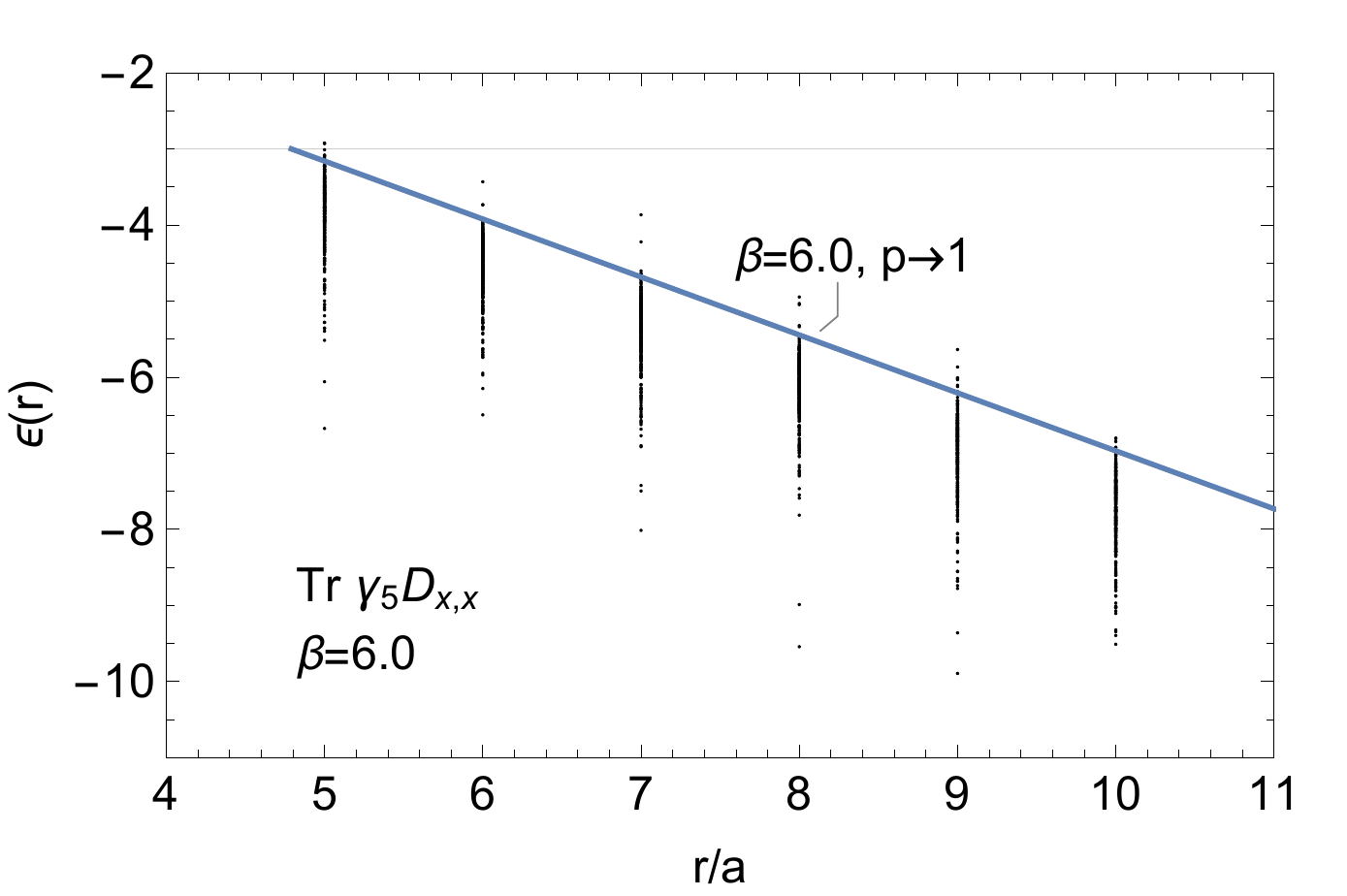}
    \hskip 0.11in
    \includegraphics[width=7.0truecm,angle=0]{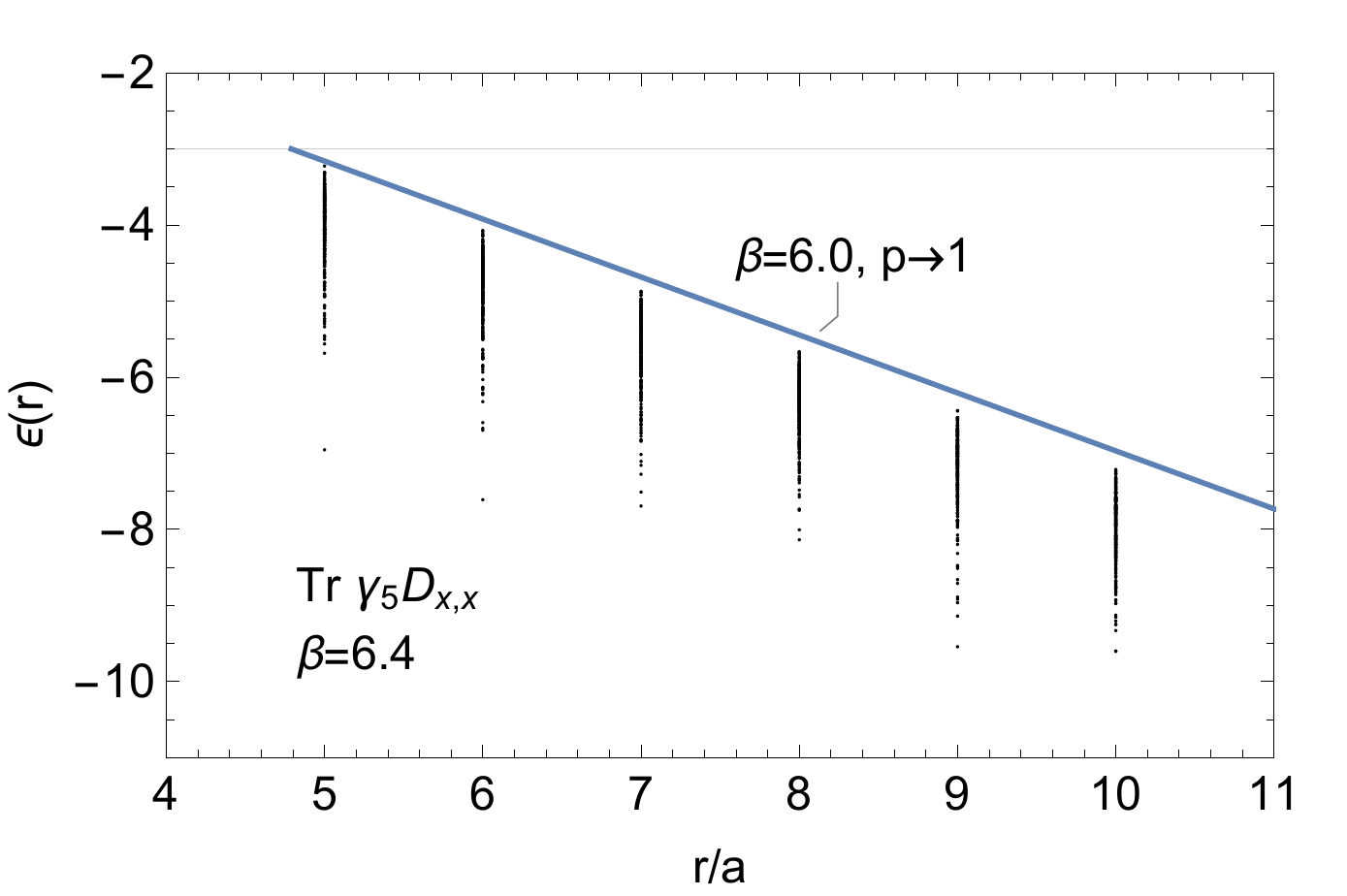}
     }
     \vskip -0.1in
     \caption{Top: continuum extrapolation to determine $\bar{R}_0/a$ and $\bar{\rc}_0/a$  
     (topological density). Bottom: $p \!\to\! 1$ bound at $\beta \!=\!6.0$
     (left) and the same bound with data at $\beta\!=\!6.4$ (right).}
     \label{fig:strong_insens}     
    \vskip -0.40in
\end{center}
\end{figure}

Assuming that the observed trends do not change dramatically closer to $p \!=\! 1$, the above 
would suggest that finite $p \!\to\! 1$ limits of parameters exists and, consequently, that 
the overlap-based gauge operators are exponentially insensitive at the lattice level. With that, 
the issues of strong insensitivity/locality in the continuum would become well-posed, and 
the corresponding tendencies to obtain $\bar{R}_0$ and $\bar{\rc}_0$ in lattice units shown  
in Fig.$\,$\ref{fig:strong_insens} (top) for topological density. Finite extrapolations 
to continuum limit are then readily concluded from the data, implying both of the above 
continuum properties.

However, the assumption about the behavior closer to $p \!\to\! 1$ does need to be scrutinized. 
Indeed, our statistics allows for reliable extraction of bound parameters up to $p \!=\! 0.95$, but 
the fluctuations in error do become somewhat larger for the ``outliers". Indeed, in 
Fig.$\,$\ref{fig:strong_insens} (bottom, left) we show the $p \!\to\! 1$ extrapolated bound at 
$\beta \!=\! 6.0$ together with actual samples of approximation differences. While the bound 
at $p\!=\! 0.95$ would leave out 16 outliers at each $r/a$, the extrapolated bound still leaves out 
about 4 on average, almost certainly pushing a small residual probability outside its reach. 
This suggests that the behavior of bound parameters is somewhat modified in the immediate 
vicinity of $p \!=\! 1$ at $\beta \!=\!6.0$, and significantly larger statistics needs to be invoked 
to extrapolate reliably.

Nevertheless, even with the current data, one can make the existence of $p$-independent bound 
sufficiently close to the continuum limit quite plausible. Indeed, if $R_0(a_0)$, $\rc_0(a_0)$ are 
the $p \!\to\! 1$ extrapolated bound parameters at lattice spacing $a_0$, let $R(a) \!=\!a R_0(a_0)/a_0$
and $\rc(a) \!=\! a \rc_0(a_0)/a_0$ be the parameters of a (non-optimal) bound at lattice spacing 
$a \!<\! a_0$. Thus, in lattice units, this bound is the same for all $a \!\le\! a_0$. Due to the observed 
decreasing trend of $\rc_0(a)/a$ for $a \!\to\! 0$ (top right in Fig.$\,$\ref{fig:strong_insens}) such bound 
may become $p$-independent sufficiently close to the continuum limit. Associating $a_0$ with 
$\beta \!=\! 6.0$ ensemble, in Fig.$\,$\ref{fig:strong_insens} (bottom, right) we show the relation 
of errors at $\beta \!=\! 6.4$ to this bound and, indeed, there are no violations at our level
of statistics. Needless to say though, the issue of strong insensitivity and locality in overlap-based
continuum operators needs to be resolved via a direct extensive calculation.

\medskip

\section{Application: Topological Structure in QCD Vacuum}
\label{sec:structure}

The computational aspect of insensitivity considerations becomes relevant in practice when 
the problem at hand significantly benefits from the use of non-ultralocal operators. One example 
is the study of QCD vacuum structure via overlap-based topological charge density $q(x)$.
This stems from the fact that $q(x)$ is topological, i.e. stable under deformations of the gauge field, 
directly on the lattice. Indeed, this property was instrumental in finding that, when fluctuations 
at all scales are included, topological charge in QCD organizes into low-dimensional global structure 
of space-filling type~\cite{Hor03A}. The structure takes the form of a double sheet formed by topological 
densities of opposite sign, and is inherently global~\cite{Hor05A}. In particular, this space-spanning 
object cannot be broken into individual pieces without severely affecting topological susceptibility. 

\begin{figure}[t]
\begin{center}
    \centerline{
    \hskip 0.00in
    \includegraphics[width=7.5truecm,angle=0]{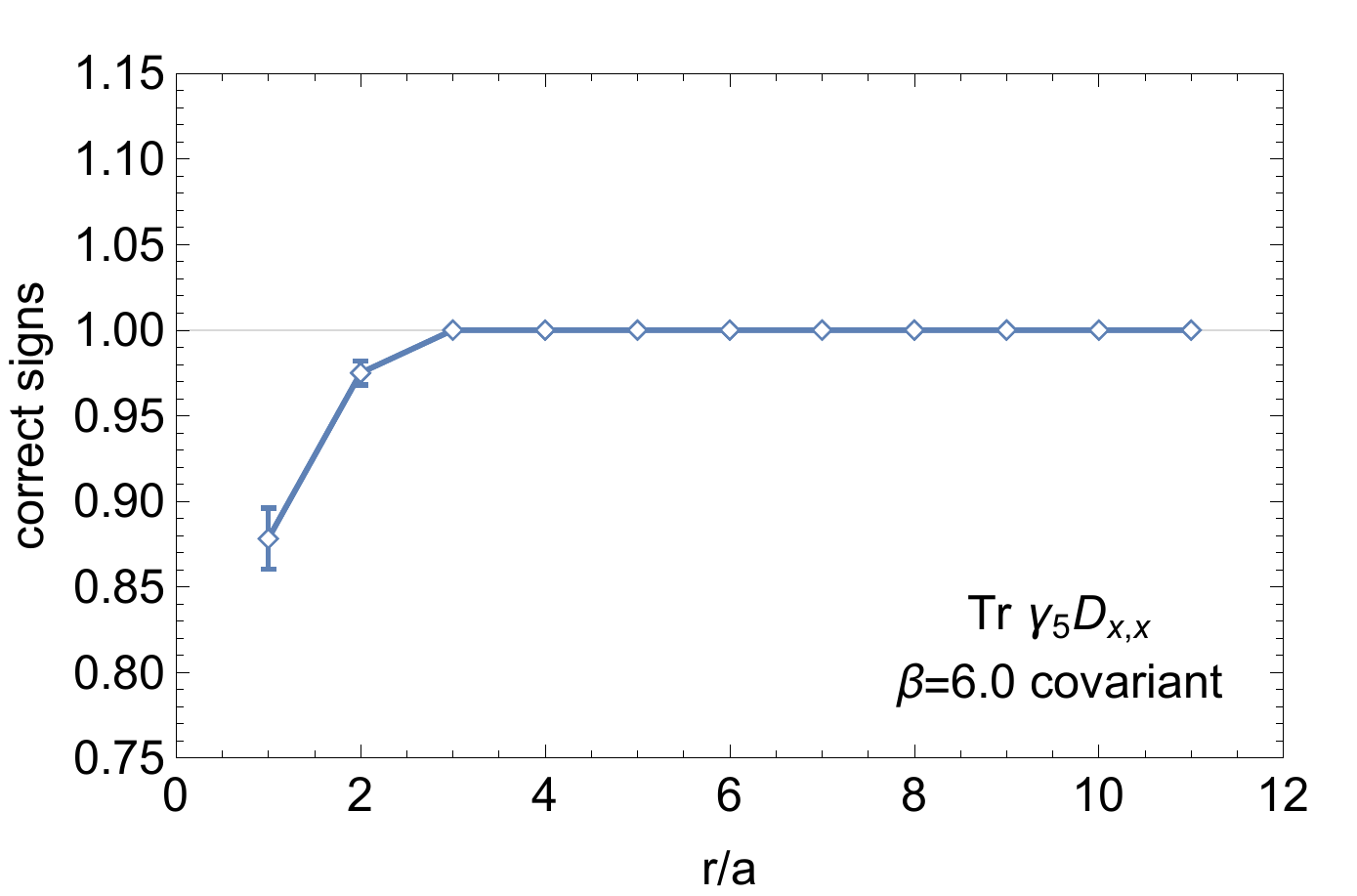}
    \hskip 0.30in
    \includegraphics[width=7.5truecm,angle=0]{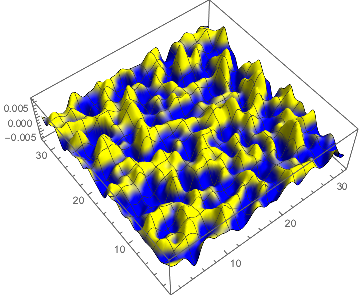}
     }
     \vskip -0.0in
     \caption{Left: the rate of correct sign when approximating topological density by covariant boundary 
     approximant at $\beta \!=\! 6.0$. Right: typical 2d slice through global topological structure on 
     $32^4$ system, computed via boundary approximant with $r\!=\!6a$.}
     \label{fig:sign_val}
    \vskip -0.40in
\end{center}
\end{figure}

Computational issue hampering extensive investigations of the above type is that, since $q(x)$ 
needs to be evaluated on entire lattices, the use of standard overlap implementations leads 
to costs that scale at least as $V^2$.  As we argued extensively, utilizing hypercubic boundary 
approximants effectively turns this into a generic $V$-problem with pre-factor depending only
logarithmically on the desired precision. One should realize in this regard that, while boundary 
insensitivity guarantees eventual fast convergence in the radius of the approximant, the sufficiency 
criteria for this radius are problem-specific. For example, if the goal were to reliably determine 
the space-time structure in the sign of topological charge (see Ref.~\cite{Hor08}), then the relevant 
criterion would be a sufficiently low rate of sign violations in the approximant. In Fig.~\ref{fig:sign_val} 
(left) we show such data for ensemble at $\beta\!=\!6.0$. Within the available statistics, no violations 
are observed already at $r\!=\! 3a$. More quantitatively, the rate of violation can be estimated 
to be significantly better than 1\% at $r\!=\! 3a$, and is likely negligible for all practical purposes 
at $r\!=\!4a$, providing for a safe choice to fix when working at arbitrarily large volume. 

If the goal is to gain more detailed information on the topological structure, we can use the values 
of bound parameters at various statistical cutoffs $p$ to obtain needed estimates.
Alternatively, one can again examine the criterion in question as a function of hypercubic 
radius $r$ in a preliminary study. In the context of topological charge, a suitable generic requirement
is that its global integer value be reproduced to a prescribed accuracy in the approximation. 
To gain some quantitative feel on this and other aspects, we have computed configurations of 
topological charge at various couplings in pure-glue lattice gauge theory with Iwasaki gauge action. 
While the analysis of these results will be given elsewhere, here we point out that these computations
for $24^4$ system at $a\!=\!0.055\,$fm yield the typical absolute error of $10^{-3}$ for global charge 
using $r\!=\!5a$ boundary approximant. The situation is quantitatively similar for $32^4$ 
system at $a\!=\!0.041\,$fm (same physical volume) using $r\!=\!6a$ approximant. The typical 
profile of topological structure on 2d slice of space-time is shown in Fig.~\ref{fig:sign_val} (right).
The properties of the structure conform to those concluded in Ref.~\cite{Hor03A}.

\section{Summary and Conclusions}
\label{sec:conclude}

Locality is an important ingredient and guiding principle in describing natural phenomena via 
quantum fields. In this paper we reexamined this notion in the context of Euclidean field theories, 
such as Euclidean QCD, with quantum continuum dynamics defined via lattice-regularized 
path integral. While the approach is entirely general with respect to the fundamental field 
content, our discussion was carried out for operators composed of gauge fields. This is partially
motivated by the fact that locality properties for useful non-ultralocal gauge operators have 
not been previously studied.

Locality is a strictly continuum notion: it is a property of continuum object $O_x^c(A)$, 
operationally defined via certain lattice $O_x(U)$ in the limiting process of driving the theory 
to the continuum limit.\footnote{Note that any lattice operator that is not point-like already extends 
over finite physical distance at finite lattice spacing, thus violating the physical meaning of locality.} 
However, there is a more general concept, meaningful already on the lattice, 
that ushers locality into the continuum as its special case. 
Indeed, universality arguments suggest that such lattice precursor of locality is the exponentially 
weak dependence of $O_x(U)$ on fields residing sufficiently far from $x$. One of the main 
results in this paper is a novel formulation of this {\em exponential insensitivity} to distant fields.

Our approach to insensitivity is information-like in that it probes how well is it possible to know 
$O_x(U)$ when $U$ itself is only known in the neighborhood of $x$ with radius $r$, i.e. when only 
the patch $U^{x,r} \!\subset\! U$ of the gauge field is accessible. This can be quantified by the accuracy 
achievable by the approximants $O_x(U) \rightarrow O_x^r(U) \!=\! O_x^r(U^{x,r})$ enforcing this 
restriction. Exponential insensitivity to distant fields is then the ability to construct the sequence 
$O_x^r$ whose accuracy can be bounded by a decaying exponential in $r$.

Making this idea precise requires us to specify the measure for ``accuracy" of a given approximant,
with subtlety being that there is, in fact, a distribution of errors generated by the underlying path integral 
distribution of gauge fields for the theory in question. To have a tunable control over the scope of these 
deviations, and to accommodate cases where violations of exponential bound form a probabilistic 
set of measure zero, we introduce the procedure of {\em statistical regularization}. 
Here boundability relates to the fraction $p$ of the path integral population with smallest errors, 
which amounts to the concept of least upper bound with probability $p$. At every statistical 
cutoff $p$, the problem of exponential boundability is well-posed, and the general framework 
describing degrees of exponential insensitivity in the process of statistical ($p \!\to\! 1$) and ultraviolet 
($a \!\to\! 0$) cutoff removals ensues. Locality, then, is simply associated with vanishing of scales 
describing the cutoff-removed bounds.
   
Classification of composite operators by degree of exponential insensitivity can in principle 
proceed by examining the properties of their optimal approximant~\cite{Hor16C}. 
However, in the realm of computable operators, the evaluation of this $O_x^{r, op}(U^{x,r})$ is much
more expensive than that of $O_x(U)$ alone, except for simple ultralocal cases. At the same time, 
constructing any approximant with required properties demonstrates insensitivity. This motivates 
our proposal to consider boundary approximants, wherein the value $O_{x,L}(U)$ on the ``large" 
system of size $L$, is replaced by the value $O_{x,2r+a}(U^{x,r})$ on the ``small" system spanned 
by the hypercubic field patch alone. Such approximant probes insensitivity to distant fields by inserting 
an artificial boundary distance $r$ away from $x$, giving rise to a somewhat stronger, but physically 
well-motivated notion of {\em boundary insensitivity} and {\em boundary locality}. 

Determining boundary insensitivity is a straightforward task for any computable $O_x(U)$. 
Moreover, if there is a program evaluating $O_{x,L}(U)$ with polynomial complexity in $L/a$ 
(typical case), then the property of boundary insensitivity makes it possible for $O_{x,L}(U)$ 
to be computed {\em efficiently}. Indeed, running the same program on input corresponding to
boundary subsystem affording the desired accuracy $\rer$, results in volume-independent 
computation whose cost only grows as a power of $\log 1/\rer$. This connection between efficient 
computation and exponential insensitivity/locality is one of the main conceptual points emphasized 
in this paper. Its full scope and ramifications will be explored elsewhere.

\smallskip

{\bf 1.} The boundary insensitivity analysis was applied to gauge operators based on the overlap 
Dirac matrix, such as topological charge density or field-strength tensor. Apart from illustrating 
the general scheme, our goal was to examine locality properties of these useful non-ultralocal 
operators on realistic equilibrium backgrounds. Invoking the hierarchy developed here, our 
results clearly indicate that these operators are weakly insensitive at the lattice level, and weakly 
local in the continuum. Quantitative evidence has also been presented favoring insensitivity on 
the lattice, and locality in the continuum. However, putting these latter findings on a firm ground 
requires more extensive simulations to be performed. Overall, our analysis shows that there 
is little doubt that these operators follow the universal behavior in the continuum.

\smallskip

{\bf 2.}
Important practical outcome of our numerical experiments with boundary approximants is that 
their power of exponential improvement in estimating $O_x(U)$ pans out already at small values 
of hypercubic radii. Our experience with overlap-based topological density in particular, suggests 
that using this technique makes extensive vacuum structure studies with such complicated 
operator no longer computationally prohibitive.

\smallskip

{\bf 3.}
For fixed $r/a$, the approximant $O_x^r$, facilitating the insensitivity of $O_x$, is simply an
ultralocal lattice operator that could be interesting in its own right. In case of gauge operators 
studied here, the covariant form of boundary $O_x^r$ (default choice) inherits all 
potential symmetries of $O_x$, making it a particularly attractive choice for standalone use. 

\smallskip

{\bf 4.}
The exponential insensitivity framework can be viewed as a tool to classify all lattice-defined 
continuum operators $O_x^c$ in terms of their reach. This is quantified by the characteristic 
length scales introduced, such as the range $R_0^c$ and threshold $\rc_0^c$.
We expect such description to be useful for defining non-local operators at fixed scale.  

\bigskip
\noindent{\bf Acknowledgments:} 
We thank Thomas Streuer for participating in the early stages of this project.
A.A. is supported in part by the National Science Foundation CAREER grant 
PHY-1151648 and Department of Energy grant DE-FG02-95ER40907. I.H. acknowledges 
the support by Department of Anesthesiology at the University of Kentucky.

\bigskip\medskip

\begin{appendix}

\section{Regularity of the Approximation}
\label{app:D}

The notion of exponential insensitivity aims at capturing the exponential (in distance) control 
over values of the composite operator. Such concept would be unsatisfactory if it admitted 
cases wherein the zero-measure sample set of violations contributed finitely to the statistics 
of $O_x$. To prevent such singular behaviors from being considered insensitive, we 
formulate here the corresponding requirement of {\em regularity}.

Strictly speaking, regularity needs to be examined in conjunction with any statistical inference 
wherein $O_x$ is replaced by its approximant $O_x^r$. Relevant situations may also involve 
arbitrary other operators but, as a defining property, it relates to $O_x$ itself. Thus, in addition 
to the error function $\aer(r,p,a)$, function ${\cal A}(r,p,a)$ is defined as a contribution to mean 
magnitude of $O_x$ due to samples whose error is larger than $\aer(r,p,a)$, i.e. due to 
``violations" at statistical cutoff $p$. If $\bar{\cal A}(r,p,a)$ is the portion governed by cutoff 
$p$, then
\begin{equation}
    \langle \, \| O_x \| \, \rangle_a \,=\, \bar{\cal A}(r,p,a)  \,+\, {\cal A}(r,p,a)
    \label{eqapp:010}      
\end{equation}
In the same way, let ${\cal A}_{\tt ap}(r,p,a)$ be the part of $\langle \, \| O_x^r \| \, \rangle_a$ 
due to violations. Approximation $O_x^r$ of lattice operator $O_x$ at fixed ultraviolet cutoff $a$ 
is regular if there exists $r_0$ such that
\begin{equation}
     \lim_{p \to 1}  {\cal A}(r,p,a) \,=\,0                 \qquad\qquad  
     \lim_{p \to 1}  {\cal A}_{\tt ap}(r,p,a) \,=\,0     \qquad,\qquad       
     \forall  \,  r  \ge  r_0 
     \label{eqapp:020}     
\end{equation}
While the first condition ensures that excluding the zero-measure set of violations doesn't 
lead to finite distortion, the second one ascertains that replacing it with zero-measure set
of approximants doesn't do that either. Regularity of the approximation is required for lattice 
operator $O_x$ to be {\em weakly insensitive}.

Additional care needs to be taken when examining exponential insensitivity in the continuum 
(Definition 4). Here $a \!\to\! 0$ limit is taken at fixed $p$ first, which may lead to a finite influence 
in subsequent $p \!\to\! 1$ limit even when lattice operator is regularly approximated at any $a$. 
Moreover, the contribution of outliers has to be considered relative to $a$-dependent average 
magnitude. In particular, the {\em continuum regularity} condition reads
\begin{equation}
     \lim_{p \to 1} \lim_{a \to 0}  \,\frac{{\cal A}(r,p,a)}{\langle \, \| O_x \| \, \rangle_a}  \,=\, 
     \lim_{p \to 1} \lim_{a \to 0}  \,\frac{{\cal A}_{\tt ap}(r,p,a)}{\langle \, \| O_x \| \, \rangle_a}  \,=\,      
     0     \qquad,\qquad     \forall  \,  r  \ge  r_0 
     \label{eqapp:030}      
\end{equation}
It is worth noting that, while lattice regularity \eqref{eqapp:020} is automatically satisfied for bounded 
pair $O_x$ and $O_x^r$, lattice boundedness at any $a$ does not necessarily guarantee 
\eqref{eqapp:030}.

\section{Parametric Freedom in Statistical Cutoff Removal}
\label{app:B}

This Appendix elaborates on point (i) of discussion in Sec.~\ref{ssec:removal}. 
In particular, we first aim to show that if $\rc_0(\kappa,\Rer) \!<\! \infty$, then 
$\rc_0(\kappa,\Rer') \!<\! \infty$ for all $\Rer' > 0$. More explicitly
\begin{equation}
     \lim_{p \to 1} \rc_0(\kappa, \Rer, p) < \infty
     \quad \Longrightarrow \quad 
     \lim_{p \to 1} \rc_0(\kappa, \Rer', p) < \infty      
     \quad,\quad \forall \; \Rer' > 0 
     \label{eqapp:810}  
\end{equation}
Before proceeding, it is useful to summarize the monotonicity properties of 
$\rc_0(\kappa, \Rer, p)$. From Eqs.~(\ref{eq:k30},\ref{eq:k50},\ref{eq:k60},\ref{eq:r20}) 
it follows that $\rc_0(\kappa, \Rer, p)$ is decreasing in $\Rer$ and non-increasing in 
$\kappa$. Moreover, since $\aer(r,p)$ is non-decreasing in $p$, so is $\Aero(R,\rc,p)$, 
and consequently $\rc_0(\kappa, \Rer, p)$. 

To show \eqref{eqapp:810}, first assume that $\Rer' \!>\! \Rer$, implying that 
$\rc_0(\kappa, \Rer', p) \!<\! \rc_0(\kappa, \Rer, p)$, $\forall \, p \!<\! 1$. Given that $\rc_0$ 
is non-decreasing in $p$, the existence of $p \!\to\! 1$ limit on the right side of this 
inequality implies the existence of the limit on the left side, as claimed by \eqref{eqapp:810}.
To include the case $\Rer' \!<\! \Rer$, first note that $\rc_0(\kappa, \Rer, p)$ is a solution
of the equation
\begin{equation}
     \Aero(R, \rc_0, p)  \,\equiv\,  
     B(R, \rc_0, p) \exp\Bigl( -\frac{\rc_0}{R} \Bigr)  \;=\;
     \Rer  \, \langle  \| O_x \| \rangle
     \qquad \text{with} \qquad
     R = \kappa R_0(p)
     \label{eqapp:820}       
\end{equation}
Combining this with the analogous equation for $\Rer'$ and taking into account that 
function $B(R, \rc_0, p)$ is non-increasing in $\rc_0$ while $\rc_0$ is itself decreasing 
in $\Rer$, we obtain the inequality
\begin{equation}
    \rc_0(\kappa,\Rer', p)   \,\le\, \rc_0(\kappa, \Rer, p)   \,+\,
    \kappa R_0(p) \log \frac{\Rer}{\Rer'}
    \qquad,\qquad \Rer' < \Rer 
    \label{eqapp:830}           
\end{equation}
Since $\rc_0$ is non-decreasing in $p$, and a finite $p \!\to\! 1$ limit on the right side of 
this inequality exists, the finite limit also exists on the left side, demonstrating 
\eqref{eqapp:810}.

The second claim we aim to substantiate here is that $\rc_0(\kappa,\Rer) \!<\! \infty$ implies
$\rc_0(\kappa',\Rer) \!<\! \infty$ for $\kappa' > \kappa$, but not for 
$1 \!<\! \kappa' \!<\! \kappa$. The former follows from monotonicity properties of $\rc_0$
in a manner analogous to that discussed in case of $\Rer$. To show the latter, it suffices to 
construct a possible $\aer(r,p)$ that exemplifies the corresponding behavior. One option is
\begin{equation}
    \aer(r,p) = \text{Taylor} \biggl[ \, \exp \Bigl( \frac{r}{2R_1} \Bigr) \,,\,
                                                      \floor \biggl( \frac{1}{1-p} \biggr)  \,\biggr] \;
                     \exp \Bigl( -\frac{r}{R_1} \Bigr)
    \label{eqapp:840}                                       
\end{equation}
where $R_1$ is a constant and $\text{Taylor}(f(x), n)$ the Taylor series of $f(x)$ around 
$x\!=\!0$ up to $n$-th order. Clearly, $\lim_{p\to 1} R_0(p) = \lim_{p\to 1}R_1 = R_1$ and, 
due to the chosen $p$-deformation of the prefactor, finite $\rc_0(\kappa,\Rer)$ is only
obtained via $\kappa R_0(p) \!\ge\! 2R_1$, i.e. with $\kappa \!\ge\! 2$.

\section{More on the Ultraviolet Cutoff Removal}
\label{app:C}

Here we elaborate on the most general form of Definition 3, specifying exponential insensitivity 
at fixed $p$ in the continuum, i.e. on the procedure of ultraviolet cutoff removal. 
This generalization requires the existence of $O_x^r$, $a_0 \!>\! 0$, $\kappa \!>\! 1$
and $\Rer \!>\! 0$ such that 
\begin{equation}
   \sup \,\{\, R_0(p,a) \,\mid\, 0 \!<\! a \!<\! a_0 \,\} < \infty  \qquad\ \qquad
   \sup \,\{\, \rc_0(\kappa,\Rer,p,a) \,\mid\, 0 \!<\! a \!<\! a_0 \,\}  <  \infty 
   \label{eqapp:700}
\end{equation}
Clearly, when $O_x(U)$ satisfies Definition 3 at given $p$, it also satisfies the above, but
not vice versa. Indeed, it is possible to construct functions satisfying bounds \eqref{eqapp:700},
for which $a \!\to\! 0$ limits \eqref{eq:160} do not exist. While such $a$-dependences are not
likely to occur for operators of practical interest, specifying a complete framework for characterizing 
exponential insensitivity to distant fields is clearly of conceptual interest at the very least. 

The question that remains to be answered in this regard is how to assign the continuum effective 
range and the threshold distance to operator-approximant combinations that behave in such 
non-standard way. Indeed, in cases covered by Definition 3, these are the scales $R_0^c(p)$ and 
$\rc_0(\kappa,\Rer,p)$ specified by the limiting procedure which is ill-defined in this situation. 
For the appropriate generalization, we denote the bounds (suprema) of \eqref{eqapp:700} as  
$R_b(p,a_0)$ and $\rc_b(\kappa,\Rer,p,a_0)$ respectively, and define
\begin{equation}
   R_b^c(p)  \,\equiv\,   \lim_{a_0\to 0} \, R_b(p,a_0) 
   \qquad\qquad
   \rc_b^c(\kappa,\Rer,p)  \,\equiv\,   \lim_{a_0\to 0} \, \rc_b(\kappa,\Rer,p,a_0) 
   \label{eqapp:720}
\end{equation}
Note that, since $R_b(p,a_0)$ and $\rc_b(\kappa,\Rer,p,a_0)$ are non-decreasing in $a_0$, 
the existence of these limits is guaranteed by \eqref{eqapp:700}. Clearly, the meaning of
these characteristics is that they specify the optimal exponential bounds \eqref{eq:r30} in 
the continuum.

Finally, we point out that the finiteness of  $\rc_b^c(\kappa,\Rer,p)$ implies finiteness of 
$\rc_b^c(\kappa',\Rer',p)$ for all $\Rer' \!>\! 0$ and $\kappa' \! \ge \! \kappa$. The proof
involves a straightforward modification of steps followed in Appendix~\ref{app:B} to 
the generalized situation described here.

\section{The General Case}
\label{app:A}

This Appendix describes exponential insensitivity to distant fields (and its boundary counterpart) 
of non-ultralocal composite fields $O_x(U)$ in general lattice setting. Here the position variables 
$x_\mu \!=\! a n_\mu$ label sites of an infinite hypercubic lattice ${\cal H}_\infty$  in $d$ 
dimensions, Eq.$\,$\eqref{eq:32}. Since non-ultralocality involves arbitrarily large lattice 
distances, the notion of ``infinite volume" is implicitly present, with 
$x \in {\cal L}_\infty \subset {\cal H}_\infty$  denoting a connected subset of points comprising 
this infinite system. Note that ${\cal L}_\infty$ can involve arbitrary boundaries. For example, one 
could be interested in $O_x$ defined at some point
$x \in {\cal L}_\infty = \{\, y \,\mid\, y_\mu \ge 0, \forall \mu \,\}$.

The definition of $O_x$ on ${\cal L}_\infty$ proceeds via infrared regularization. The procedure 
specifies sequence of nested finite lattices (finite connected sets of points) ${\cal L}^x_k$ 
\begin{equation}
    x \in {\cal L}^x_1 \,\subset\,  {\cal L}^x_2  \,\subset\,  {\cal L}^x_3 \ldots  \,\subset\,  {\cal L}_\infty
    \label{eqapp:10}
\end{equation}
such that every point in ${\cal L}_\infty$ also belongs to ${\cal L}^x_k$ for sufficiently large values
of $k$, i.e. ${\cal L}_\infty$ is the ``infinite-volume limit" of ${\cal L}^x_k$. For every $k$ there 
is an action $S_k\!=\!S_k(U)$ defining a theory on ${\cal L}^x_k$, and thus probability distribution 
for the associated gauge fields, as well as the prescription $O_{x,k}\!=\!O_{x,k}(U)$ for the operator 
in question. Here ``theory" is any model where $S_k$ is guaranteed to depend on all link variables 
connecting a pair of nearest neighbors from ${\cal L}^x_k$, and to not depend on link variables with 
both endpoints outside ${\cal L}^x_k$. Operator $O_{x,k}$ depends on the same set of variables 
at most. If $N_k$ is the number of lattice points in ${\cal L}^x_k$, the effective length scale $L_k$ and 
the associated discrete variables $L$ and $\ell$ are specified by
\begin{equation}
    L_k \equiv a \, (N_k)^{1/d}     \qquad\qquad   L, \ell \in \{ \, L_k \, \mid \, k=1,2,\ldots  \, \}
    \label{eqapp:20}
\end{equation}
The defining sequence $( {\cal L}^x_k, S_k, O_{x,k} )$ can then be equivalently written e.g. as 
$( {\cal L}^x_L, S_L, O_{x,L} )$.

\subsection{Exponential Insensitivity to Distant Fields}

The finite-volume setup for studying exponential insensitivity involves fixing $L$, which is treated 
as ``large", and examining possible approximations $O_x^r(U)$ of $O_x(U) \!\equiv\! O_{x,L}(U)$, 
that only involve gauge fields $U^{x,r}$ within increasing hypercubic distance $r$ away from $x$. 
In this general case, the collection of link variables $U_{y,\mu}$ included in $U^{x,r}$ is defined as
\begin{equation}
    U^{x,r} \equiv \{\; U_{y,\mu}  \; \mid\; \; 
    y,y\!+\!\muhat \, \in \,  {\cal L}^x_L \intersection {\cal H}^{x,r}  \;\}
    \label{eqapp:30}
\end{equation}
i.e. $U^{x,r}$ doesn't include any links that ``dangle" with respect to 
${\cal L}^x_L \intersection {\cal H}^{x,r}$. Note that even though $L$ is fixed, $U^{x,r}$ is formally 
defined for arbitrarily large $r$, and $U^{x,r} \subseteq U$ for any $r$. 

With the above specifics in place, one can now proceed to investigate exponential insensitivity 
as described in Sec.\ref{sec:insensitivity}. In other words, for any approximant $O_x^r(U)=O_x^r(U^{x,r})$ 
one can compute statistically regularized error function $\delta(r,p,L)$ and its infinite-volume limit 
$\delta(r,p)$. Various degrees of exponential insensitivity are then uniquely defined depending 
on the existence of an approximant with required exponential behaviors.

\subsection{Boundary Approximants}

As an intermediate step toward boundary insensitivity to distant fields, we first construct 
specific computable approximants of $O_x(U)$ that are interesting in their own right. Given 
a configuration $U\equiv U_L$ on ${\cal L}^x_L$, a sequence of nested configurations
\begin{equation}
     U_{L_1} \,\subset\,  U_{L_2}  \,\subset\,  \ldots U_{\ell} \ldots  \,\subset\,  U
    \label{eqapp:40}
\end{equation}
is defined by restricting $U$ to a valid field $U_\ell$ on ${\cal L}^x_\ell$,  i.e. $\ell < L$.  Note that 
a precise link content of $U_\ell$ depends not only on ${\cal L}^x_\ell$ but also on boundary 
conditions used in specifying $S_\ell$. This setup provides for a sequence of 
{\em boundary approximants}
\begin{equation}
     O_x(U)  \;\longrightarrow\; O_x^\ell(U)   \,\equiv\, O_{x,\ell}(U_\ell) \;, \qquad   \ell < L
     \tag{\ref{eq:290}}    
\end{equation}
depending only on variables contained in $U_\ell$. Here ``boundary" refers to the boundary of 
${\cal L}^x_\ell$ in ${\cal L}^x_L$, artificially invoked by restricting ${\cal L}^x_L$ to ${\cal L}^x_\ell$ for 
this purpose. The existence and computability of $O_x^\ell(U)$ follow from the very definition of $O_x$
and its computability.  

For physically relevant operator it is usually useful if its transformation properties are matched 
by those of the approximant.  However, gauge covariance of $O_x(U)$, if any, will not automatically 
transfer to $O_x^\ell(U)$ if $U_\ell$ contains dangling links with respect to ${\cal L}^x_\ell$.
Indeed, for covariantly defined $O_x(U)$, each operator $O_{x,\ell}(U_\ell)$ is covariant in theory
$(\, {\cal L}^x_\ell, S_\ell \,)$. As such, it only involves a combination of closed loops on ${\cal L}^x_\ell$, 
and dangling links can participate in such loops via periodicity. However, loops with danglers 
generically translate into open line segments on ${\cal L}^x_L$, and spoil gauge 
covariance of $O^\ell_x(U)$ in theory $(\, {\cal L}^x_L, S_L \,)$.

Given the importance of gauge covariance, we formulate the version of boundary approximant
that automatically retains this feature. To that end, consider the partition of $U_\ell$ 
into non-dangling ($U_\ell^{nd}$) and dangling ($U_\ell^d$) subsets of links
\begin{equation}
     U_{\ell} \,=\,  U_\ell^{nd}  \, \union \,  U_\ell^d    \qquad\qquad
     U_\ell^{nd} \equiv \{\, U_{y,\mu}  \,\mid\,  y,y\!+\!\mu \, \in \, {\cal L}^x_\ell  \,\} 
      \label{eqapp:50}
\end{equation}
If a given link is set to ${\mathbf 0}$ ($3\times 3$ matrix of zero elements), then any Wilson loop 
containing it becomes ${\mathbf 0}$ as well. We can thus prevent non-covariant terms from 
occurring in the boundary approximant by applying the replacement
\begin{equation}
    U_\ell^d  \;\longrightarrow\; \bar{U}_\ell^d  \;=\;    
    \{\, U_{y,\mu}  \rightarrow \mathbf{0} \;\mid\; U_{y,\mu} \in U_\ell^d \;\}
     \label{eqapp:52}
\end{equation} 
in order to construct a modified ``configuration" $U_\ell \rightarrow U_\ell^c$ on ${\cal L}^x_\ell$ via
\begin{equation}
     U \quad\longrightarrow\quad 
     U_\ell \,=\,  U_\ell^{nd}  \, \union \,  U_\ell^d    \quad \longrightarrow\quad
     U_{\ell}^{c} \,\equiv\,  U_\ell^{nd}  \, \union \,  \bar{U}_\ell^d  
    \label{eqapp:54}
\end{equation}
and defining the {\em covariant form} of the boundary approximant as
\begin{equation}
     O_x(U)  \;\longrightarrow\; O_x^\ell (U)  \,\equiv\, O_{x,\ell}(U_\ell^c)  \;, \qquad   \ell < L
     \tag{\ref{eq:290}a} 
     \label{eq:290a}
\end{equation}
Note that, while the standard form \eqref{eq:290} of boundary approximant is apriori well-defined, the covariant 
form \eqref{eq:290a} may not be since ${\mathbf 0} \!\notin\! SU(3)$. In particular, $O_{x,\ell}$ could become 
singular as  generic $U_\ell$ is deformed to $U_\ell^c$. Such operators can be constructed but are highly 
contrived. 

\subsection{Hypercubic Boundary Approximants and Insensitivity}

We now use the idea of boundary approximant to arrive at the notion of boundary exponential insensitivity 
to distant fields in this general setting. The basic step is to introduce distance variable $r$ so that gauge 
fields in $U_\ell$ can be restricted accordingly.  Thus, we define function $\ell=\ell(r)$ as the smallest $\ell$ 
for which ${\cal L}^x_\ell$ covers all points of ${\cal L}^x_L$ belonging to ${\cal H}^{x,r}$, namely
\begin{equation}
    \ell(r)  \,\equiv\, \min \, \{\; \ell \; \mid\; 
    {\cal L}^x_\ell \intersection {\cal H}^{x,r} =  {\cal L}^x_L \intersection {\cal H}^{x,r} \;\}
     \label{eqapp:60}
\end{equation}
This guarantees that $U^{x,r}$ defined by \eqref{eqapp:30} is a subset of $U_{\ell(r)}$, thus
inducing its partition
\begin{equation}
     U_{\ell(r)} \,=\,  U^{x,r}  \, \union \,  U_{\ell(r)}^b    \qquad\qquad
     U_{\ell(r)}^b \equiv U_{\ell(r)} \smallsetminus U^{x,r}
    \label{eqapp:70}
\end{equation}
If non-empty, we treat entire $U_{\ell(r)}^b$ as a ``boundary" in that its links will be frozen to values 
independent of underlying $U$. Each choice $\bar{U}_{\ell(r)}^b$ of this fixing entails the assignement 
\begin{equation}
     U \quad\longrightarrow\quad 
     U_{\ell(r)} \,=\,  U^{x,r}  \, \union \,  U_{\ell(r)}^b    \quad \longrightarrow\quad
     U_{\ell(r)}^{h} \,\equiv\,  U^{x,r}  \, \union \,  \bar{U}_{\ell(r)}^b  
    \label{eqapp:80}
\end{equation}
where the field $U_{\ell(r)}^h$ on ${\cal L}^x_{\ell(r)}$ depends on $U^{x,r}$ only, and 
the {\em hypercubic boundary approximant} 
\begin{equation}
      O_x(U)  \;\longrightarrow\;  O_x^r(U) \,\equiv\, O_{x,\ell(r)} \bigl( U_{\ell(r)}^h \bigr)  
      \,=\, O_x^r(U^{x,r}) 
      \tag{\ref{eq:340}}  
\end{equation}

If there is a systematic choice of boundary values such that $\delta(p,r)=\lim_{L\to \infty}\delta(p,r,L)$
for the associated $O_x^r(U)$ exhibits exponential behaviors described in Sec.\ref{sec:insensitivity}, 
then the corresponding types of {\em boundary insensitivity} to distant fields occur. 
Here we wish to highlight two practical choices (constant values) to test this property.

\medskip
\noindent {\bf (i) Universal Form}:   
    $\qquad\quad U_{\ell(r)}^b  \;\longrightarrow\; \bar{U}_{\ell(r)}^b  \;=\;    
    \{\, U_{y,\mu}  \rightarrow \mathbf{1} \;\mid\; U_{y,\mu} \in U_{\ell(r)}^b \;\}$
\medskip

\noindent The associated hypercubic boundary approximant $O_x^r(U)$ is always well-defined and
computable, but gauge covariance of $O_x(U)$, if any, is not automatically inherited.

\medskip
\noindent {\bf (ii) Covariant Form}:   
    $\qquad\;\; U_{\ell(r)}^b  \;\longrightarrow\; \bar{U}_{\ell(r)}^b  \;=\;    
    \{\, U_{y,\mu}  \rightarrow \mathbf{0} \;\mid\; U_{y,\mu} \in U_{\ell(r)}^b \;\}$
\medskip

\noindent This hypercubic boundary approximant may be ill-defined for some artificial operators since 
${\mathbf 0} \notin \text{SU(3)}$, but gauge covariance of $O_x(U)$, if any, transfers to $O_x^r(U)$. 
Whenever it exists, the covarint form is computable, and is taken to be the default approximant of 
$O_x(U)$.
\smallskip

This Appendix specifies the framework for studying exponential insensitivity to distant fields in arbitrary 
lattice hypercubic geometry. As such, it can be used to define the corresponding continuum notions, as 
described in Sec.$\,$\ref{ssec:continuum}. It should be remarked though that the position of $x$ within 
${\cal L}^x_L$ (and ${\cal L}_\infty$) has to maintain its continuum geometric meaning throughout 
the process of ultraviolet cutoff removal. This is automatic in the symmetric setup discussed 
in the body of the paper, but requires some care in general case. 

Finally, the notion of boundary insensitivity can be used to define {\em boundary locality} following 
the formalism described in Sec.$\,$\ref{ssec:locality}

\section{Overlap Dirac Operator}
\label{app:E}

The overlap Dirac operator used in the study of Sec.~\ref{sec:overlapops} is based on massless
Wilson-Dirac matrix in four space-time dimensions, namely
\begin{equation}
      (D_W)_{x,y} \;=\; 4\,\delta_{x,y} \,-\, 
      {1\over 2}\sum_{\mu=1}^4\, \Bigl[ \, (1-\gmu)\,U_{x,\mu}\,\delta_{x+\mu,y} \,+\,
                         (1+\gmu)\, U_{x-\mu,\mu}^+\, \delta_{x-\mu,y}  \,\Bigr]
    \label{eqapp:500}                         
\end{equation} 
The associated 1-parameter family of massless overlap Dirac operators~\cite{Neu98BA} is then defined as
\begin{equation}
      D(\rho) \;=\; \rho\, \Biggl[\; 1 \,+\, (D_W-\rho) \, {1 \over \sqrt{(D_W-\rho)^+(D_W-\rho)}}\;\Biggr]
      \quad ,\quad 0 < \rho < 2
       \label{eqapp:510}                                   
\end{equation}
Note that the interval of allowed values for negative mass parameter $\rho$ is dictated by spectral 
properties of $D_W$, and the space-time range of $D(\rho)$ has been shown to depend on it~\cite{Her98A}.

\end{appendix}

\end{document}
\bye